%% file: Paper.tex
\newcommand*{\ATLASLATEXPATH}{latex/}
\documentclass[cernpreprint,texlive=2011,txfonts,UKenglish]{\ATLASLATEXPATH atlasdoc}

\usepackage[full]{\ATLASLATEXPATH atlaspackage}
\usepackage{\ATLASLATEXPATH atlasbiblatex}

\usepackage{\ATLASLATEXPATH atlascontribute}

\usepackage{\ATLASLATEXPATH atlasphysics}
\DeclareFieldFormat[phdthesis]{citetitle}{#1}
\DeclareFieldFormat[phdthesis]{title}{#1}

\graphicspath{{logos/}{figures/}}

\usepackage{enumitem}
\usepackage{epstopdf}

\def\ipb {\mbox {$~{\rm pb}^{-1}$}}
\def\ifb {\mbox {$~{\rm fb}^{-1}$}}
\def\lumiA{\mbox{$2.1\ifb$}}

\def\lumiB{\mbox{$11.4\ifb$}}

\def\pt{\mbox{$p_{\text{T}}$}}

\newcommand{\psiprime}{\ensuremath{\psi(2\mathrm{S})}}

\newcommand{\apsi}{\ensuremath{\psi}}

\hypersetup{pdftitle={Measurement of the differential cross-sections of prompt and non-prompt production of $\jpsi$ and $\psiprime$ in $pp$ collisions at $\sqrt{s} = 7$ and $8$~TeV with the ATLAS detector},pdfauthor={The ATLAS Collaboration}}
\input{Paper-metadata}

\AtlasTitle{Measurement of the differential cross-sections of prompt and non-prompt production of $\jpsi$ and $\psiprime$ in $pp$ collisions at $\sqrt{s} = 7$ and $8$~\TeV\ with the ATLAS detector}
\AtlasNote{ATL-COM-PHYS-2015-161}
\AtlasJournal{Eur.\ Phys.\ J.\ C.}
\PreprintIdNumber{CERN-PH-EP-2015-292}
\AtlasAbstract{The production rates of prompt and non-prompt \jpsi\ and \psiprime\ mesons in their dimuon decay modes are measured using \lumiA\ and \lumiB\ of data collected 
with the ATLAS experiment at the Large Hadron Collider, in proton--proton collisions at
$\sqrt{s}=7$ and $8$~\TeV\ respectively. Production cross-sections for prompt as well as non-prompt sources, ratios of \psiprime\ to \jpsi\ production, 
and the fractions of non-prompt production for \jpsi\ and \psiprime\
are measured as a function of meson transverse momentum and rapidity. 
The measurements are compared to theoretical predictions.}

\begin{document}

\maketitle

\tableofcontents

\clearpage

\input{Introduction}

\input{ATLAS}

\input{DataMC}

\input{Methodology}

\input{Systematics}

\input{Results}

\FloatBarrier

\input{Conclusions}

\section*{Acknowledgements}
\input{Acknowledgements}

\clearpage
\printbibliography
\clearpage
\appendix
\part*{Appendix}
\addcontentsline{toc}{part}{Appendix}
\input{SpinAlignmentFactors}

\newpage 
\input{atlas_authlist}

\end{document}

%% file: Paper-metadata.tex
\AtlasTitle{AtlasTitle: Bare bones ATLAS document}

\author{The ATLAS Collaboration}

\AtlasRefCode{GROUP-2014-XX}

\AtlasAbstract{%
  This is a bare bones ATLAS document. Put the abstract for the document here.
}

%% file: Introduction.tex
\section{Introduction}
\label{intro}

Measurements of heavy quark--antiquark bound states (quarkonia) production processes provide 
an insight into the nature of quantum
chromodynamics (QCD) close to the boundary between the perturbative and non-perturbative regimes. 
More than forty years since the discovery of the $\jpsi$, the investigation of
hidden heavy-flavour production in hadronic collisions 
still presents significant challenges to both theory and experiment.

In high-energy hadronic collisions, charmonium states can be produced either directly by 
short-lived QCD sources (``prompt'' production), or by long-lived sources in the decay chains of beauty hadrons (``non-prompt'' production). 
These can be separated experimentally using the
distance between the proton--proton primary interaction and the decay vertex of the quarkonium state.
While {\em Fixed-Order with Next-to-Leading-Log} (FONLL) calculations~\cite{FONLL_2001,Cacciari:2012ny}, made within the framework of perturbative QCD, have
been quite successful in describing non-prompt production of various quarkonium states, a satisfactory understanding of the prompt production mechanisms is still to be achieved.

The $\psiprime$ meson is the only vector charmonium state that 
is produced with no significant  contributions from decays of higher-mass quarkonia, 
referred to as feed-down contributions. This
provides a unique opportunity to study production mechanisms specific to
$J^{PC}=1^{--}$ states~\cite{CDFjpsianomaly1,Abulencia:2007us,Abelev:2014qha,Chatrchyan:2011kc,Chatrchyan:2013cla,Aaij:2012ag,Aad:2014fpa,Aaij:2013jxj,Khachatryan:2015rra,Aaij:2013yaa}.
Measurements of the production of $J^{++}$ states with $J=0, 1, 2$,
\cite{CDFjpsianomaly2,ATLAS:2014ala,Chatrchyan:2012ub,Aaij:2013yaa,Aaij:2013dja,LHCb:2012ac},
strongly coupled to the two-gluon channel, allow similar studies in the $CP$-even sector, complementary to the $CP$-odd vector sector. 
Production of $\jpsi$ mesons
\cite{Aad:2011sp,Abulencia:2007us,Aaij:2011jh,Chatrchyan:2011kc,Abelev:2012gx,Chatrchyan:2013cla,Aaij:2013jxj,Abelev:2014qha,CDFjpsianomaly3,CDFjpsianomaly1,D0jpsi1,D0jpsi2,Khachatryan:2015rra,Aad:2014fpa,CDFjpsianomaly2,LHCb:2012af}
arises from a mixture of different sources, receiving contributions from the production of $1^{--}$ and $J^{++}$ states in comparable amounts.

Early attempts to describe the formation of charmonium
\cite{CSM1,CSM2,CSM3,CSM4,CSM5,CSM7,CEM1,CEM2}
using leading-order perturbative QCD 
gave rise to a variety of models, none of which could explain the
large production cross-sections measured at the Tevatron
\cite{CDFjpsianomaly1,CDFjpsianomaly2,CDFjpsianomaly3,D0jpsi1,D0jpsi2}.
Within the colour-singlet model (CSM) \cite{Lansberg:2008gk}, next-to-next-to-leading-order (NNLO) contributions to the hadronic production of S-wave
quarkonia were calculated without introducing any new phenomenological parameters. However,
technical difficulties have so far made it impossible to perform the full NNLO calculation, or to 
extend those calculations to the P-wave states. 
So it is not entirely surprising that the predictions of the model underestimate the experimental data for inclusive production of
\jpsi\ and $\Upsilon$ states, where the feed-down is significant, but offer a better description  for \psiprime\ production \cite{Aad:2011sp,Aad2012dlq}.

Non-relativistic QCD (NRQCD) calculations that include colour-octet (CO) contributions 
\cite{Bodwin:1994jh} introduce a number of phenomenological parameters --- long-distance matrix elements
(LDMEs) --- which are determined from fits to the experimental data, and can hence
describe the cross-sections and differential spectra satisfactorily 
{\cite{CO_LDME1}}. 
However, the attempts to describe the polarization of S-wave quarkonium states using this approach
have not been so successful~\cite{Gong:2012ug},
prompting a suggestion~\cite{Faccioli:2014cqa} that a more coherent approach is needed for the treatment of polarization
within the QCD-motivated models of quarkonium production. 

Neither the CSM nor the NRQCD model gives a satisfactory explanation for the measurement of prompt
\jpsi\ production in association with the $W$~\cite{Aad:2014rua} and $Z$~\cite{Aad:2014kba} bosons: in both cases,
the measured differential cross-section is larger than theoretical expectations~\cite{Li:2010hc,Lansberg:2013wva,Gong:2012ah, Mao:2011kf}.
It is therefore  important to broaden the scope of comparisons 
between theory and experiment by providing a variety of experimental information
about quarkonium production across a wider kinematic range.
In this context, ATLAS has measured the inclusive differential cross-section of \jpsi\  production, with $2.3$~pb$^{-1}$ of integrated luminosity~\cite{Aad:2011sp}, 
at $\sqrt{s} = 7$~\TeV\ using the data collected in 2010, 
as well as the differential cross-sections of the production of  $\chi_c$ states~(4.5~fb$^{-1}$)~\cite{ATLAS:2014ala}, and of the \psiprime\ in its $J/\psi\pi\pi$ decay mode (2.1 fb$^{-1})$~\cite{Aad:2014fpa}, at $\sqrt{s} = 7$~\TeV\ with data collected in 2011.
The cross-section and polarization measurements 
from CDF~\cite{Abulencia:2007us},
CMS~\cite{Chatrchyan:2013cla,Khachatryan:2010yr,CMS},
LHCb~\cite{Aaij:2012ag,Aaij:2013jxj,Aaij:2013yaa,Aaij:2012asz,Aaij:2013nlm,Aaij:2014qea}  
and ALICE~\cite{Abelev:2014qha,Abelev:2011md,Aamodt:2011gj},
cover a considerable variety of charmonium production characteristics in a wide 
kinematic range (transverse momentum $\pt\leq 100$ \GeV\ and rapidities $|y|<5$), thus providing a wealth of
information for a new generation of theoretical models.

This paper presents a precise measurement of
\jpsi\ and  \psiprime\ production in the dimuon decay mode, both at $\sqrt{s} = 7$~TeV
and at $\sqrt{s} = 8$~TeV.
It is presented as a double-differential measurement in transverse momentum and rapidity of the quarkonium state, 
separated into prompt and non-prompt contributions,
covering a range of transverse momenta $8 < \pt\leq 110$ \GeV\ and rapidities $|y|<2.0$. 
The ratios of $\psiprime$ to $\jpsi$  cross-sections for prompt and non-prompt processes are also reported, as well as the 
non-prompt fractions of $\jpsi$ and $\psiprime$.

%% file: ATLAS.tex
\section{The ATLAS detector}
\label{sec:atlas}

The ATLAS experiment~\cite{ATLASdetector} is a general-purpose detector consisting of an inner tracker, a calorimeter and a muon spectrometer. 
The inner detector (ID) directly surrounds the interaction point; 
it consists of a silicon pixel detector, a semiconductor tracker and a transition radiation tracker, and is embedded in an axial~$2$ T magnetic field.
The ID covers the pseudorapidity\footnote{ATLAS uses a right-handed coordinate system with its origin at the nominal interaction point (IP) 
in the centre of the detector and the $z$-axis along the beam pipe.
The $x$-axis points from the IP to the centre of the LHC ring, and the $y$-axis points upward. Cylindrical 
coordinates $(r,\phi)$ are used in the transverse plane,
$\phi$ being the azimuthal angle around the beam pipe. The pseudorapidity $\eta$ is defined in terms of the 
polar angle $\theta$ as $\eta=-\ln\tan(\theta/2)$ and the
transverse momentum $p_{\rm T}$ is defined as $p_{\rm T}=p\sin\theta$. The rapidity is defined as 
$y=0.5\ln\left[\left( E + p_z \right)/ \left( E - p_z \right)\right]$,
where $E$ and $p_z$ refer to energy and longitudinal momentum, respectively. The $\eta$--$\phi$ distance 
between two particles is defined as
 $\Delta R=\sqrt{(\Delta\eta)^2 + (\Delta\phi)^2}$.}
range $|\eta| = $ 2.5 and is enclosed by a calorimeter system containing
electromagnetic and hadronic sections.
The calorimeter is surrounded by a large muon spectrometer (MS) in a toroidal magnet system.
The MS consists of monitored drift tubes and cathode strip chambers, designed to provide
precise position measurements in the bending plane in the range $|\eta| <$ 2.7.
Momentum measurements in the muon spectrometer are based on track segments formed in at least two of the three precision chamber planes.

The ATLAS trigger system \cite{ATLAS:trig} is separated into three levels: the hardware-based Level-1 trigger
and the two-stage High Level Trigger (HLT), comprising the Level-2 trigger and Event
Filter, which reduce the 20~MHz proton--proton collision rate to several-hundred Hz of events of interest for data recording to mass storage. 
At Level-1, the muon trigger searches for patterns of hits satisfying different transverse momentum thresholds with a coarse 
position resolution but a fast response time using resistive-plate chambers and thin-gap chambers in the ranges $|\eta| <$ 1.05 and $1.05 <|\eta| < 2.4$, respectively.
Around these Level-1 hit patterns ``Regions-of-Interest'' (RoI) are defined that
serve as seeds for the HLT muon reconstruction. 
The HLT uses dedicated algorithms to incorporate information from both the 
MS and the ID, achieving position and momentum resolution close to that provided by the offline muon reconstruction.

%% file: DataMC.tex
\section{Candidate selection}
\label{sec:data}

The analysis is based on data recorded at the LHC in 2011 and 2012 during proton--proton collisions at centre-of-mass energies of $7$~\TeV\ and $8$~\TeV, respectively.
This data sample corresponds to a total integrated luminosity of \lumiA\ and \lumiB\ for  $7$~\TeV\ data and $8$~\TeV\ data, respectively.

Events were selected  using a trigger requiring two oppositely charged muon candidates, each passing the requirement $\pt>4$~\GeV. 
The muons are constrained to originate from a common vertex, which is fitted with the track parameter uncertainties taken into account. 
The fit is required to satisfy $\chi^2 < 20$ for the one degree of freedom.

For $7$~\TeV\ data, the Level-1 trigger required only spatial coincidences in the MS~\cite{ATLAS:2010kba}. For $8$~\TeV\ data, a $4$~\GeV\ muon $\pt$ threshold was also applied at Level-1, which reduced the trigger efficiency for low-$\pt$ muons.

The offline analysis requires events to have at least two muons, identified by the muon spectrometer and with matching tracks reconstructed in the ID~\cite{muons}.
Due to the ID acceptance, muon reconstruction is possible only for $|\eta| <$ 2.5. 
The selected muons are further restricted to $|\eta| <$ 2.3 to ensure high-quality tracking and triggering, and to reduce the contribution from misidentified muons.
For the momenta of interest in this analysis (corresponding to muons with a transverse momentum of at most $O(100)$ \GeV), measurements of the
muons are degraded by multiple scattering within the MS and so only the ID tracking
information is considered. 
To ensure accurate ID measurements, each muon track must fulfil muon reconstruction and selection requirements~\cite{muons}.
The pairs of muon candidates satisfying these quality criteria are required to have opposite charges.

In order to allow an accurate correction for trigger inefficiencies, each reconstructed muon candidate is required to match a trigger-identified muon
candidate within a cone of $\Delta R = \sqrt{(\Delta\eta)^2 + (\Delta\phi)^2}=0.01$.
Dimuon candidates are obtained from muon pairs, constrained to originate from a common vertex using ID track parameters and uncertainties, with a 
requirement of $\chi^2 < 20$ of the vertex fit for the one degree of freedom.
All dimuon candidates with an invariant mass
within $2.6 < m(\mu\mu) < 4.0$ \GeV\ and within the kinematic range $\pt(\mu\mu) > 8$ \GeV, $|y(\mu\mu)| < 2.0$ are retained for the analysis.
If multiple candidates are found in an event (occurring in approximately $10^{-6}$ of selected events), all candidates are retained.
The properties of the dimuon system, such as invariant mass $m(\mu\mu)$, transverse momentum $\pt(\mu\mu)$, and rapidity $|y(\mu\mu)|$ are determined from the result of the vertex fit.

%% file: Methodology.tex
\section{Methodology}
\label{sec:method}

The measurements are performed in intervals of dimuon $\pt$ and absolute value of the rapidity ($|y|$). 
The term ``prompt'' refers to the $\jpsi$ or $\psiprime$ states --- hereafter called $\apsi$ to refer to either --- are produced 
from short-lived QCD decays, including feed-down from other charmonium states as long as they are also produced
from short-lived sources. If the decay chain producing a $\apsi$ state includes long-lived particles such as
$b$-hadrons, then such $\apsi$ mesons are labelled as ``non-prompt''. Using a simultaneous fit to the invariant mass of the dimuon and its ``pseudo-proper decay time'' 
(described below), prompt and non-prompt signal and background contributions can be extracted from the data.

The probability for the decay of a particle as a function of proper decay time $t$ follows an exponential distribution, 
$p(t) =  1/\tau_{B}\cdot e^{-t/\tau_{B}}$ where $\tau_{B}$ is the mean lifetime of the particle. For each decay, the proper decay time 
can be calculated as $t = L m/p$,
where $L$ is the distance between the particle production and decay vertices, $p$ is the momentum of the particle, and $m$ is its invariant mass.
As the reconstruction of non-prompt $\apsi$ mesons, such as $b$-hadrons, does not fully describe the properties of the parent,
the transverse momentum of the 
dimuon system and the reconstructed dimuon invariant mass are used to construct the
``pseudo-proper decay time'', $\tau = L_{xy} m(\mu\mu)/\pt(\mu\mu)$, where $L_{xy} \equiv \vec{L} \cdot \vec{\pt}(\mu\mu)/\pt(\mu\mu)$ 
is the signed projection of the distance of the dimuon decay vertex 
from the primary vertex, $\vec{L}$, onto its transverse momentum, $\vec{\pt}(\mu\mu)$.
This is a good approximation of using the parent $b$-hadron information when the $\psi$ and parent momenta are closely aligned, which is the case for the values of $\psi$
transverse momenta considered here, and  $\tau$
therefore can be used to distinguish statistically between the non-prompt and prompt processes (in which the latter are assumed to decay with vanishingly small lifetime).
If the event contains multiple primary vertices~\cite{ATLASdetector}, the primary vertex closest in $z$ to the dimuon decay vertex is selected.
The effect of selecting an incorrect vertex has been shown~\cite{BJpsiPhi} to have a negligible impact on the extraction of prompt and non-prompt contributions.
If any of the muons in the dimuon candidate contributes to the construction of the primary vertex, the corresponding tracks are removed and the vertex is refitted.

\subsection{Double differential cross-section determination}
\label{sec:s:diff_xSecDet}

The double differential dimuon prompt and non-prompt production cross-sections times branching ratio 
are measured separately for \jpsi\ and \psiprime\ mesons according to the equations:

\begin{equation}
\frac{\mathrm{d}^2\sigma(pp \rightarrow \psi)}{\mathrm{d}\pt\mathrm{d}y} \times \mathcal{B} (\psi \rightarrow \mu^+\mu^- ) = \frac{N_{\psi}^{\mathrm{p}}}{\Delta \pt \Delta y \times \int\mathcal{L} \mathrm{d}t},
\label{equ:xSecP}
\end{equation}
\begin{equation}
\frac{\mathrm{d}^2\sigma(pp \rightarrow b\bar{b} \rightarrow \psi)}{\mathrm{d}\pt\mathrm{d}y} \times \mathcal{B} (\psi \rightarrow \mu^+\mu^- ) = \frac{N_{\psi}^{\mathrm{np}}}{\Delta \pt \Delta y \times \int\mathcal{L} \mathrm{d}t},
\label{equ:xSecNP}
\end{equation}

\noindent where $\int\mathcal{L} dt$ is the integrated luminosity, $\Delta \pt $ and $ \Delta y$ are the interval sizes in terms of dimuon transverse momentum and
rapidity, respectively, and $N_{\psi}^{\mathrm{p(np)}}$ is the number of observed prompt (non-prompt) $\psi$ mesons in
the slice under study, corrected for acceptance, trigger and reconstruction efficiencies. 
The intervals in $\Delta y$ combine the data from negative and positive rapidities.

The determination of the cross-sections proceeds in several steps. First, a weight is determined for each
selected dimuon candidate equal to the inverse of the total efficiency for each candidate. 
The total weight, $w_\mathrm{tot}$, for each dimuon candidate includes three factors: the fraction of produced $\psi \rightarrow \mu^+\mu^-$ decays with both
muons in the fiducial region $\pt (\mu) > 4$ \GeV\ and $|\eta(\mu)| <$ 2.3 (defined as acceptance, $\mathcal{A}$), the probability that a candidate
within the acceptance satisfies the offline reconstruction selection ($\epsilon_\mathrm{reco}$), and 
the probability that a reconstructed event satisfies the trigger selection 
($\epsilon_\mathrm{trig}$). The weight assigned to a given candidate when calculating the cross-sections is therefore  given by:

\begin{equation*}
w_{\mathrm{tot}}^{-1} = \mathcal{A} \cdot \epsilon_{\mathrm{reco}} \cdot \epsilon_{\mathrm{trig}}.
\end{equation*}

After the weight determination, an unbinned maximum-likelihood fit is performed to these weighted events in each ($\pt (\mu\mu), \ |y(\mu\mu)|$) interval using the 
dimuon invariant mass, $m(\mu\mu)$, and pseudo-proper decay time, $\tau(\mu\mu)$, observables. 
The fitted yields of $\jpsi \rightarrow \mu^+\mu^-$ and $\psiprime \rightarrow \mu^+\mu^-$ are determined separately for prompt and non-prompt processes. 
Finally, the differential cross-section times the  $\psi \rightarrow \mu^+\mu^-$ branching fraction is calculated for each
state by including the integrated luminosity and the $\pt$ and rapidity interval widths as shown in Eqs. (\ref{equ:xSecP}) and (\ref{equ:xSecNP}).

\subsection{Non-prompt fraction}
\label{sec:s:NPFDet}

The non-prompt fraction $f_{b}^{\psi}$ is defined as the number of non-prompt $\psi$ (produced
via the decay of a $b$-hadron) divided by the number of inclusively produced $\psi$ decaying to muon pairs after applying weighting corrections:

\begin{equation*}
f_{b}^{\psi} \equiv \frac{pp \rightarrow b + X \rightarrow \psi + X'}{pp \xrightarrow{\mathrm{Inclusive}} \psi + X'} = \frac{N^{\mathrm{np}}_{\psi}}{N^{\mathrm{np}}_{\psi} + N^{\mathrm{p}}_{\psi}}, 
\end{equation*}

\noindent where this fraction is determined separately for \jpsi\ and \psiprime. 
Determining the fraction from this ratio is advantageous since acceptance and efficiencies largely cancel and the systematic uncertainty is reduced.

\subsection{Ratio of \psiprime\ to \jpsi\ production}
\label{sec:s:PNPRatioDet}

The ratio of \psiprime\ to \jpsi\ production, in their dimuon decay modes, is defined as:

\begin{equation*}
R^{\mathrm{p(np)}} =\frac{N^{\mathrm{p(np)}} _{\psi(2\mathrm{S})}}{N^{\mathrm{p(np)}} _{J/\psi}},
\end{equation*}

\noindent where $N_{\psi}^{\mathrm{p(np)}}$ is the number of prompt (non-prompt) \jpsi\ or \psiprime\ mesons decaying into a 
muon pair in an interval of $\pt$ and $y$, corrected for selection efficiencies and acceptance.

For the ratio measurements, similarly to the non-prompt fraction, the acceptance and efficiency corrections 
largely cancel, thus allowing a more precise measurement.
The theoretical uncertainties on such ratios are also smaller, as several dependencies, such as
parton distribution functions and $b$-hadron production spectra, largely cancel in the ratio.

\subsection{Acceptance}
 
The kinematic acceptance $\mathcal{A}$ for a $\psi \rightarrow \mu^+\mu^-$ decay with $\pt$ and $y$ is given by the 
probability that both muons pass the fiducial selection ($\pt(\mu)>4$ \GeV\ and $|\eta(\mu)|<2.3$).
This is calculated using generator-level ``accept-reject'' simulations, based on the analytic formula described below. 
Detector-level corrections, such as bin migration effects due to detector resolution, 
are found to be small. They are applied to the results and are also considered as part of the systematic uncertainties. 

The acceptance $\mathcal{A}$ depends on five independent variables (the two muon momenta are constrained by the
$m(\mu\mu)$ mass condition), chosen as the $\pt$, $|y|$ and azimuthal angle $\phi$ of the $\psi$ meson in the laboratory frame,
and two angles
characterizing the $\psi \rightarrow \mu^+\mu^-$ decay, $\theta^{\star}$ and $\phi^{\star}$, described in detail in Ref. \cite{Faccioli:2010kd}.
The angle $\theta^{\star}$ is the angle between the direction of the positive-muon momentum in the $\psi$ rest frame
and the momentum of the $\psi$ in the laboratory frame, while $\phi^{\star}$ is defined as the angle between the dimuon 
production and decay planes in the laboratory frame. 
The $\psi$ production plane is defined by the momentum of the $\psi$ in the laboratory frame and the positive $z$-axis direction.
The distributions in $\theta^{\star}$ and $\phi^{\star}$ 
differ for various possible spin-alignment scenarios of the dimuon system.

The spin-alignment of the $\psi$ may vary depending on the production mechanism, which in turn affects the angular distribution of the dimuon decay.
Predictions of various theoretical models are quite contradictory, while the recent experimental measurements~\cite{Chatrchyan:2013cla} indicate that the angular dependence of $\jpsi$ and $\psiprime$
decays is consistent with being isotropic.  

The coefficients $\lambda_{\theta}, \lambda_{\phi}$ and~$\lambda_{\theta\phi}$ in
\begin{equation}
\frac{\mathrm{d}^2N}{\mathrm{d}\cos\theta^{\star}\mathrm{d}\phi^{\star}} \propto 1 + \lambda_{\theta} \cos^2\theta^{\star} + \lambda_{\phi} \sin^2\theta^{\star}\cos2\phi^{\star} +  \lambda_{\theta\phi} \sin2\theta^{\star}\cos\phi^{\star}
\label{equ:acc}
\end{equation}
\noindent are related to the spin-density matrix elements of the dimuon spin wave function. 

Since the polarization of the $\psi$ state may affect acceptance, seven extreme cases that lead to the largest possible variations of 
acceptance within the phase space of this measurement are identified.
These cases, described in Table~\ref{tab:spin}, are used to define a range in which the results may vary under any physically allowed spin-alignment assumptions.
The same technique has also been used in other measurements~\cite{Aad:2014fpa,ATLAS:2014ala,Aad2012dlq}.
This analysis adopts the isotropic distribution in both $\cos\theta^{\star}$ and $\phi^{\star}$ as  nominal,
and the variation of the results for a  number of extreme spin-alignment scenarios is studied and presented as sets of correction factors, detailed further in Appendix~\ref{sec:spincorrection}.
 
\begin{table}[htbp]
\begin{center}
\caption{Values of angular coefficients describing the considered spin-alignment scenarios.}
\vspace{2mm}
\begin{tabular}[h]{r|ccc}
\hline\hline
      & \multicolumn{3}{c}{Angular coefficients} \\ 
      & $\lambda_{\theta}$ & $\lambda_{\phi}$ & $\lambda_{\theta\phi}$ \\ \hline
Isotropic {\em (central value)}            & $0$ & $0$ & $0$ \\
Longitudinal         & $-1$ & $0$ & $0$ \\
Transverse positive  & $+1$ & $+1$ & $0$ \\
Transverse zero      & $+1$                & $0$ & $0$ \\
Transverse negative  & $+1$            & $-1$ & $0$ \\
Off-($\lambda_{\theta}$--$\lambda_{\phi}$)-plane positive   & $0$ & $0$ & $+0.5$ \\
Off-($\lambda_{\theta}$--$\lambda_{\phi}$)-plane negative   & $0$ & $0$ & $-0.5$ \\
\hline\hline
\end{tabular}
\label{tab:spin}
\end{center}
\end{table}

For each of the two mass-points (corresponding to the \jpsi\ and \psiprime\ masses), two-dimensional maps are produced
as a function of dimuon $\pt(\mu\mu)$ and $|y(\mu\mu)|$ for the set of spin-alignment hypotheses. 
Each point on the map is determined from a uniform sampling over~$\phi^{\star}$ and~$\cos\theta^{\star}$, 
accepting those trials that pass the fiducial selections. To account for various spin-alignment scenarios, all trials are weighted according to  Eq.~\ref{equ:acc}.
Acceptance maps are defined within the range $8 < \pt(\mu\mu) < 110$~\GeV and $|y(\mu\mu)| < 2.0$, corresponding to the data considered in the analysis. 
The map is defined by 100 slices in $|y(\mu\mu)|$ and 4400 in $\pt(\mu\mu)$, using 200k trials for each point, resulting in sufficiently high precision that the statistical uncertainty can be neglected.
Due to the contributions of background, and the detector resolution of the signal, the acceptance for each candidate is determined from a linear interpolation of the two maps, which are generated for the \jpsi\ and \psiprime\ known masses, as a function of the reconstructed mass $m(\mu\mu)$.

\begin{figure}[ht]
  \begin{center}
    \includegraphics[width=0.8\textwidth]{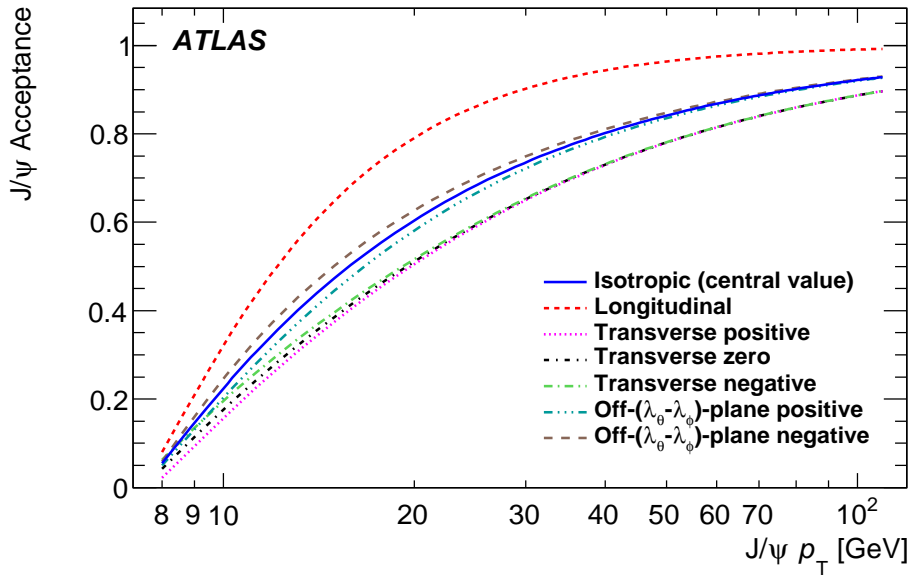}
     \caption{Projections of the acceptance as a function of $\pt$ for the \jpsi\ meson for various spin-alignment hypotheses.}
     \label{fig:accMapUnpolBoth}
  \end{center}
\end{figure}

Figure~\ref{fig:accMapUnpolBoth} shows the acceptance, projected in $\pt$ for all the spin-alignment hypotheses for the \jpsi\ meson.
The differences between the acceptance of the $\psiprime$ and $\jpsi$ meson, are independent of rapidity, except near $|y|\approx2$
at low $\pt$. Similarly, the only dependence on $\pt$ is found below $\pt\approx9$~\GeV.
The correction factors (as given in Appendix.~\ref{sec:spincorrection}) vary most at low $\pt$, ranging from $-35\%$ under longitudinal, to $+100\%$ for transverse-positive scenarios.
At high $\pt$, the range is between $-14\%$ for longitudinal, and $+9\%$ for transverse-positive scenarios.
For the fraction and ratio measurements, the correction factor is determined from the appropriate ratio of the individual correction factors. 

\subsection{Muon reconstruction and trigger efficiency determination}

\noindent The technique for correcting the 7~\TeV\ data for trigger and reconstruction inefficiencies is described in detail in Ref.~\cite{Aad:2014fpa,Aad2012dlq}.
For the 8~\TeV\ data, a similar technique is used, however different efficiency maps are required for each set of data, and the 8~\TeV\ corrections are detailed briefly below.

The single-muon reconstruction efficiency is determined from a tag-and-probe study in dimuon decays~\cite{Aad:2014kba}.
The efficiency map is calculated as a function of $\pt (\mu)$ and  $q\times \eta(\mu)$, where $q=\pm1$ is the electrical charge of the muon, expressed in units of $e$.

The trigger efficiency correction consists of two components. The first part represents the trigger efficiency for a 
single muon in intervals of $\pt (\mu)$ and $q\times \eta(\mu)$.
For the dimuon system there is a second correction to account for reductions in efficiency due to closely spaced
muons firing only a single RoI, 
vertex-quality cuts, and opposite-sign requirements.
This correction is performed in three rapidity intervals: 0--1.0, 1.0--1.2 and 1.2--2.3. The correction is a function of $\Delta R(\mu\mu)$ 
in the first two rapidity intervals and a function of $\Delta R(\mu\mu)$ and $|y(\mu\mu)|$ in the last interval.

\begin{figure} [!ht]
   \begin{center}
    \includegraphics[scale=0.6]{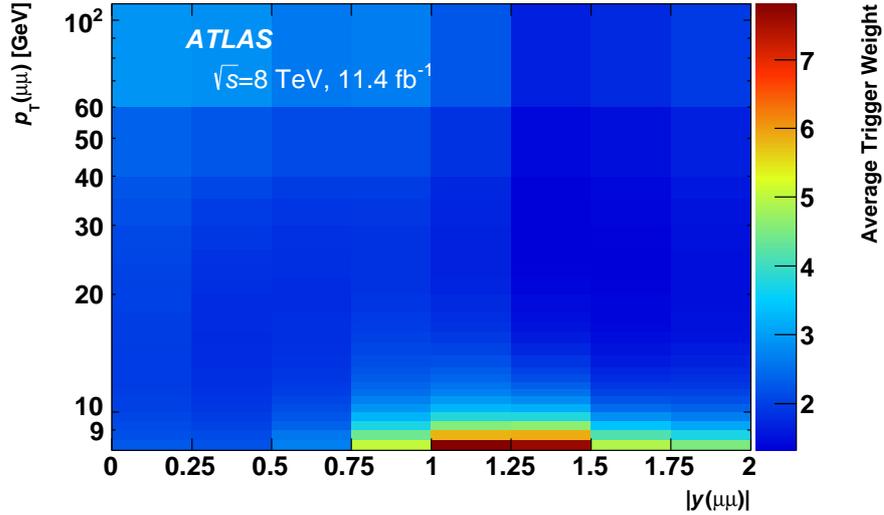}
    \caption{Average dimuon trigger-weight in the intervals of $\pt (\mu\mu) $ and $ |y(\mu\mu)|$ studied in this set of measurements. 
    }
    \label{fig:EFmu4DataJpsiTandPCmumupolyL2StarBMaxpt50}
   \end{center}
\end{figure}

The combination of the two components (single-muon efficiency map and dimuon corrections) is illustrated 
in Figure~\ref{fig:EFmu4DataJpsiTandPCmumupolyL2StarBMaxpt50} by plotting the average trigger-weight correction for the events in this analysis
in terms of $\pt(\mu\mu)$ and $|y(\mu\mu)|$.
The increased weight at low $\pt$ and $|y|\approx 1.25$ is caused by the geometrical acceptance of the muon trigger system and the turn-on threshold behaviour of the muon trigger. At high $\pt$ the weight is increased due to the reduced opening angle between the two muons.

\subsection{Fitting technique}
\label{sec:method:fit}
To extract the corrected yields of prompt and non-prompt \jpsi\ and \psiprime\ mesons, two-dimensional
weighted unbinned maximum-likelihood fits are performed on the dimuon invariant mass, $m(\mu\mu)$,
and pseudo-proper decay time, $\tau(\mu\mu)$, in intervals of $\pt(\mu\mu)$ and~$|y(\mu\mu)|$.
Each interval is fitted independently from all the others.
In $m(\mu\mu)$, signal processes of $\apsi$ meson decays are statistically distinguished  as narrow peaks convolved with the detector resolution, at their respective mass positions, on top of background continuum.
In $\tau(\mu\mu)$, decays originating with zero pseudo-proper decay time and those following an exponential decay distribution (both convolved with a detector resolution function) statistically distinguish prompt and non-prompt signal processes, respectively.
Various sources of background processes include Drell-Yan processes, mis-reconstructed muon pairs from prompt and non-prompt sources,
and semileptonic decays from separate $b$-hadrons. 

The probability density function (PDF) for each fit is defined as a normalized sum, 
where each term represents a specific signal or background contribution, with a physically motivated mass and $\tau$ dependence.
The PDF can be written in a compact form as
\begin{equation}
\label{eqn:pdf}
\mathrm{PDF}(m,\tau) = \sum_{i=1}^{7} \kappa_i f_i(m) \ \cdot h_i(\tau) \otimes R(\tau),
\end{equation}
where $\kappa_i$ represents the relative normalization of the $i^\mathrm{th}$ term of the seven considered signal and background contributions (such that $\sum_i \kappa_i = 1$), 
$f_i(m)$ is the mass-dependent term, and $\otimes$ represents the convolution of 
the $\tau$-dependent function $h_i(\tau)$ with the $\tau$ resolution term, $R(\tau)$. 
The latter is modelled by a double Gaussian distribution with both means fixed to zero and widths determined from the fit.

Table \ref{table:fitModel} lists the contributions to the overall PDF with the corresponding $f_i$ and
$h_i$ functions. Here $G_1$ and $G_2$ are Gaussian functions, $B_1$ and $B_2$ are Crystal Ball\footnote{The Crystal Ball function is given by:
\\$B(x;\alpha,n,\bar x,\sigma) = N \cdot \begin{cases} \exp\left(- \frac{(x - \bar x)^2}{2 \sigma^2}\right), & \mbox{for }\frac{x - \bar x}{\sigma} > -\alpha \\
 A \cdot \left(A' - \frac{x - \bar x}{\sigma}\right)^{-n}, & \mbox{for }\frac{x - \bar x}{\sigma} \leqslant -\alpha \end{cases}$ \\where $A = \left(\frac{n}{\left| \alpha \right|}\right)^n \cdot \exp\left(- \frac {\left| \alpha \right|^2}{2}\right),
A' = \frac{n}{\left| \alpha \right|}  - \left| \alpha \right|$}
distributions~\cite{CB1}, 
while F is a uniform distribution and $C_1$ a first-order Chebyshev polynomial. 
The exponential functions $E_1$, $E_2$, $E_3$, $E_4$ and $E_5$
have different decay constants, where $E_5(|\tau|)$ is a double-sided exponential with the same decay constant on
either side of $\tau = 0$. The parameter $\omega$ represents the fractional contribution 
of the $B$ and $G$ mass signal functions, while the Dirac delta function, $\delta(\tau)$,
is used to represent the pseudo-proper decay time distribution of the prompt candidates.

\begin{table}[h!]
  \centering
  \caption{Description of the fit model PDF in Eq.~\ref{eqn:pdf}. Components of the probability density function used to extract the prompt (P) and
non-prompt (NP) contributions for \jpsi\ and \psiprime\ signal and the P, NP, and incoherent or mis-reconstructed  background (Bkg) contributions.}
    \begin{tabular}{ l  c  c  c  c  }
      \hline \hline
      $i$ & Type & Source & $f_i(m)$ & $h_i(\tau)$ \\ \hline \hline
      1 & \jpsi\     & P     & $\omega B_1(m) + (1-\omega) G_1(m)$   & $\delta (\tau)$ \\
      2 & \jpsi\     & NP    & $\omega B_1(m) + (1-\omega) G_1(m)$   & $ E_1(\tau)$ \\
      3 & \psiprime\    & P    & $\omega B_2(m) + (1-\omega) G_2(m)$   & $\delta (\tau)$ \\
      4 & \psiprime\   & NP    & $\omega B_2(m) + (1-\omega) G_2(m)$   & $E_2(\tau)$ \\ \hline
      5 & Bkg          & P     & $ F$     & $\delta (\tau)$ \\
      6 & Bkg          & NP    & $ C_1(m)$    & $E_3 (\tau)$ \\ 
      7 & Bkg          & NP    & $ E_4(m)$   & $E_5 (|\tau|)$ \\
      \hline  
    \end{tabular}
  \label{table:fitModel}
\end{table}

In order to make the fitting procedure more robust and to reduce the number of free parameters, a number of component terms share common parameters, 
which led to 22 free parameters per interval.
In detail, the signal mass models are described by the sum of a Crystal Ball shape ($B$) and a Gaussian shape ($G$). For each
of \jpsi\ and \psiprime, the $B$ and $G$ share a common mean, and freely determined widths, 
with the ratio of the $B$ and $G$ widths common to \jpsi\ and \psiprime.
The $B$ parameters $\alpha$, and $n$, describing the transition point of the low-edge from a Gaussian to a power-law 
shape, and the shape of the tail, respectively, are fixed, and variations are considered as part of the fit model
systematic uncertainties.
The width of $G$ for \psiprime\ is set to the width for \jpsi\ multiplied by a free parameter scaling term.
The relative fraction of $B$ and $G$ is left floating, but common to \jpsi\ and \psiprime.

The non-prompt signal decay shapes ($E_1$,$E_2$) are described by an exponential function (for positive $\tau$ only) convolved with a
double Gaussian function, $R(\tau)$ describing the pseudo-proper decay time resolution for the non-prompt component, 
and the same Gaussian response functions to
describe the prompt contributions. 
Each Gaussian resolution component has its mean fixed at $\tau$ = 0 and a free width.
The decay constants of the \jpsi\ and \psiprime\ are separate free parameters in the fit.

The background contributions are described by a prompt and non-prompt component, as well as a double-sided exponential function
convolved with a double Gaussian function describing mis-reconstructed or non-coherent muon pairs. 
The same resolution function as in signal is used to describe the background. For the non-resonant mass
parameterizations, the non-prompt contribution is modelled by a first-order Chebyshev polynomial. The prompt mass
contribution follows a flat distribution and the double-sided background uses an exponential function.
Variations of this fit model are considered as systematic uncertainties.

The following quantities are extracted directly from the fit in each interval:
the fraction of events that are signal (prompt or non-prompt \jpsi\ or \psiprime); the fraction of signal events that are prompt;
the fraction of prompt signal that is \psiprime; and the fraction of non-prompt signal that is \psiprime. From these
parameters, and the weighted sum of events, all measured values are calculated.

For $7$~\TeV\ data, 168 fits are performed across the range of $8<\pt<100$~\GeV\ ($8<\pt<60$~\GeV) for $\jpsi$ ($\psiprime$) and $0<|y|<2$.
For $8$~\TeV\ data, 172 fits are performed across the range of $8<\pt<110$~\GeV\ and $0<|y|<2$,
excluding the area where $\pt$ is less than 10 \GeV\ and simultaneously $|y|$ is
greater than 0.75. This region is excluded due to a steeply changing low trigger efficiency 
causing large systematic uncertainties in the measured cross-section.

\begin{figure} [!ht]
  \begin{center} 
   \includegraphics[width=0.51\textwidth]{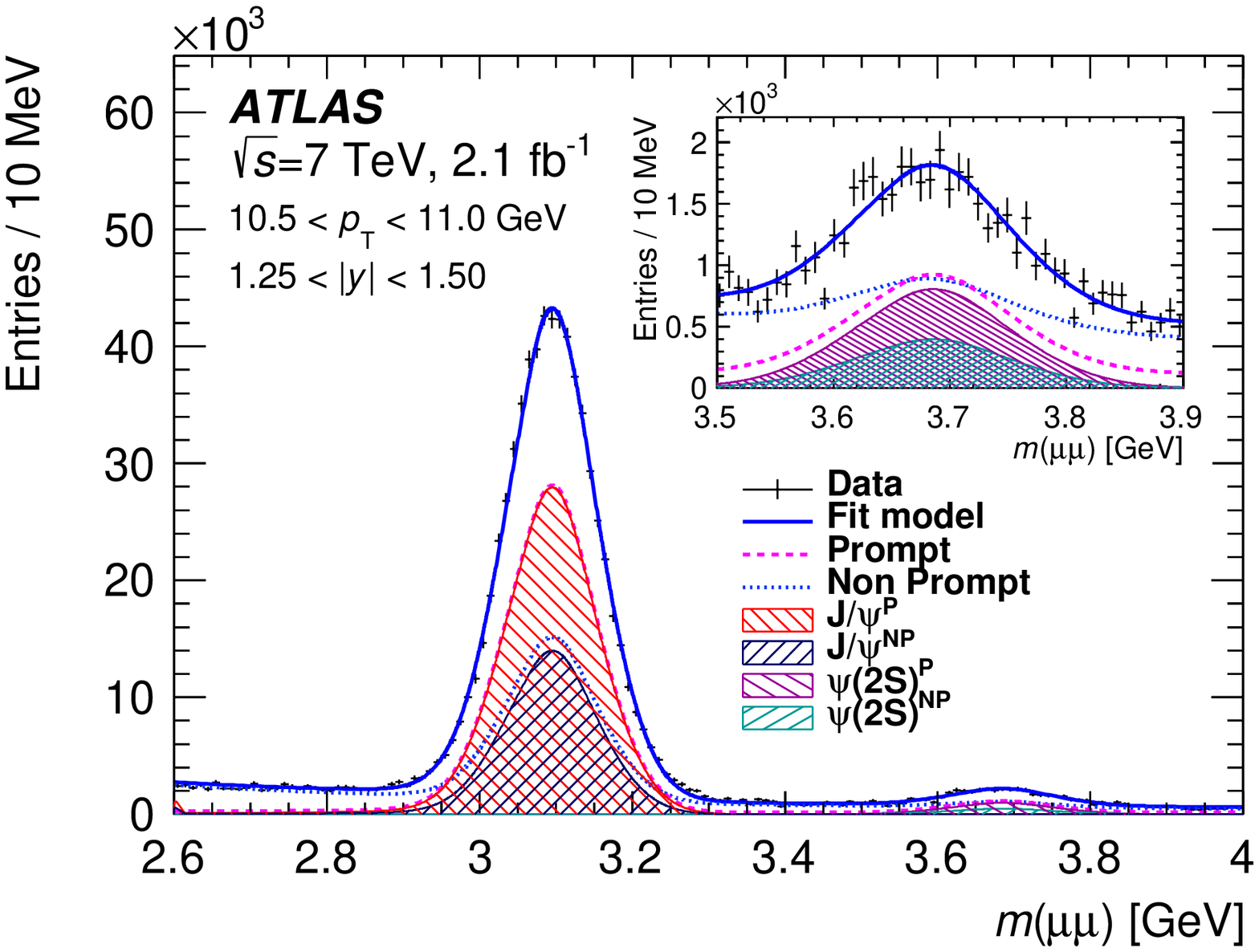}
   \hspace{-0.55cm}
   \includegraphics[width=0.51\textwidth]{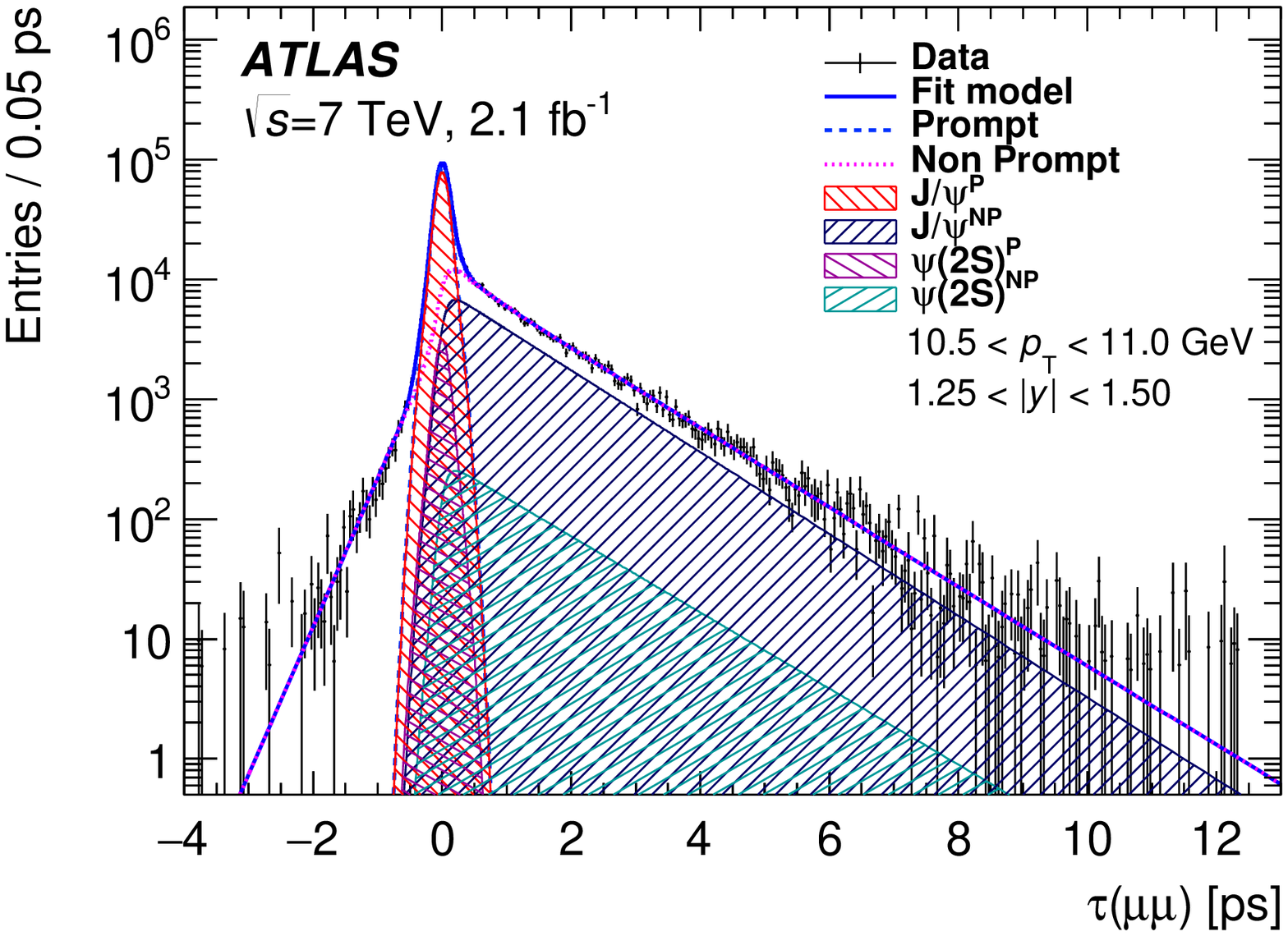}
      \caption{Projections of the fit result over the mass (left) and pseudo-proper decay time (right) distributions for data collected at $7$~\TeV\  for one typical interval. The data are shown with error bars in black, superimposed with the individual components of the fit result projections, where the total prompt and non-prompt components are represented by the dashed and dotted lines, respectively, and the shaded areas show the signal $\psi$ prompt and non-prompt contributions.\label{fig:fitprojmain}}
  \end{center}
\end{figure}

Figure~\ref{fig:fitprojmain} shows the fit results for one of the intervals considered in the analysis, projected onto the invariant mass and pseudo-proper decay time distributions, for  $7$~\TeV\ data, weighted according to the acceptance and efficiency corrections.
The fit projections are shown for the total prompt and total non-prompt contributions (shown as curves), and also for the individual contributions of the $\jpsi$ and $\psiprime$ prompt and non-prompt signal yields (shown as hashed areas of various types). 

\begin{figure} [!ht]
  \begin{center} 
   \includegraphics[width=0.51\textwidth]{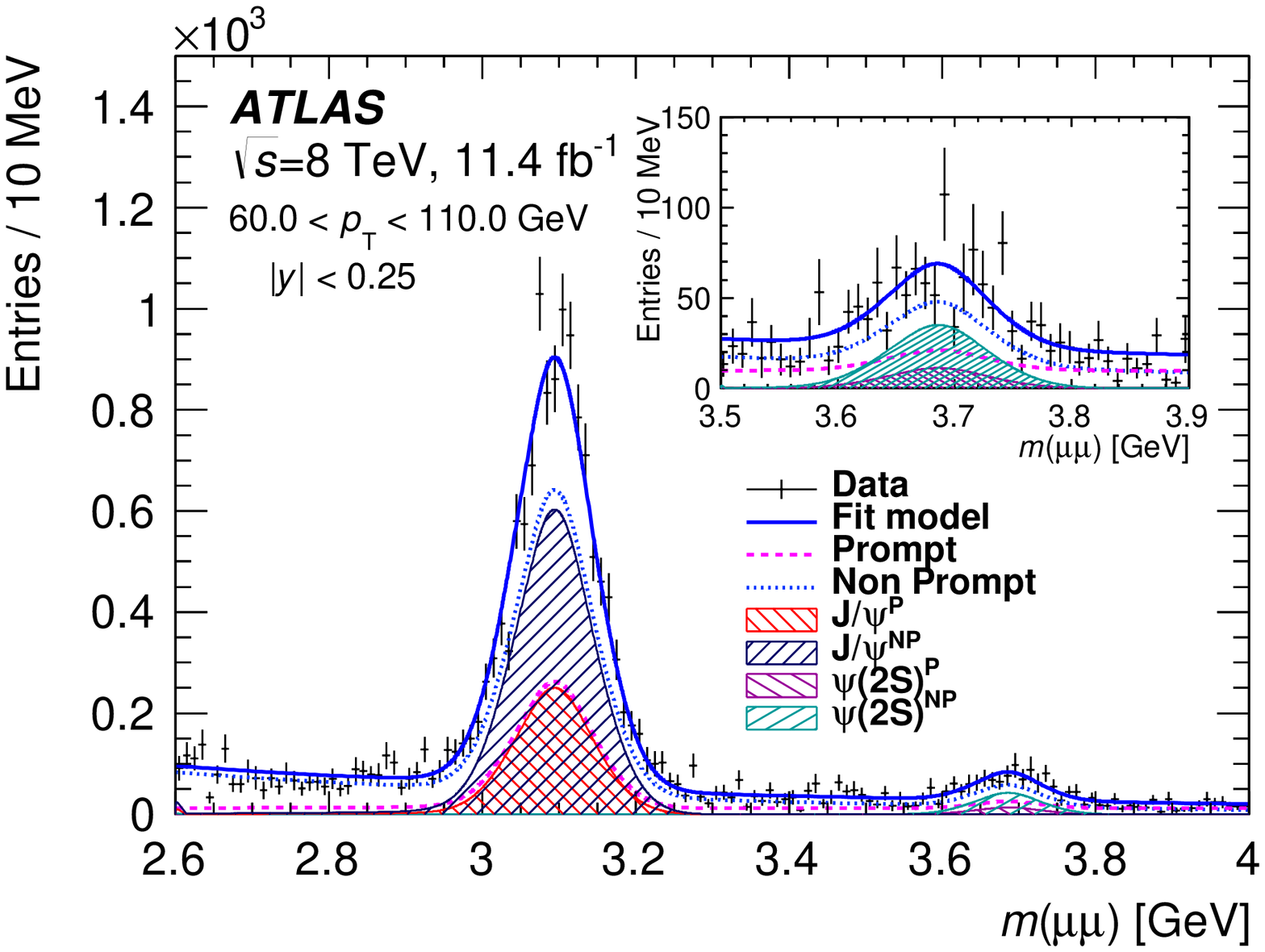}
   \hspace{-0.55cm}
   \includegraphics[width=0.51\textwidth]{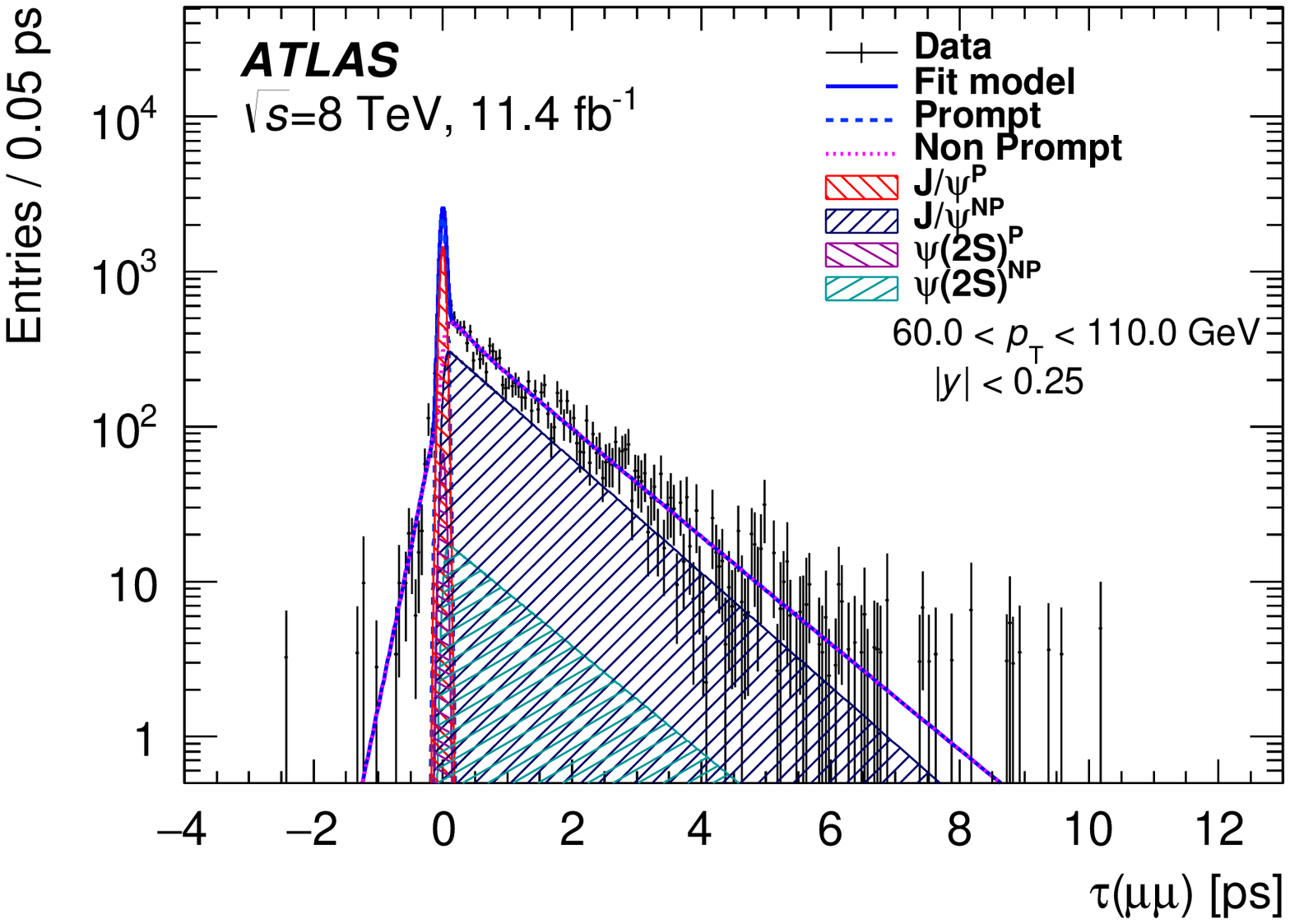}
      \caption{Projections of the fit result over the mass (left) and pseudo-proper decay time (right) distributions for data collected at $8$~\TeV\ for one high-$\pt$ interval. The data are shown with error bars in black, superimposed with the individual components of the fit result projections, where the total prompt and non-prompt components are represented by the dashed and dotted lines, respectively, and the shaded areas show the signal $\psi$ prompt and non-prompt contributions.\label{fig:fitprojmainVIII}}
  \end{center}
\end{figure}

In Figure~\ref{fig:fitprojmainVIII} the fit results are shown for one high-$\pt$ interval of $8$~\TeV\  data.

\subsection{Bin migration corrections}
To account for bin migration effects due to the detector resolution, 
which results in decays of~$\apsi$ in one bin, being identified and accounted for in another,
the numbers of acceptance- and efficiency-corrected dimuon decays extracted from
the fits in each interval of $\pt(\mu\mu)$ and rapidity are corrected for the differences
between the true and reconstructed values of the dimuon $\pt$.
These corrections are derived from data by comparing analytic functions that are 
fitted to the $\pt(\mu\mu)$ spectra of dimuon events with and without
convolution by the experimental resolution in $\pt(\mu\mu)$ (as determined from the fitted mass resolution and measured muon angular resolutions), as described in Ref.~\cite{Aad2012dlq}.

The correction factors applied to the fitted yields deviate from unity by no more than $1.5\%$, and for the majority of slices are smaller than $1\%$.
The ratio measurement and non-prompt fractions are corrected by the corresponding ratios of
bin migration correction factors.
Using a similar technique, bin migration corrections as a function of $|y|$ are found 
to differ from unity by negligible amounts.

%% file: Systematics.tex
\section{Systematic uncertainties}
\label{sec:syst}

\begin{figure} [!h]
  \begin{center}
    \includegraphics[width=0.49\textwidth]{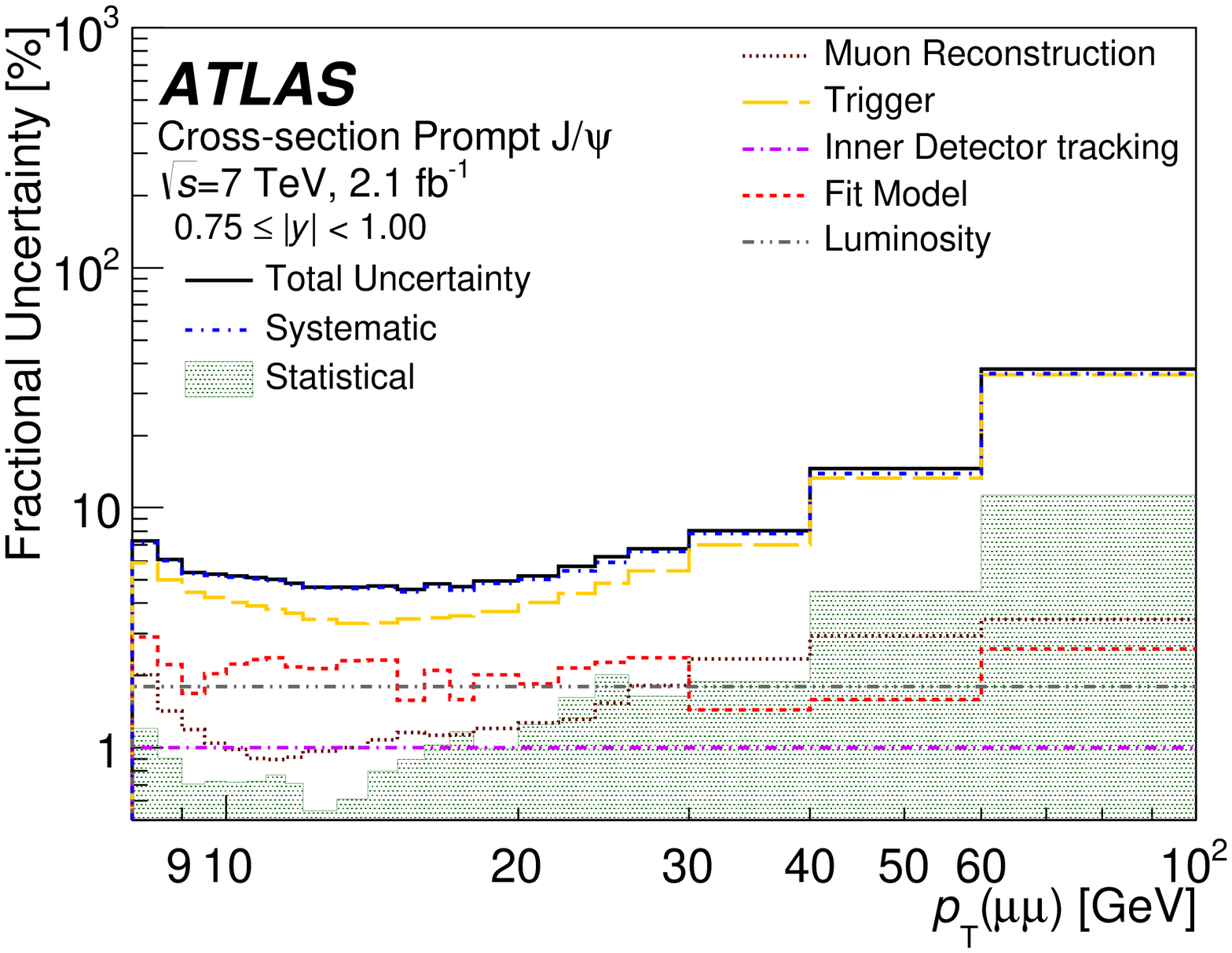}
    \includegraphics[width=0.49\textwidth]{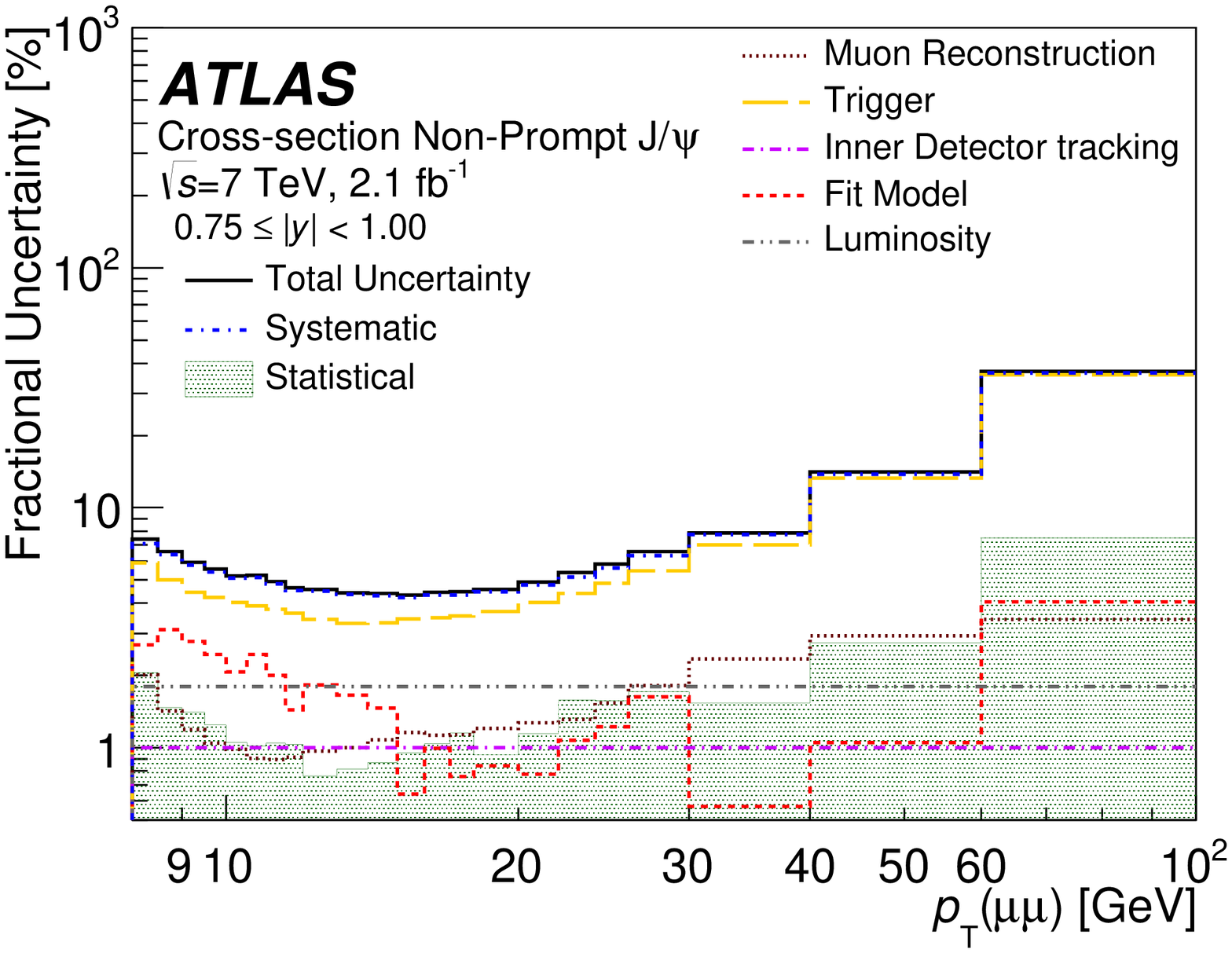}\\
    \includegraphics[width=0.49\textwidth]{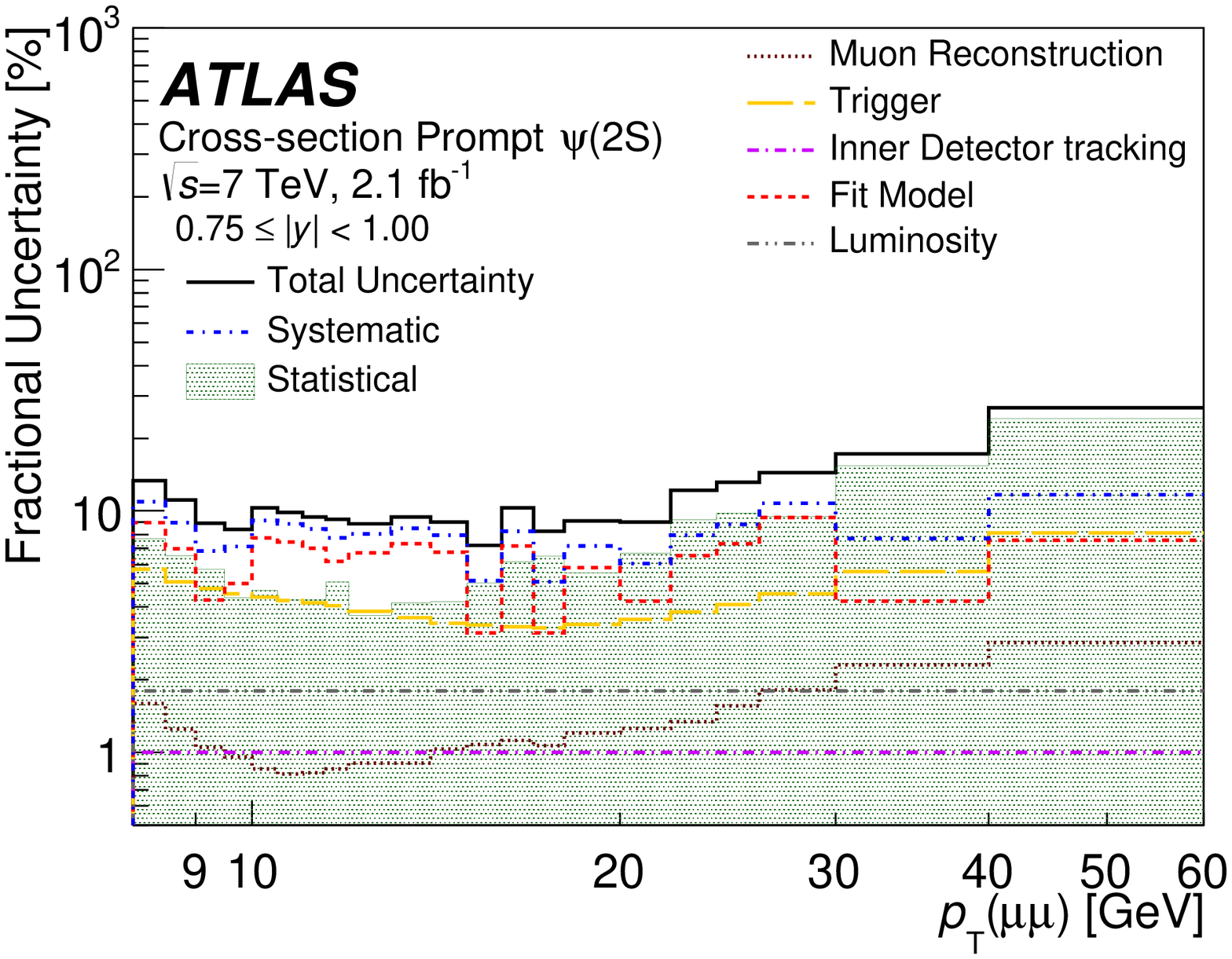}
    \includegraphics[width=0.49\textwidth]{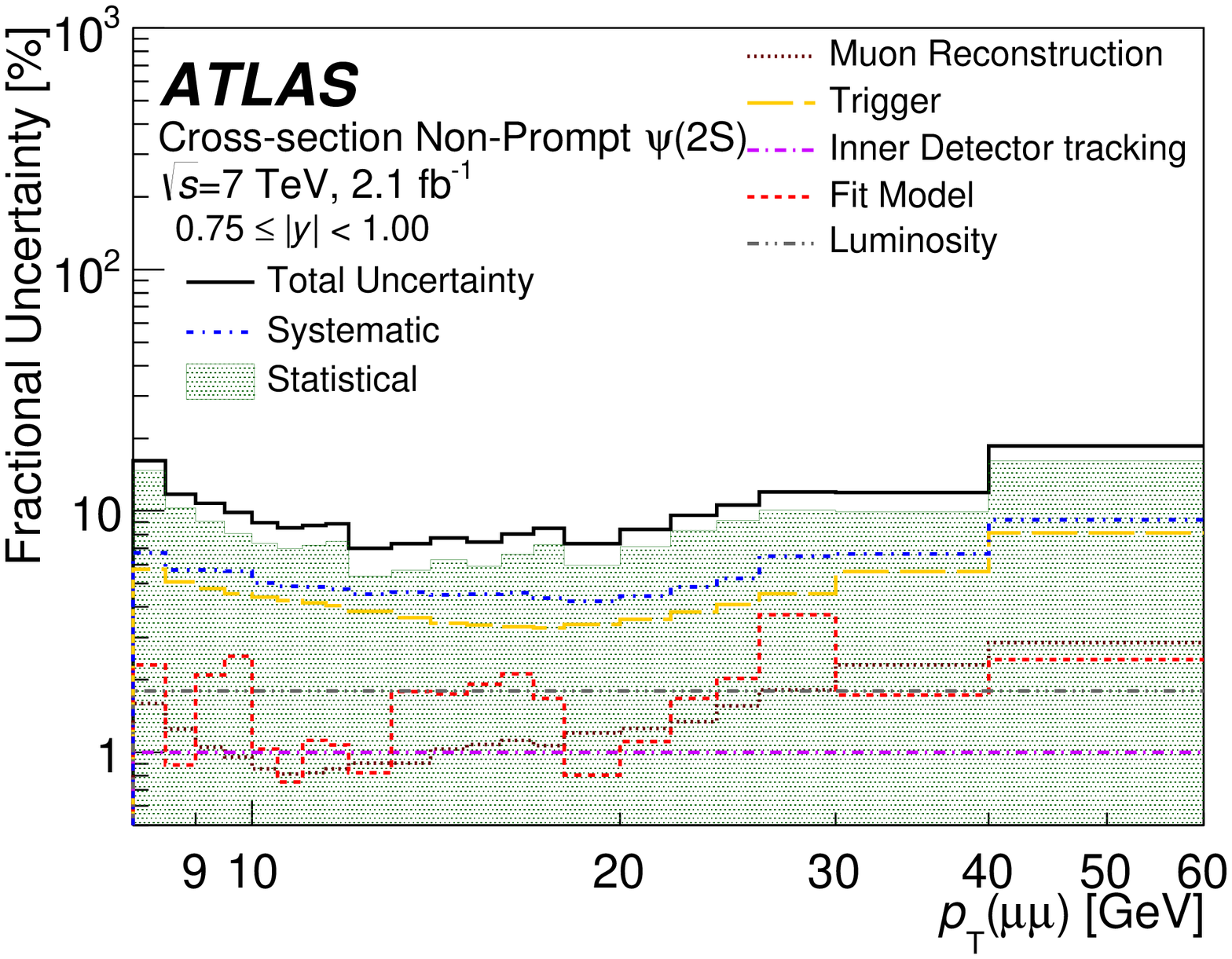}    
    \caption{Statistical and systematic contributions to the fractional uncertainty on the prompt (left column) and non-prompt (right column)
    $\jpsi$ (top row) and $\psiprime$ (bottom row) cross-sections for 7 \TeV, 
    shown for the region $0.75<|y|<1.00$.}
    \label{fig:fracsyst_xs}
  \end{center}
\end{figure}

\begin{figure} [!h]
  \begin{center}
    \includegraphics[width=0.49\textwidth]{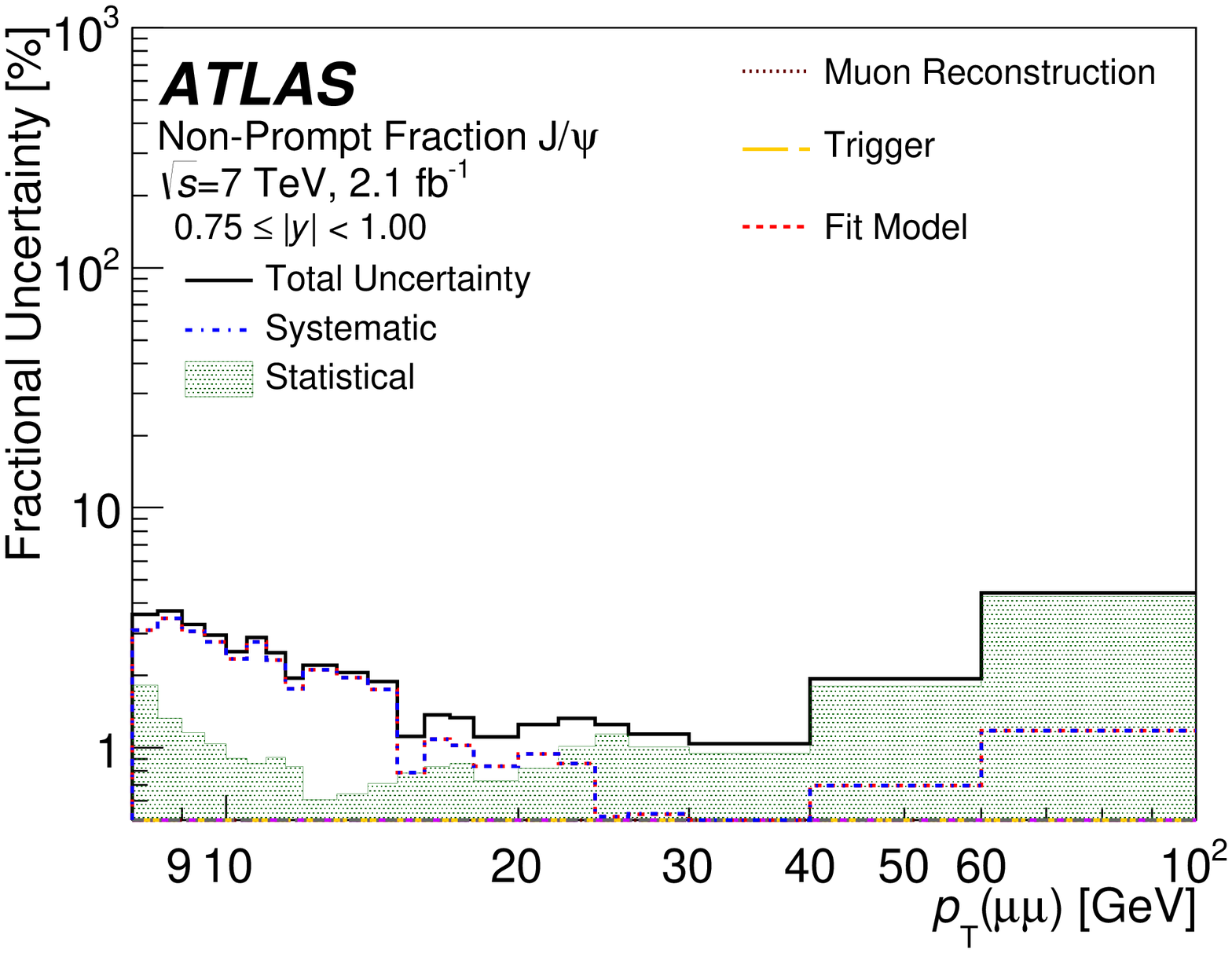}
    \includegraphics[width=0.49\textwidth]{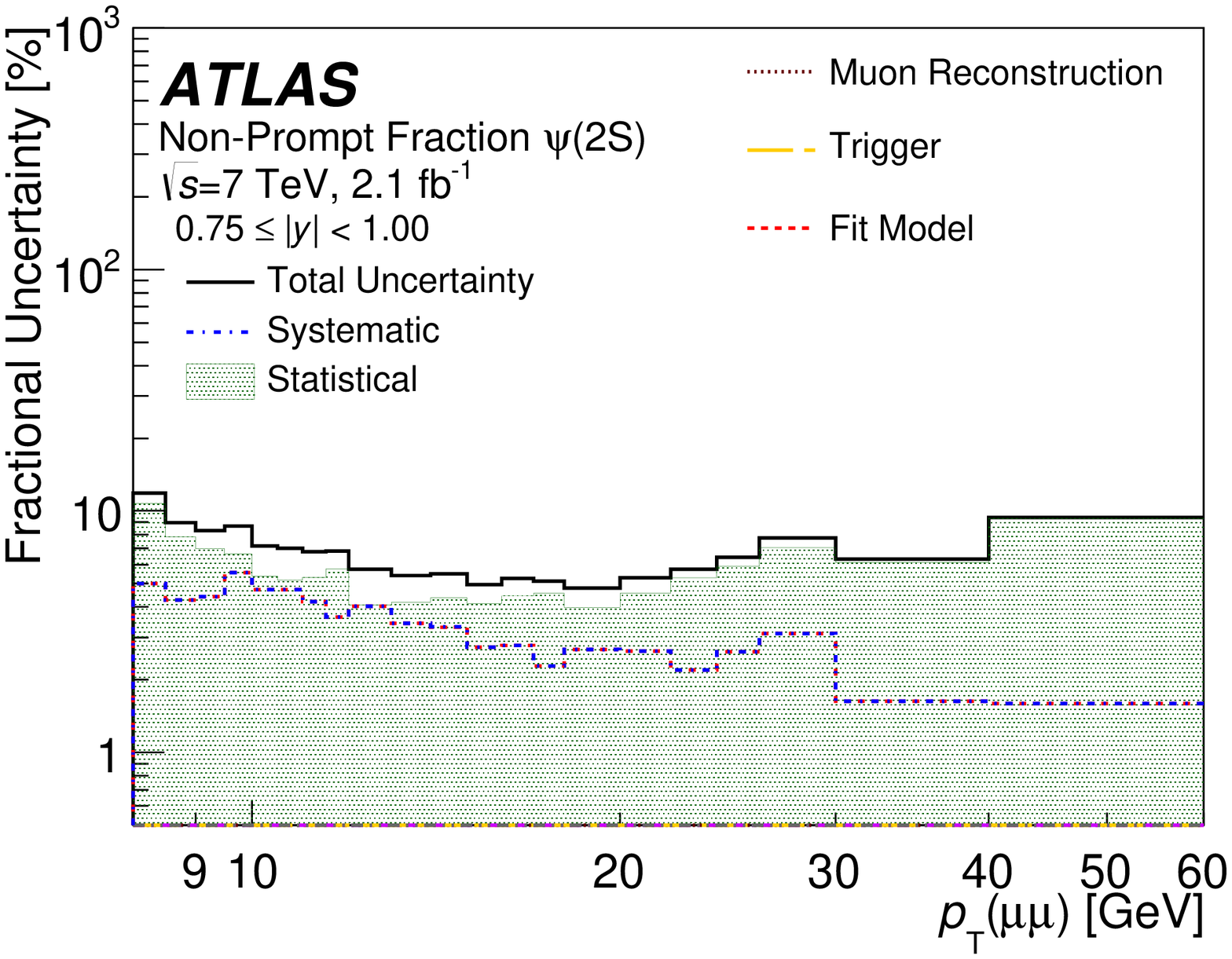}\\
    \includegraphics[width=0.49\textwidth]{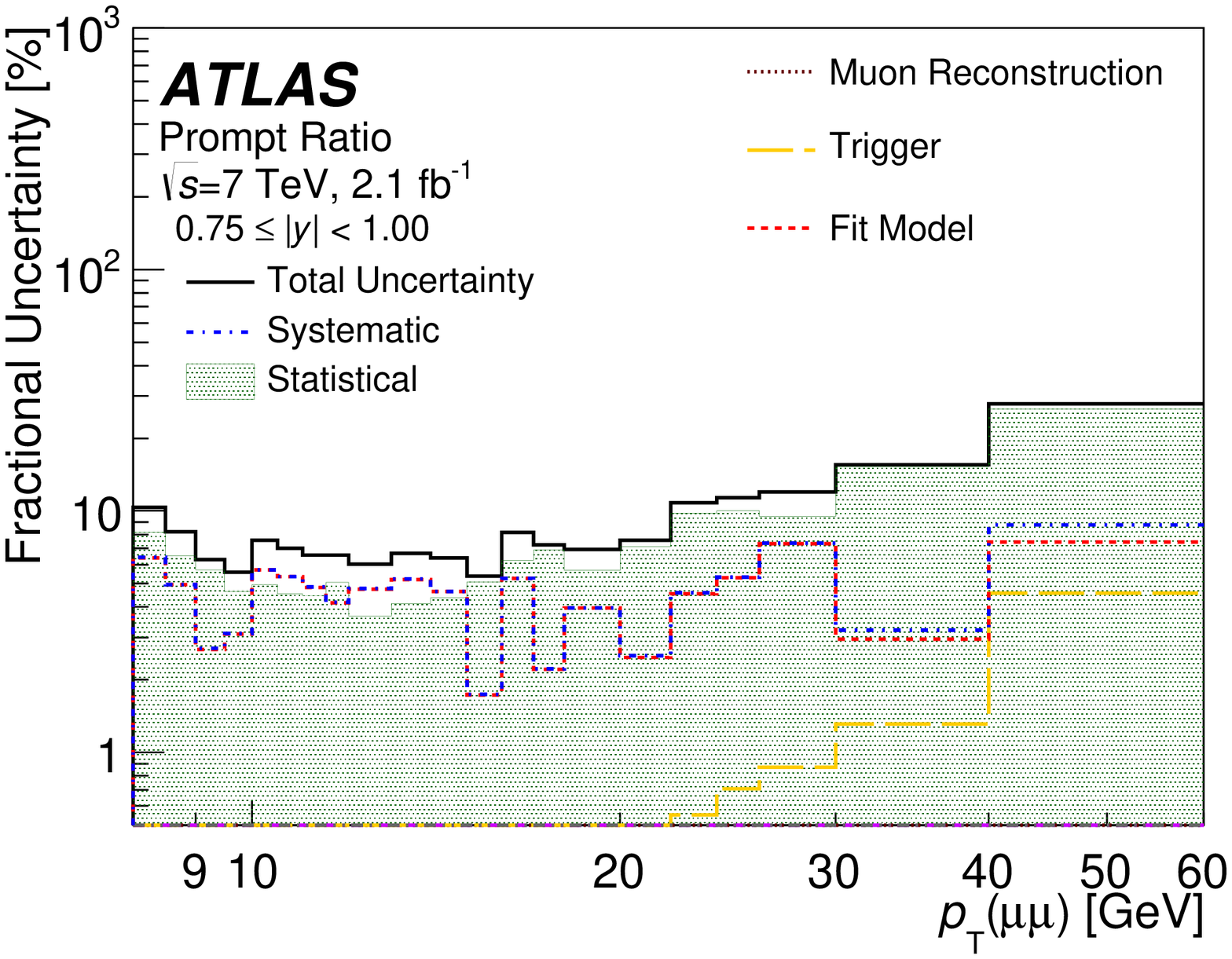}
    \includegraphics[width=0.49\textwidth]{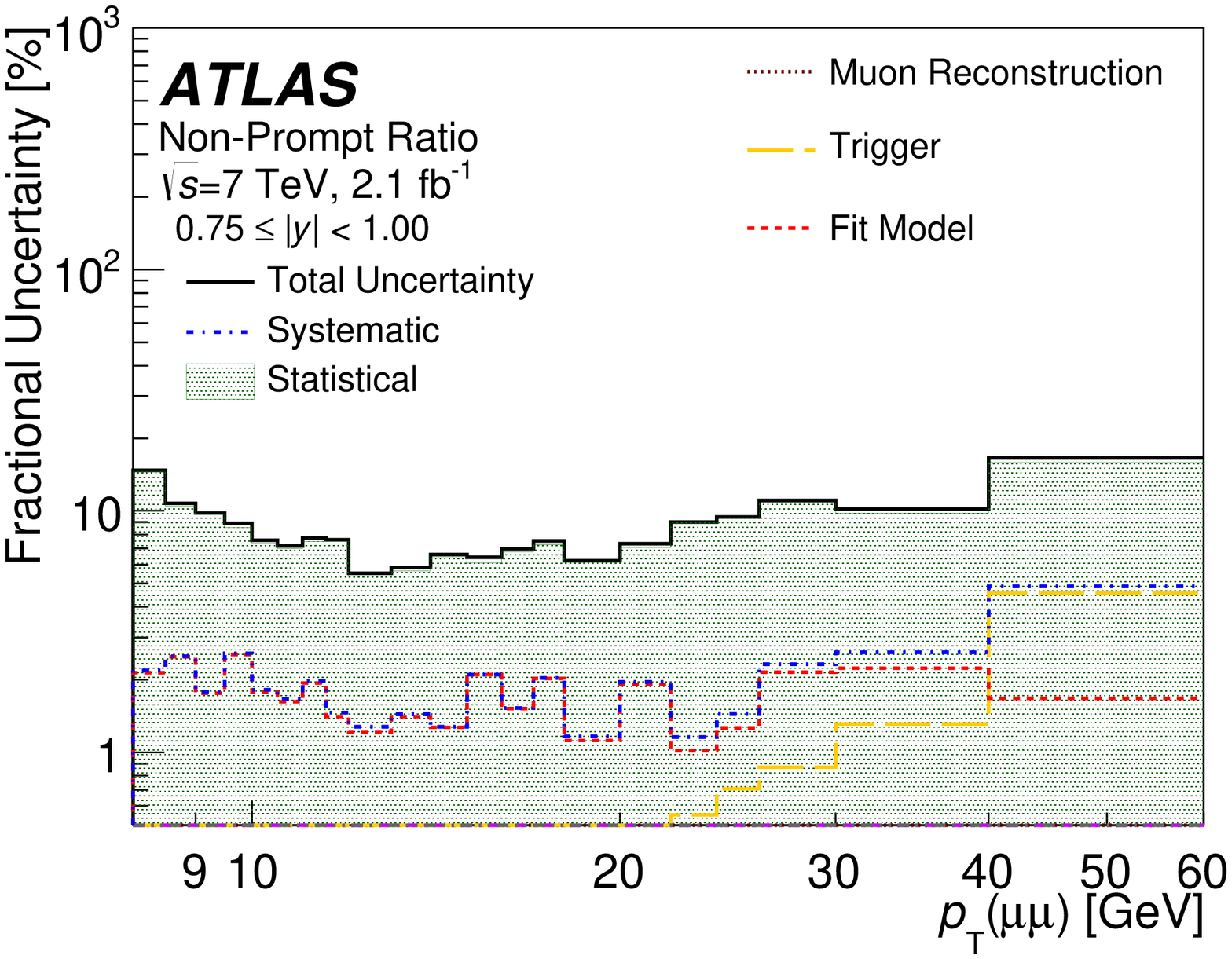} 
    \caption{Breakdown of the contributions to the fractional uncertainty on the non-prompt fractions for $\jpsi$ (top left) and $\psiprime$ (top right), 
    and the prompt (bottom left) and non-prompt (bottom right) ratios for 7 \TeV, 
    shown for the region $0.75<|y|<1.00$.}
    \label{fig:fracsyst_ratio}
  \end{center}
\end{figure}

The sources of systematic uncertainties that are applied to the $\psi$ double differential cross-section
measurements are from uncertainties in: the luminosity determination; muon and trigger efficiency corrections; inner detector tracking efficiencies; the fit model parametrization; and due to bin migration corrections.
For the non-prompt fraction and ratio measurements the systematic uncertainties are assessed in the same 
manner as for the uncertainties on the cross-section, 
except that in these ratios some systematic uncertainties, such as the luminosity uncertainty, cancel out.
The sources of systematic uncertainty evaluated for the prompt and non-prompt $\apsi$ cross-section measurements, along with the minimum, maximum and median values, are listed in Table~\ref{table:systlist}.
The largest contributions, which originate from the trigger and fit model uncertainties, are typically for the 
high $\pt$ intervals and are due to the limited statistics of the efficiency maps (for the trigger), and the data sample (for the fit model).

Figures~\ref{fig:fracsyst_xs} and~\ref{fig:fracsyst_ratio} show, for a representative interval,
the impact of the considered uncertainties
on the production cross-section, as well as the non-prompt fraction and ratios for $7$~\TeV\ data.
The impact is very similar at 8 \TeV.

\begin{table}[h!]
 \caption{Summary of the minimum and maximum contributions along with the median value of the systematic uncertainties as percentages for the prompt and non-prompt $\psi$ cross-section results.
  Values are quoted for 7 and 8 \TeV\ data.}
 \centering
    \begin{tabular}{l || c c c | c c c  }
    \hline 
     & \multicolumn{3}{c|}{7 \TeV\ [\%]} & \multicolumn{3}{c}{8 \TeV\ [\%]} \\
    \hline 
    Source of systematic uncertainty                   & Min & Median & Max & Min & Median & Max \\
    Luminosity                         & 1.8 & 1.8 & 1.8    & 2.8 & 2.8 & 2.8 \\
    Muon reconstruction efficiency     & 0.7 & 1.2 & 4.7    & 0.3 & 0.7 & 6.0 \\
    Muon trigger efficiency            & 3.2 & 4.7 & 35.9   & 2.9 & 7.0 & 23.4 \\
    Inner detector tracking efficiency & 1.0 & 1.0 & 1.0    & 1.0 & 1.0 & 1.0 \\
    Fit model parameterizations        & 0.5 & 2.2 & 22.6   & 0.26 & 1.07 & 24.9 \\
    Bin migrations                     & 0.01 & 0.1 & 1.4   & 0.01 & 0.3 & 1.5 \\ \hline
    Total                              & 4.2 & 6.5  & 36.3  & 4.4 & 8.1 & 27.9 \\
    \hline  
      \hline
    \end{tabular}
\label{table:systlist}
\end{table}

\setdescription{font=\normalfont\itshape}

\begin{description}[style=unboxed,leftmargin=0cm]
\item[Luminosity] \hfill \\
The uncertainty on the integrated luminosity is $1.8\%$ ($2.8\%$) for the 7~\TeV\ (8~\TeV) data-taking period.  
The methodology used to determine these uncertainties is described in Ref.~\cite{Aad:2013ucp}.
The luminosity uncertainty is only applied to the \jpsi\ and \psiprime\ cross-section results.

\item[Muon reconstruction and trigger efficiencies] \hfill \\
To determine the systematic uncertainty on the muon reconstruction and trigger efficiency maps, each of the maps is reproduced in 100 pseudo-experiments.
The dominant uncertainty in each bin is statistical and hence any bin-to-bin correlations are neglected.
For each pseudo-experiment a new map is created by varying independently each bin content according to a Gaussian distribution about its estimated value, 
determined from the original map.
In each pseudo-experiment, the total weight is recalculated for each dimuon $\pt$ and $|y|$ interval of the analysis.
The RMS of the total weight pseudo-experiment distributions for each efficiency type is used as the systematic uncertainty,
where any correlation effects between the muon and trigger efficiencies can be neglected.

The ID tracking efficiency is in excess of $99.5\%$ \cite{Aad2012dlq}, and
an uncertainty of 1\% is applied to account for the ID dimuon reconstruction inefficiency (0.5\% per muon, added coherently). This
uncertainty is applied to the differential cross-sections and is assumed to cancel in the fraction of non-prompt to inclusive production for \jpsi\ and 
\psiprime\ and in the ratios of \psiprime\ to \jpsi\ production.

For the trigger efficiency  $\epsilon_\mathrm{trig}$, in addition to the trigger efficiency map, 
there is an additional correction term that accounts for inefficiencies due to correlations between the two trigger muons, such as the dimuon opening angle.
This correction is varied by its uncertainty, and the shift in the resultant total weight relative to its central value is added in quadrature to the uncertainty from the map.
The choice of triggers is known \cite{Aad:2014cqa} to introduce a small lifetime-dependent efficiency loss but it is determined to have a negligible effect on the prompt and non-prompt yields and no correction is applied in this analysis.
Similarly, the muon reconstruction efficiency corrections of prompt and non-prompt signals are found to be consistent within the statistical uncertainties of the efficiency measurements, and no additional uncertainty is applied.

  \item[Fit model uncertainty] \hfill \\
The uncertainty due to the fit procedure is determined by varying one component at a
time in the fit model described in Section \ref{sec:method:fit}, creating a set of new fit models. For each new
fit model, all measured quantities are recalculated, and in each $\pt$ and $|y|$ interval the spread of
variations around the central fit model is used as its systematic uncertainty.
The variations of the fit model also account for possible uncertainties due to final-state radiation.
The following variations to the central model fit are evaluated: 

\begin{itemize}

  \item  signal mass model --- using double Gaussian models in place of the Crystal Ball plus Gaussian model;  variation of the
 $\alpha$ and $n$ parameters of the $B$ model, which are originally fixed;

    \item  signal pseudo-proper decay time model --- a double exponential function is used to describe the pseudo-proper decay time distribution for the 
    $\apsi$ non-prompt signal;

  \item  background mass models --- variations of the mass model using exponentials functions, or quadratic Chebyshev polynomials to describe the components of prompt,
 non-prompt and double-sided background terms;

  \item  background pseudo-proper decay time model --- a single exponential function was considered for the non-prompt component;

  \item  pseudo-proper decay time resolution model --- using a single Gaussian function in place of the double Gaussian function to model the lifetime resolution
 (also prompt lifetime model); and variation of the mixing terms for the two Gaussian components of this term.

\end{itemize}

Of the variations considered, it is typically the parametrizations of the signal mass model and pseudo-proper decay time resolution model that dominate the contribution to the fit model uncertainty.

  \item[Bin migrations] \hfill \\
As the corrections to the results due to bin migration effects are factors close to unity in all regions, the difference between the correction factor and unity is applied as the uncertainty. 

\end{description}

The variation of the acceptance corrections with spin-alignment is treated
separately, and scaling factors supplied in Appendix~\ref{sec:spincorrection}.

%% file: Results.tex
\section{Results}
\label{sec:results}

The $\jpsi$ and $\psiprime$ non-prompt and prompt production cross-sections are presented, corrected for acceptance and detector efficiencies 
while assuming isotropic decay, as described in Section~\ref{sec:s:diff_xSecDet}. 
Also presented are the ratios of non-prompt production relative to the inclusive production for $\jpsi$ and $\psiprime$ mesons separately, described in Section~\ref{sec:s:NPFDet},
and the ratio of $\psiprime$ to $\jpsi$ production for prompt and non-prompt components separately, described in Section~\ref{sec:s:PNPRatioDet}.
Correction factors for various spin-alignment hypotheses for both 7 and 8~\TeV\  data can be found in 
Tables~\ref{tab:sa_long_jpsi}--\ref{tab:sa_offN_psi2s} (in Appendix) and Tables~\ref{tab:sa_long_jpsi8}--\ref{tab:sa_offN_psi2s8} (in Appendix) respectively, in terms of 
\pt\ and rapidity intervals.

\setdescription{font=\normalfont\itshape}

\begin{description}[style=unboxed,leftmargin=0cm]

\item[Production cross-sections] \hfill

Figures \ref{fig:res:xSecP} and \ref{fig:res:xSecNP} show respectively the prompt and non-prompt 
differential cross-sections of \jpsi\ and \psiprime\ as functions of $\pt$ and $|y|$, together with the relevant theoretical predictions, 
which are described below.

\begin{figure} [!ht]
  \begin{center}
    \includegraphics[width=0.44\textwidth]{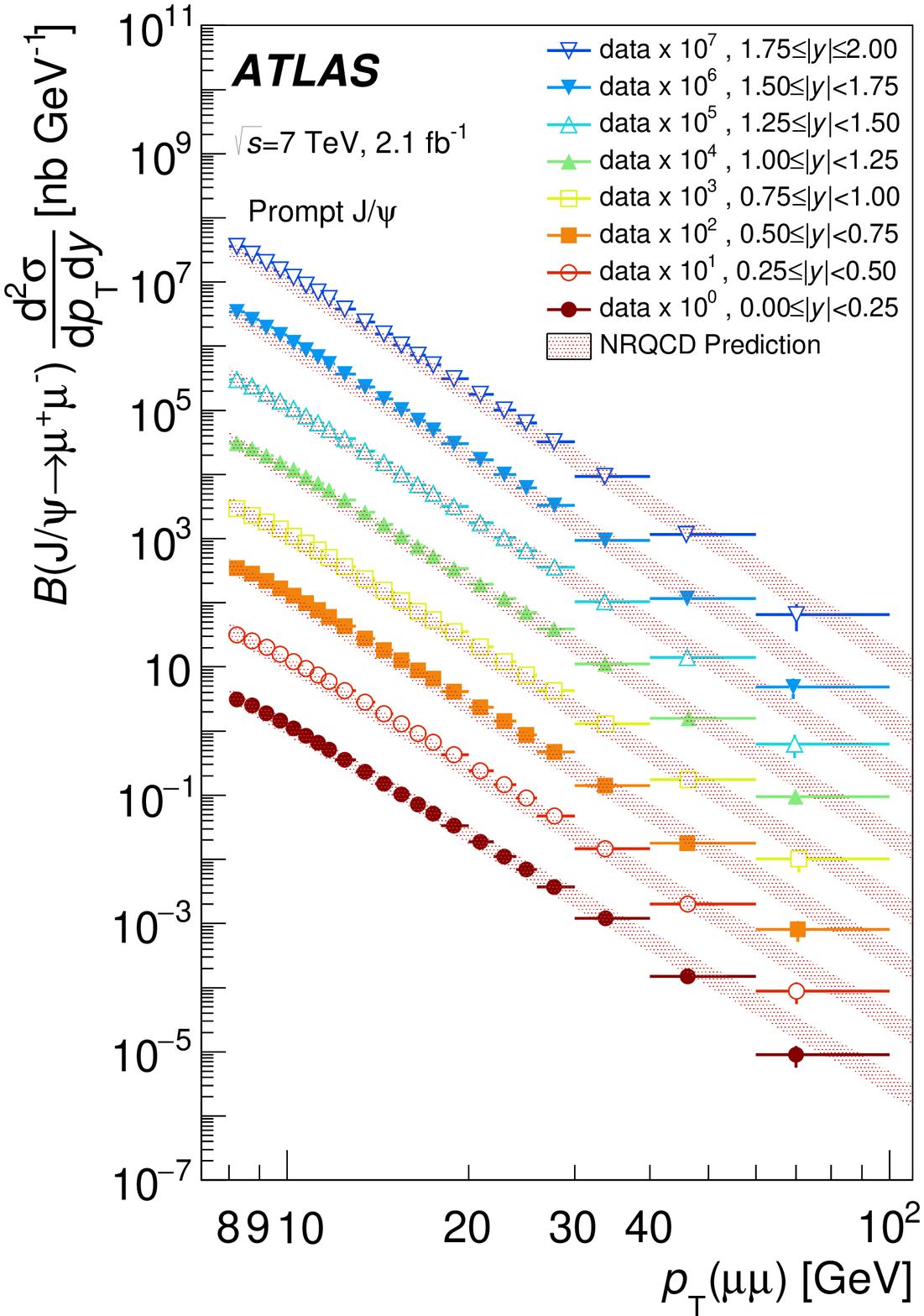} 
    \includegraphics[width=0.44\textwidth]{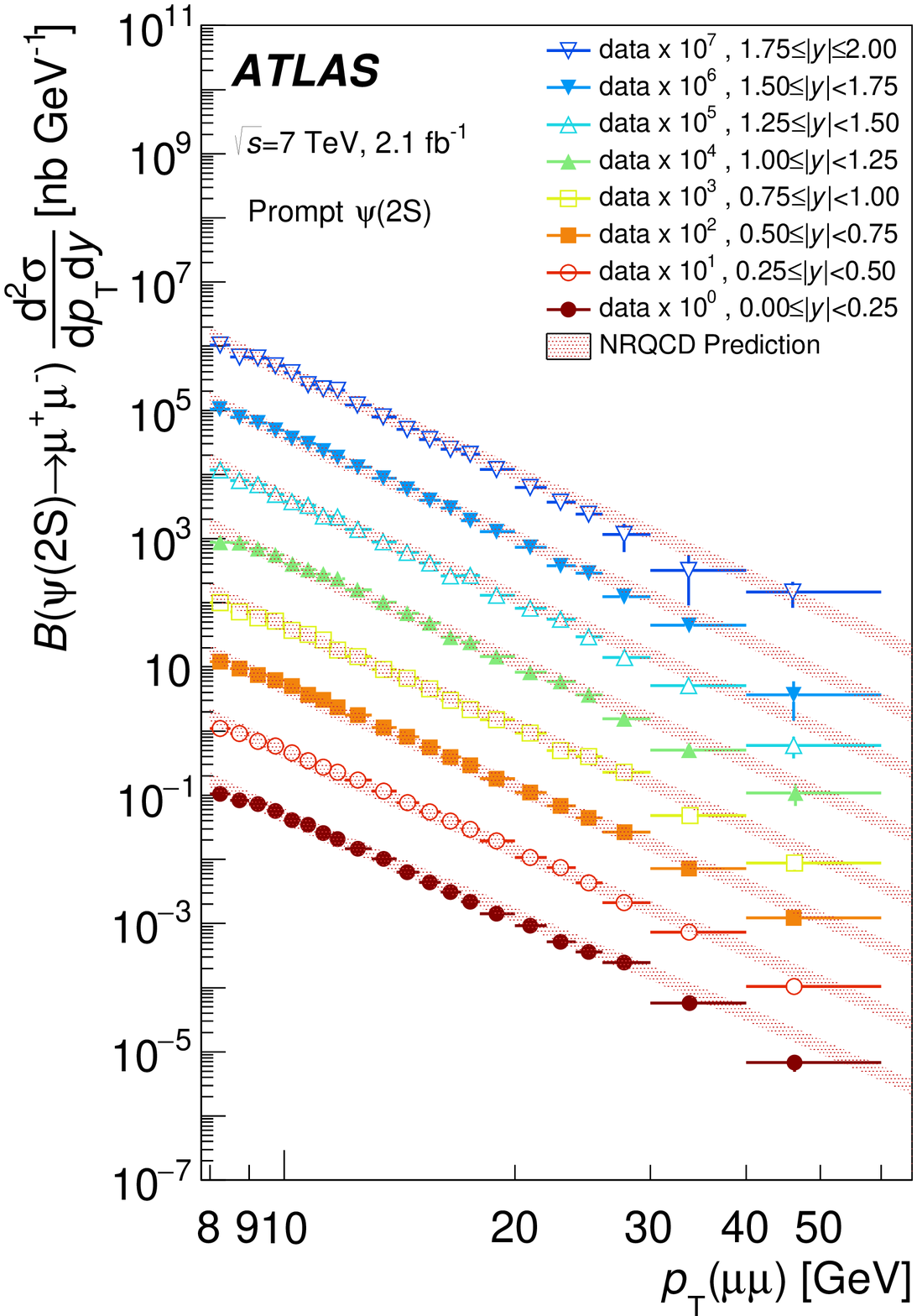}\hfil\\
    \includegraphics[width=0.44\textwidth]{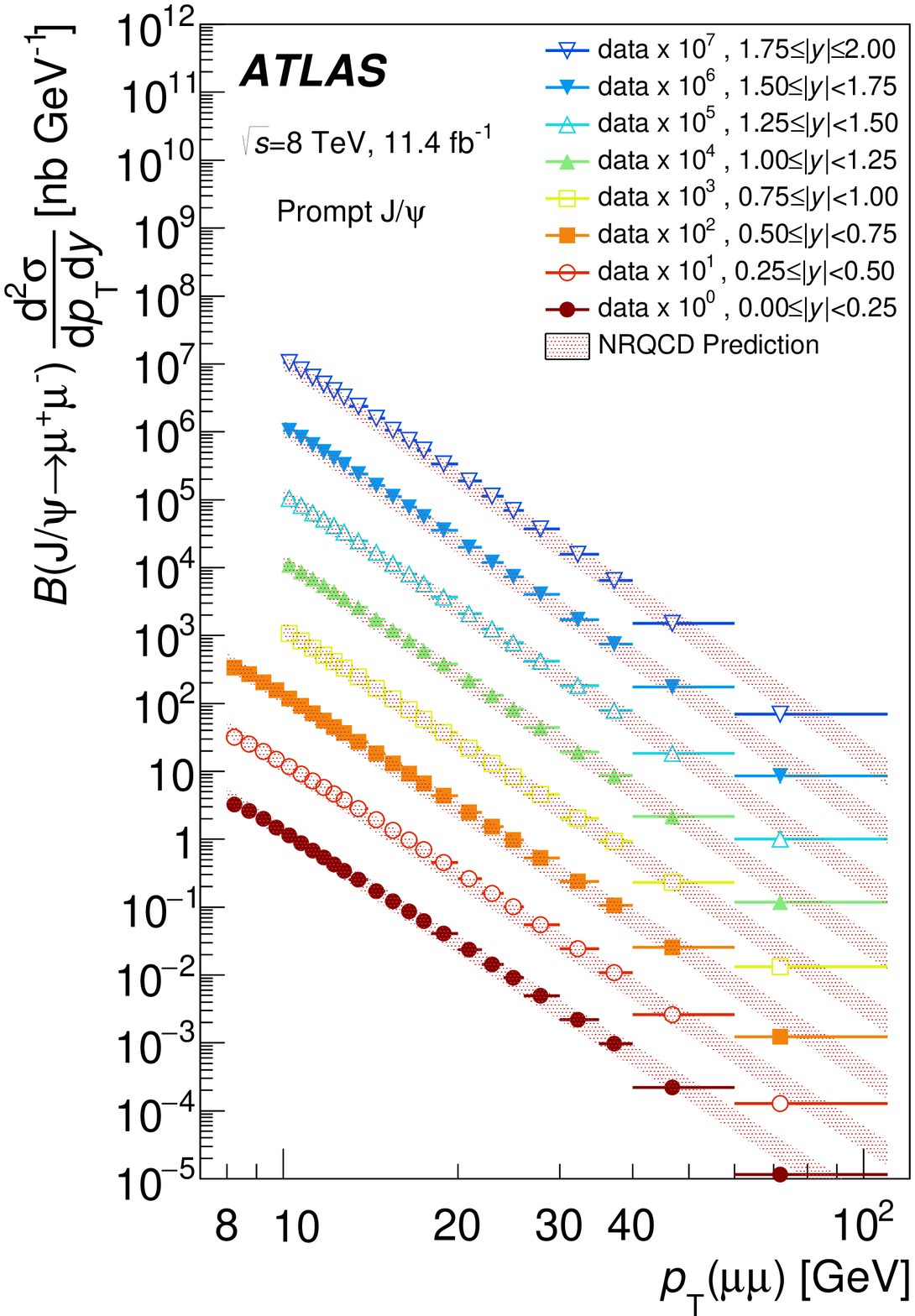}
    \includegraphics[width=0.44\textwidth]{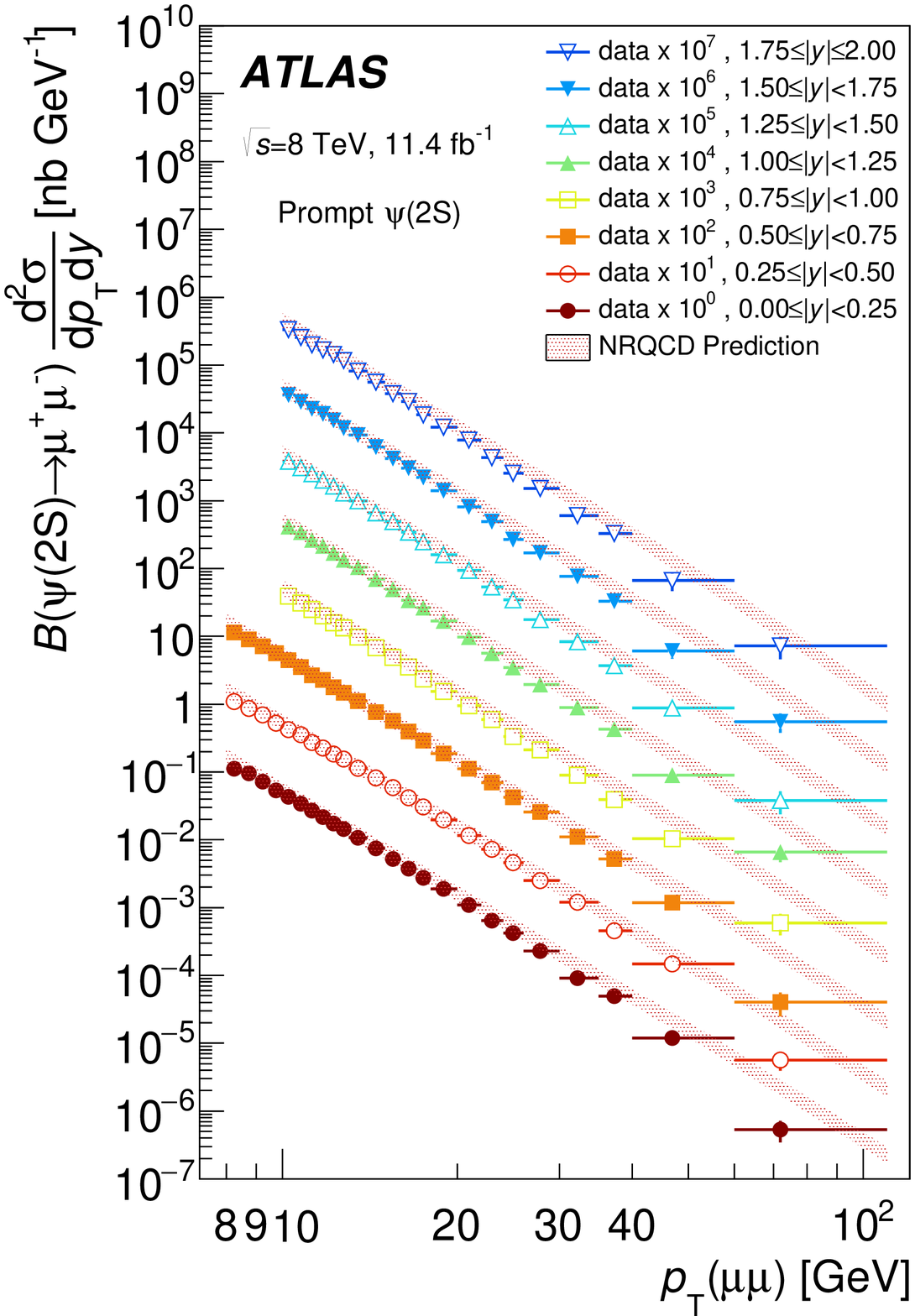}\hfil
    \caption{The differential prompt cross-section times dimuon branching fraction of \jpsi\ (left) and \psiprime\ (right) as a function 
    of $\pt(\mu\mu)$ for each slice of rapidity. 
    The top (bottom) row shows the 7~\TeV\ (8~\TeV) results.
    For each increasing rapidity slice, an additional scaling factor of 10 is applied to the plotted points for visual clarity. The
      centre of each bin on the horizontal axis represents the mean of the weighted $\pt$ distribution. The
      horizontal error bars represent the range of $\pt$ for the bin, and the vertical error bar covers 
      the statistical and systematic 
      uncertainty (with the same multiplicative scaling applied). 
      The NLO NRQCD theory predictions are also shown.}
    \label{fig:res:xSecP}
  \end{center}
\end{figure} 

\begin{figure} [!ht]
  \begin{center} 
    \includegraphics[width=0.44\textwidth]{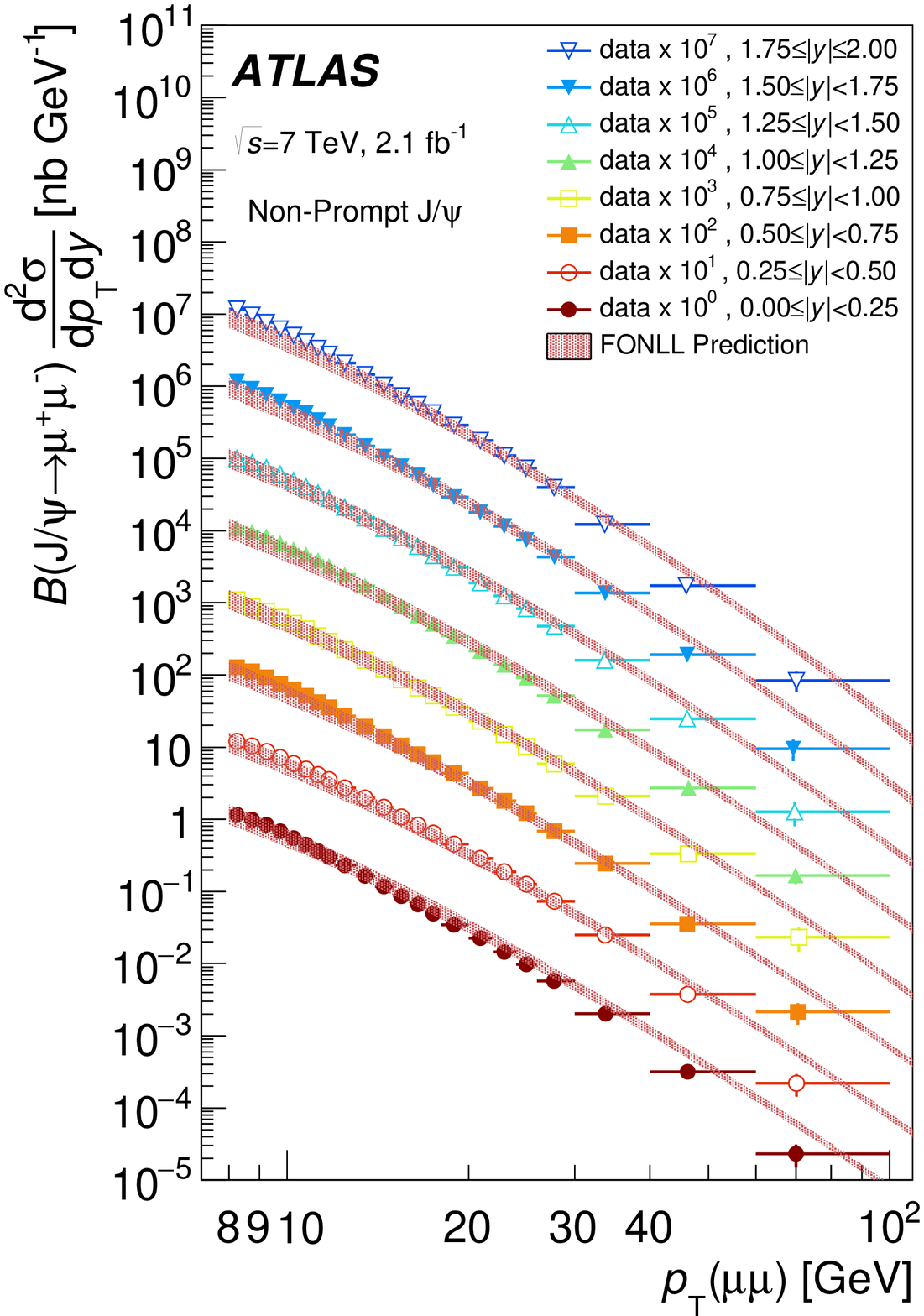}
    \includegraphics[width=0.44\textwidth]{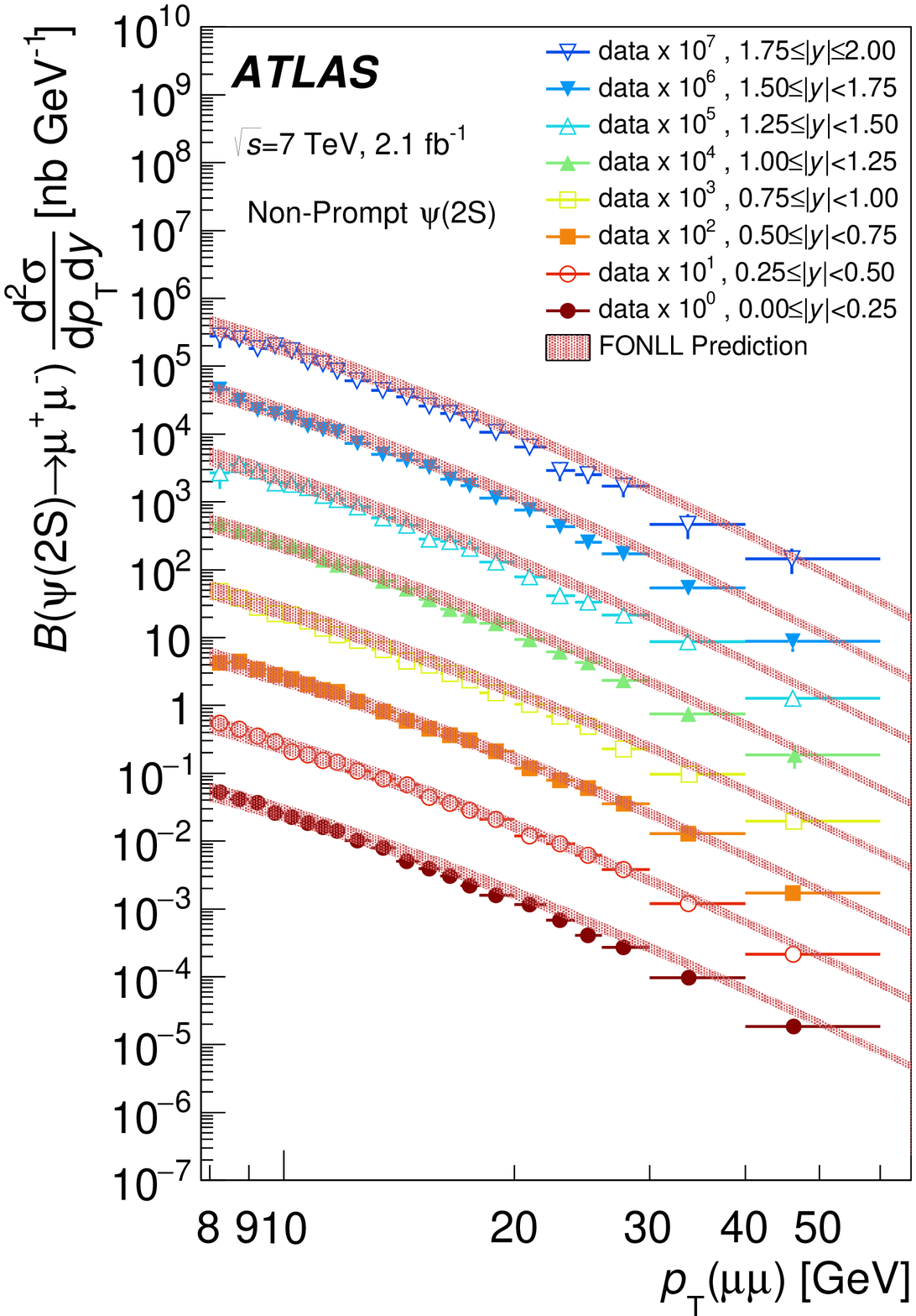}\hfil
    \includegraphics[width=0.44\textwidth]{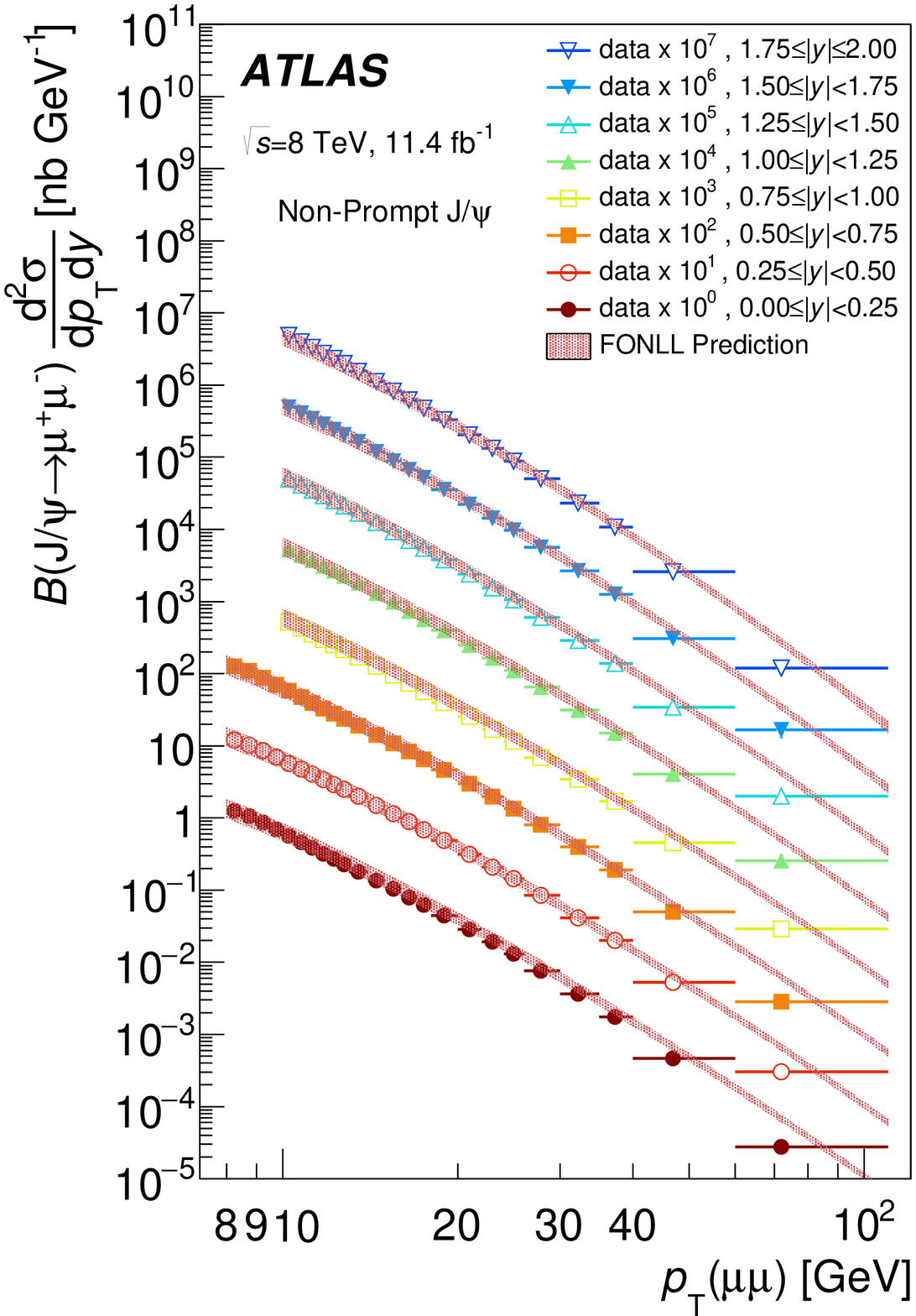}
    \includegraphics[width=0.44\textwidth]{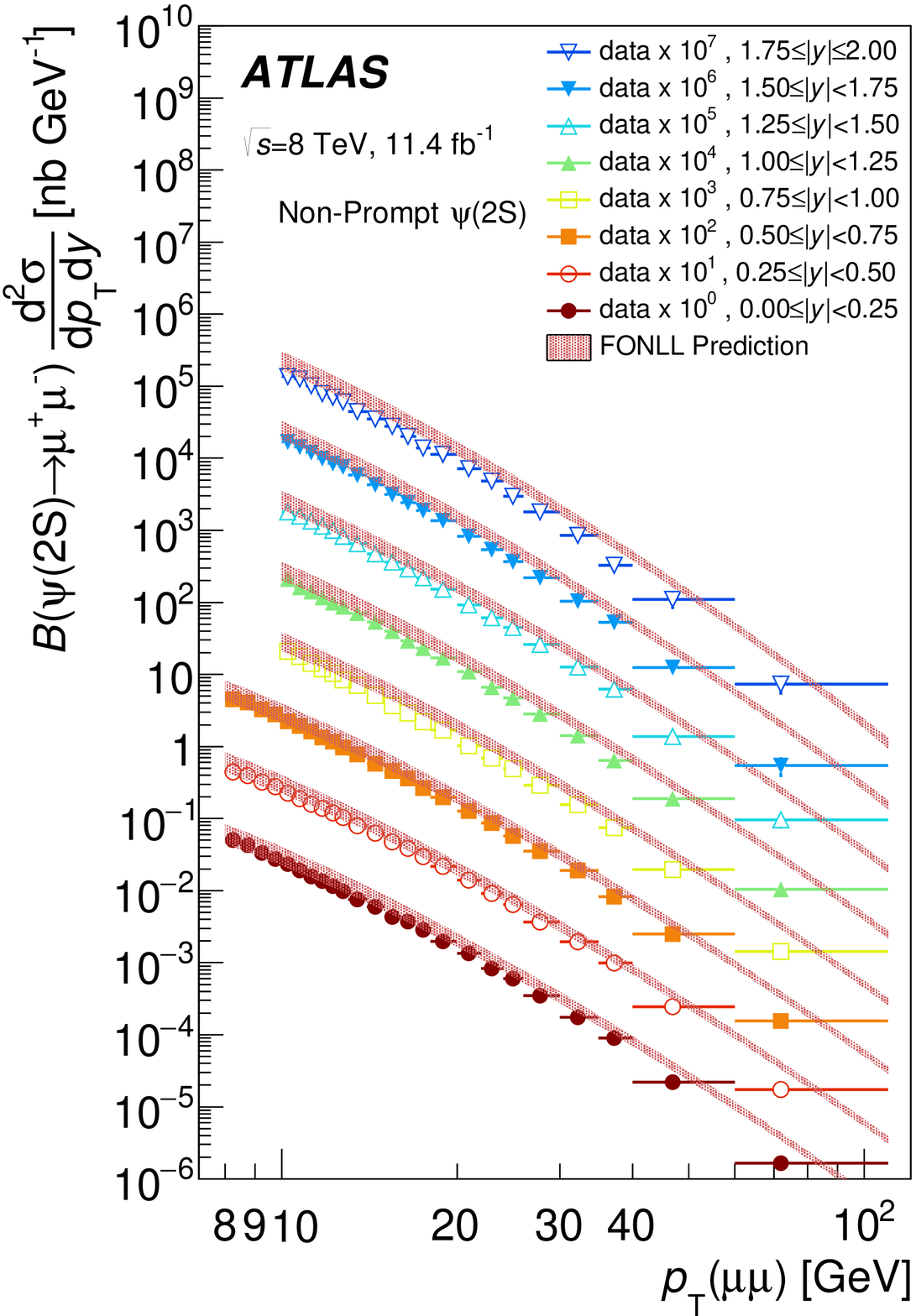}\hfil
    \caption{The differential non-prompt cross-section times dimuon branching fraction of \jpsi\ (left) and \psiprime\ (right) as a function 
    of $\pt(\mu\mu)$ for each slice of rapidity. 
    The top (bottom) row shows the 7~\TeV\ (8~\TeV) results.
    For each increasing rapidity slice, an additional scaling factor of 10 is applied to the plotted points for visual clarity. The
      centre of each bin on the horizontal axis represents the mean of the weighted $\pt$ distribution. The
      horizontal error bars represent the range of $\pt$ for the bin, and the vertical error bar covers the statistical
      and systematic uncertainty (with the same multiplicative scaling applied).
      The FONLL theory predictions are also shown.}
    \label{fig:res:xSecNP}
  \end{center}
\end{figure}

\item[Non-prompt production fractions] \hfill 

The results for the fractions of non-prompt production relative to the inclusive production of \jpsi\ and \psiprime\, are presented as a function of $\pt$ for slices of rapidity in Figure~\ref{fig:res:NPF}. 
In each rapidity slice, the non-prompt fraction is seen to increase as a function of $\pt$ and has no strong dependence on either rapidity or centre-of-mass energy.

\begin{figure} [!ht]
  \begin{center}
    \includegraphics[width=0.44\textwidth]{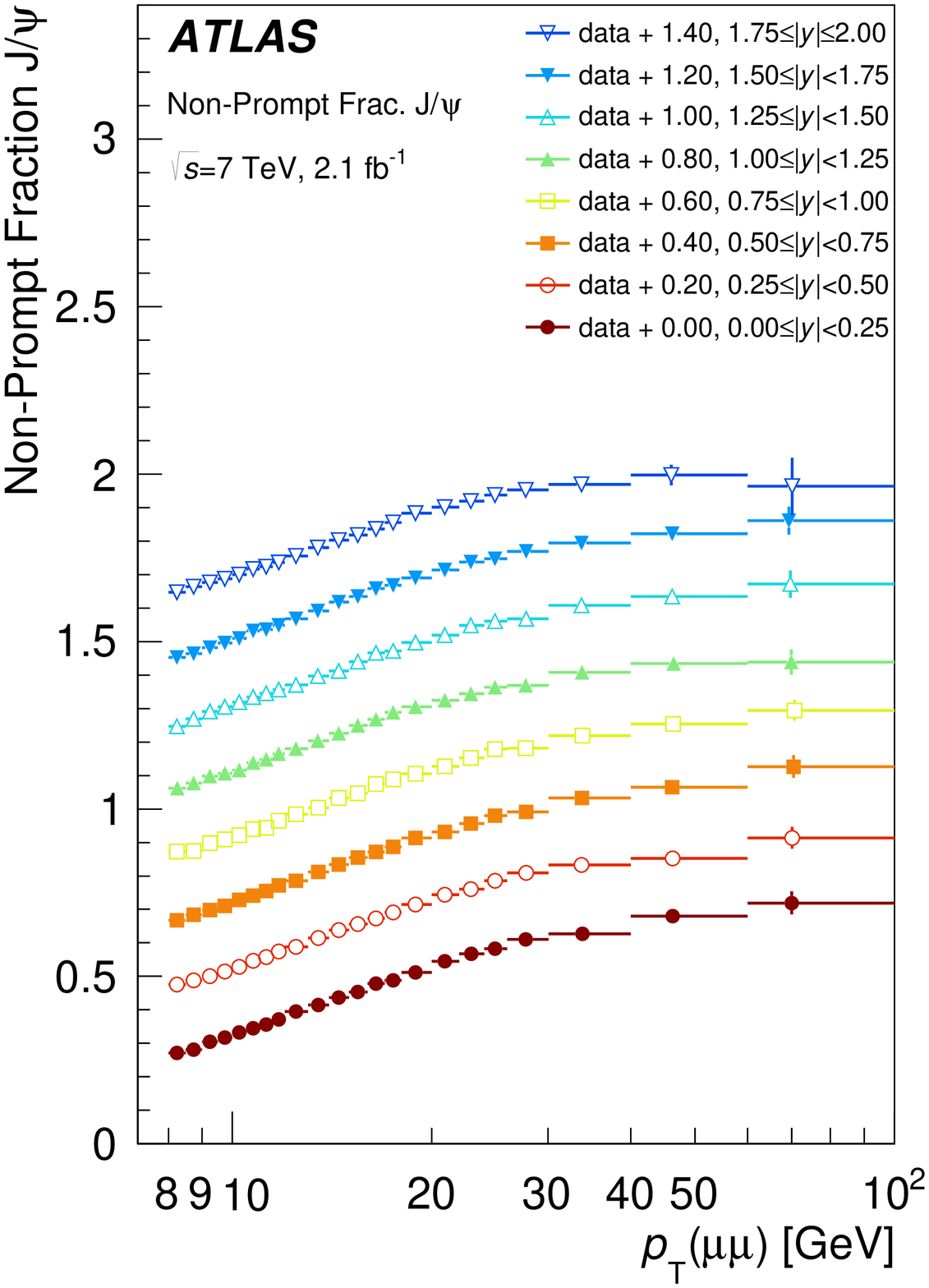} 
    \includegraphics[width=0.44\textwidth]{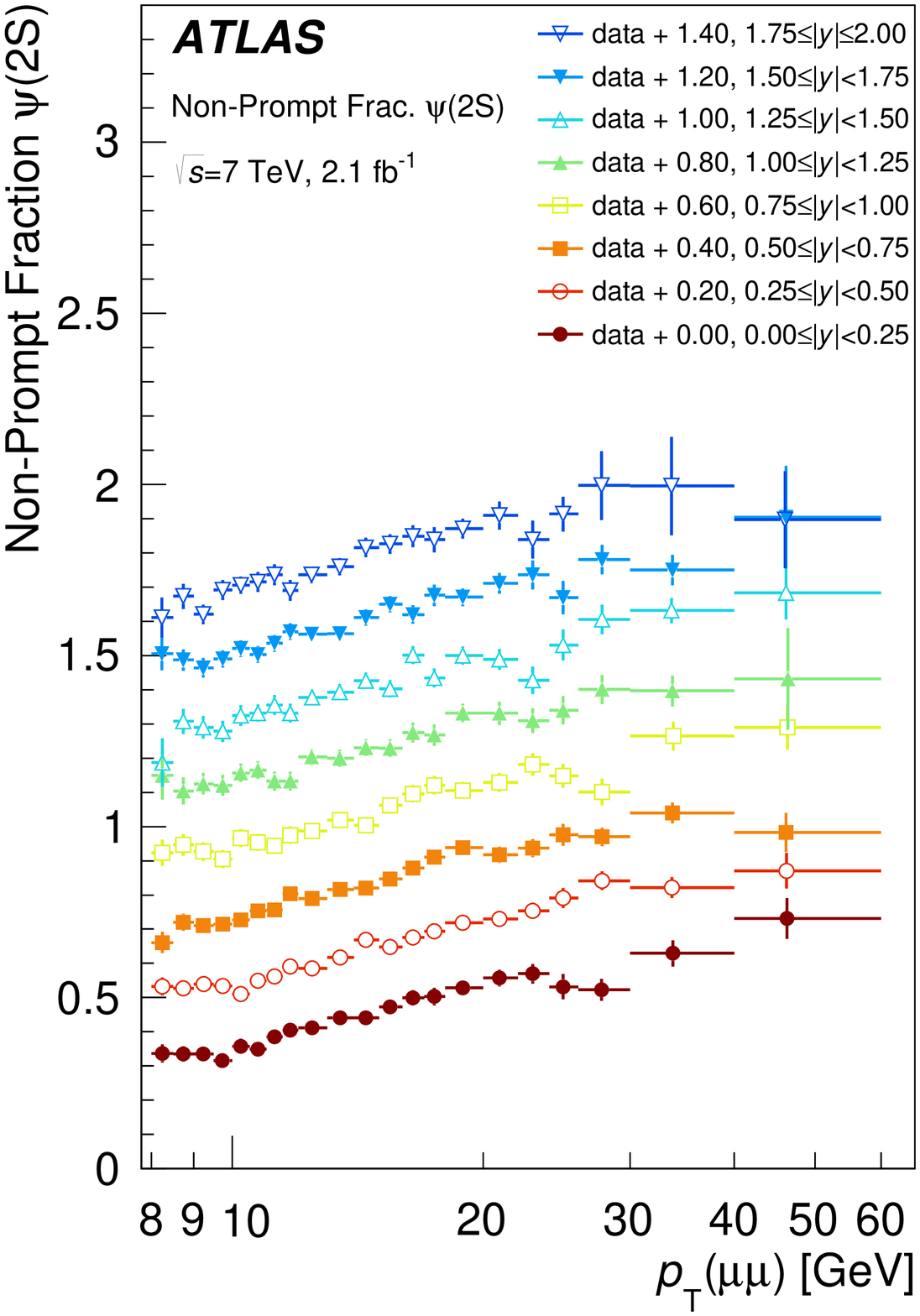}\hfil\\
    \includegraphics[width=0.44\textwidth]{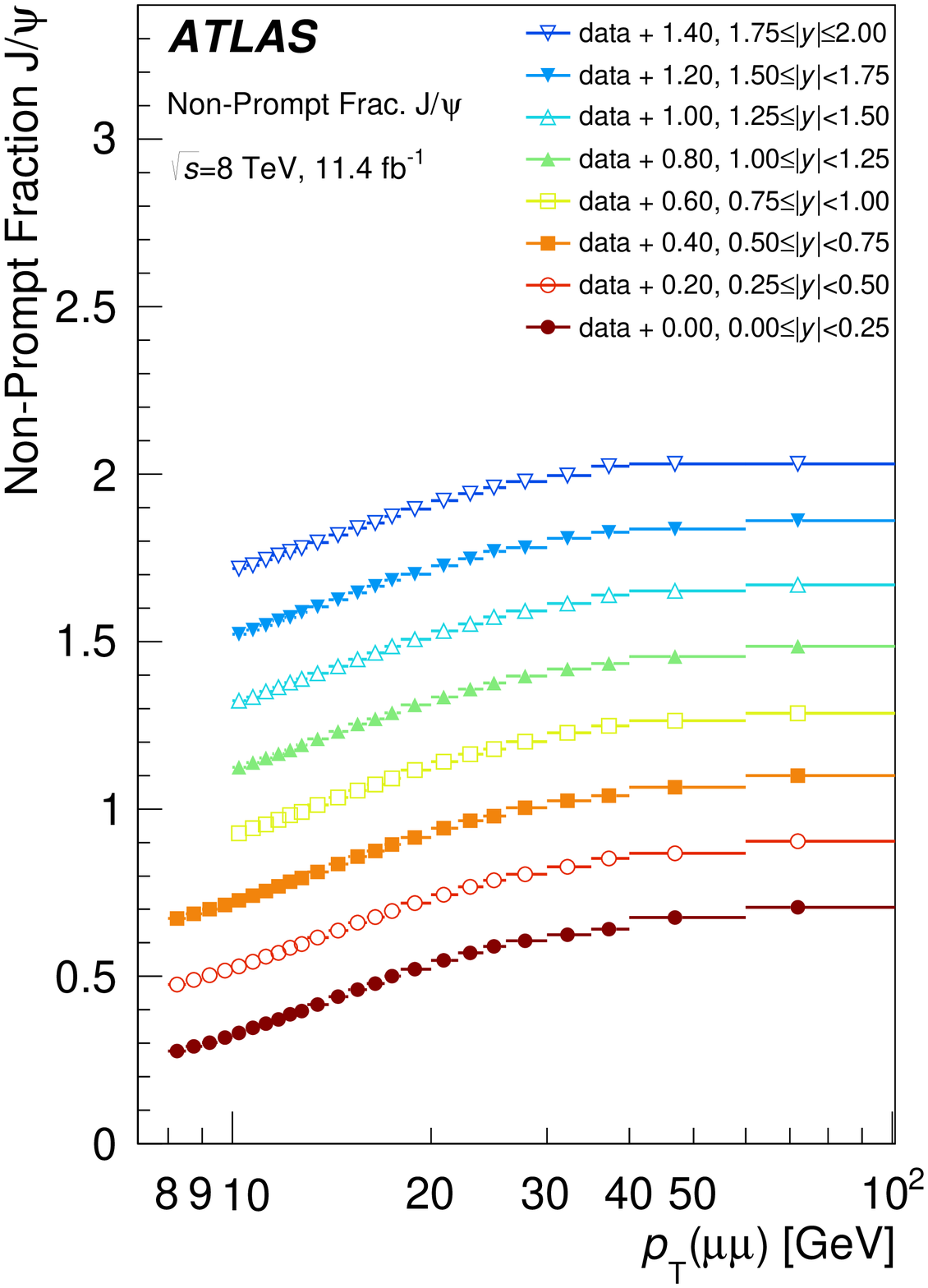}
    \includegraphics[width=0.44\textwidth]{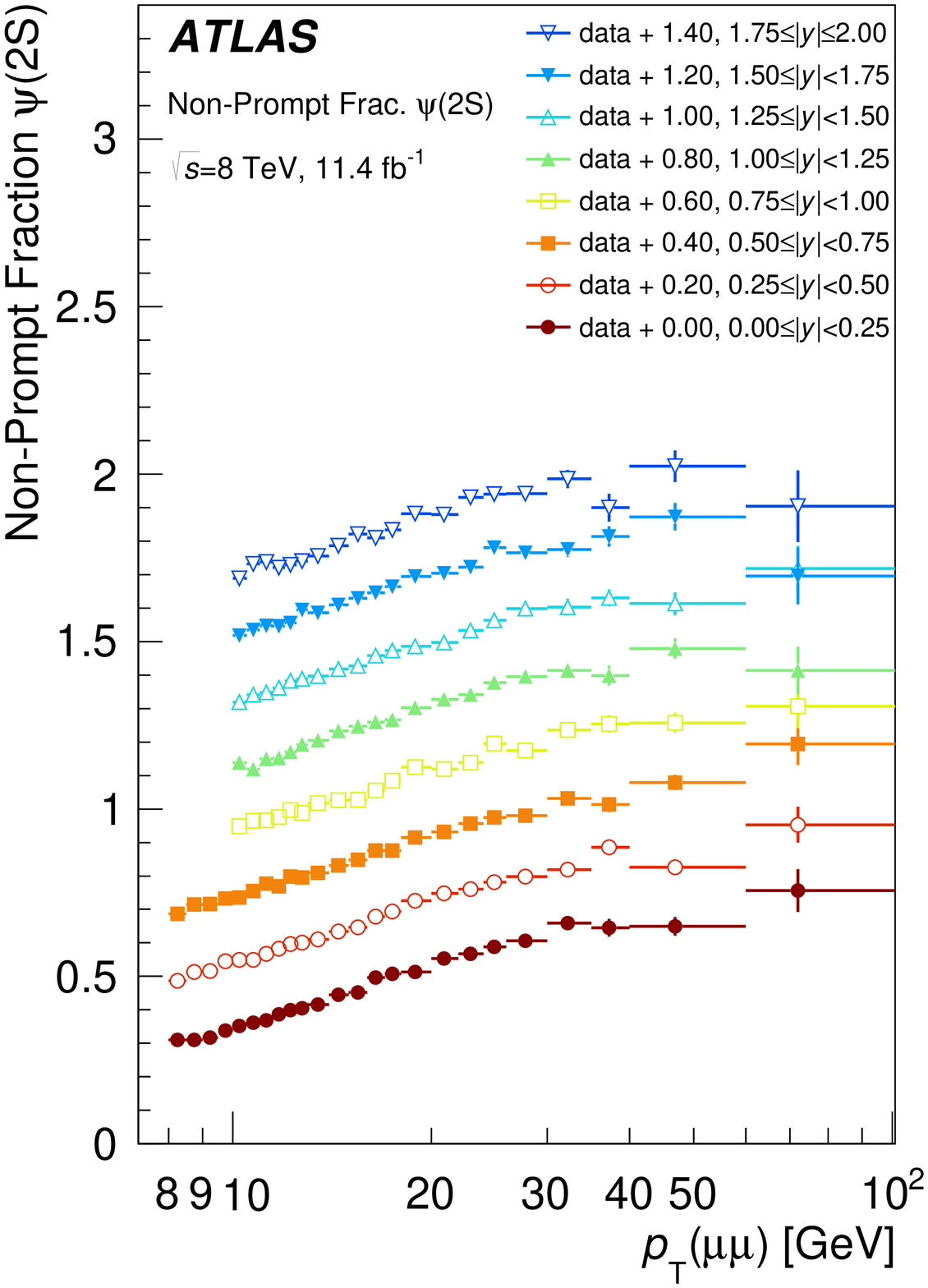}\hfil
    \caption{The non-prompt fraction of \jpsi\ (left) and \psiprime\ (right), as a function of $\pt(\mu\mu)$ for each slice of rapidity. 
    The top (bottom) row shows the 7~\TeV\ (8~\TeV) results.
      For each increasing rapidity slice, an additional factor of 0.2 is applied to the plotted points for visual clarity. The
      centre of each bin on the horizontal axis represents the mean of the weighted $\pt$ distribution. The
      horizontal error bars represent the range of $\pt$ for the bin, and the vertical error bar covers the statistical
    and systematic  uncertainty (with the same multiplicative scaling applied).}
    \label{fig:res:NPF}
  \end{center}
\end{figure}

\textit{Production ratios of \textmd{\psiprime} to \textmd{\jpsi} }

Figure~\ref{fig:res:PNP_Ratio} shows the ratios of \psiprime\ to \jpsi\ decaying to a muon pair in prompt and non-prompt processes,
 presented as a function of $\pt$ for slices of rapidity. The non-prompt ratio is shown to be relatively flat across the considered range of $\pT$,
for each slice of rapidity.
For the prompt ratio, a slight increase as a function of $\pt$ is observed, with no strong dependence on rapidity or centre-of-mass energy.

\begin{figure} [!ht]
  \begin{center}
    \includegraphics[width=0.44\textwidth]{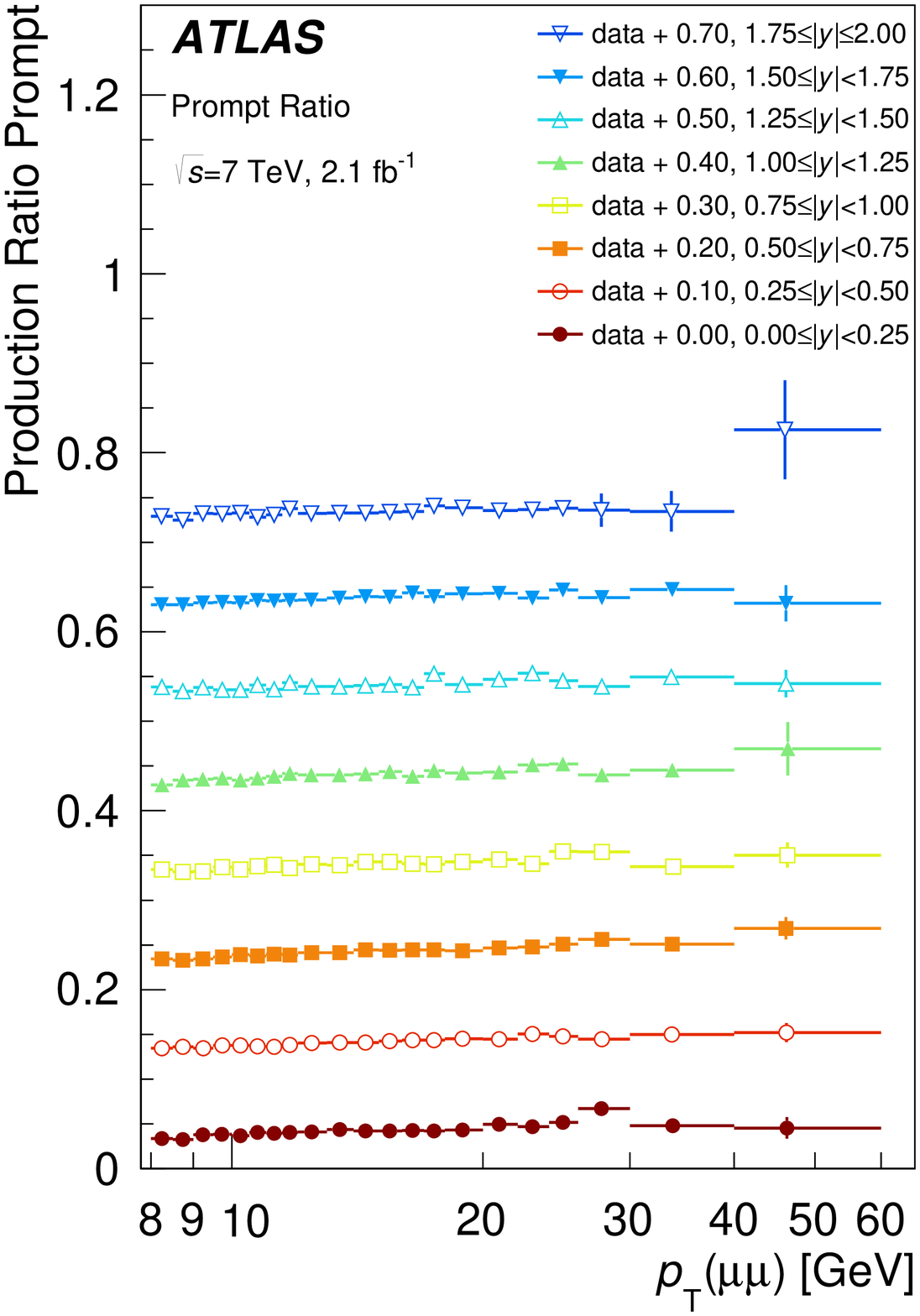} 
    \includegraphics[width=0.44\textwidth]{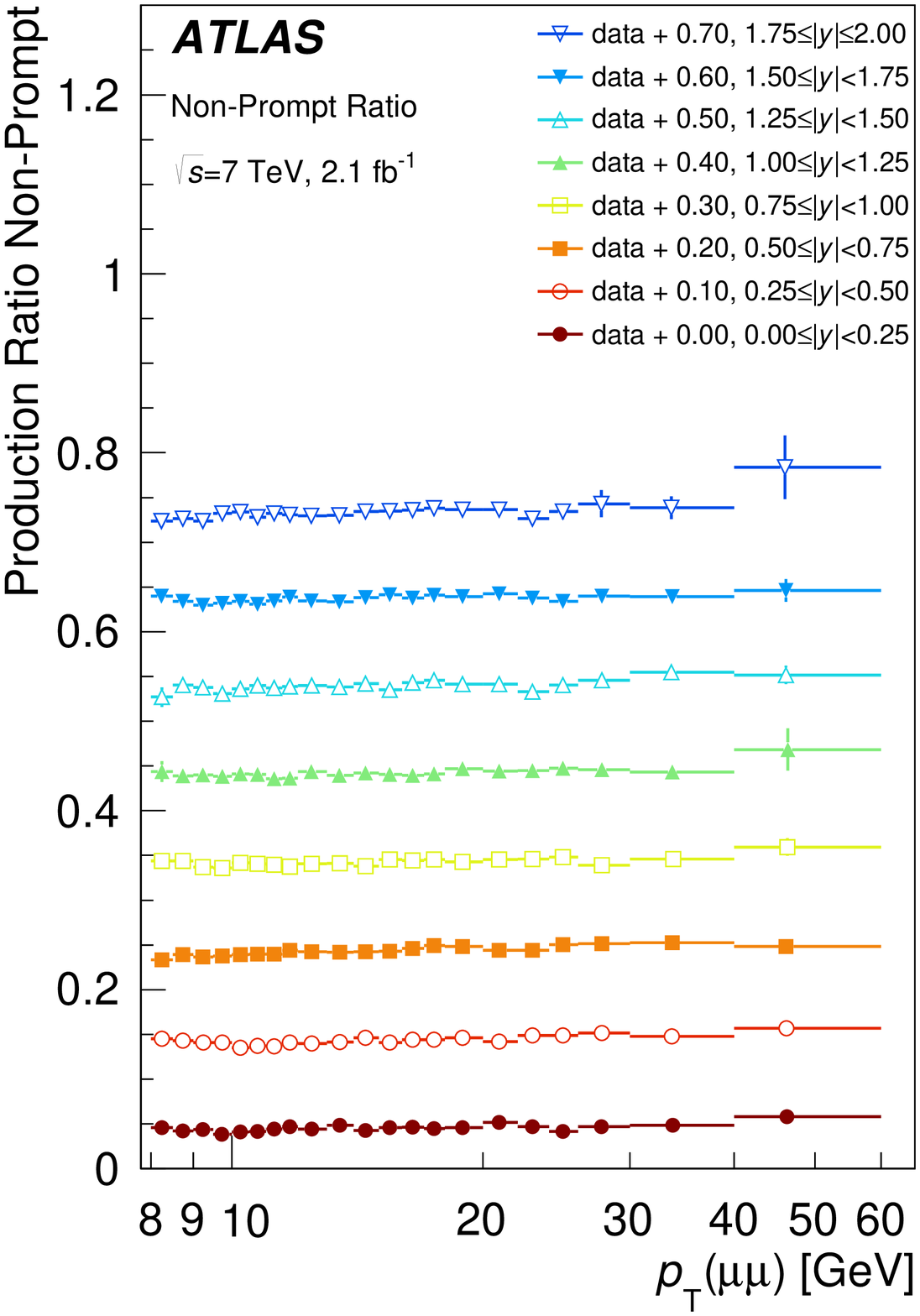}\hfil\\
    \includegraphics[width=0.44\textwidth]{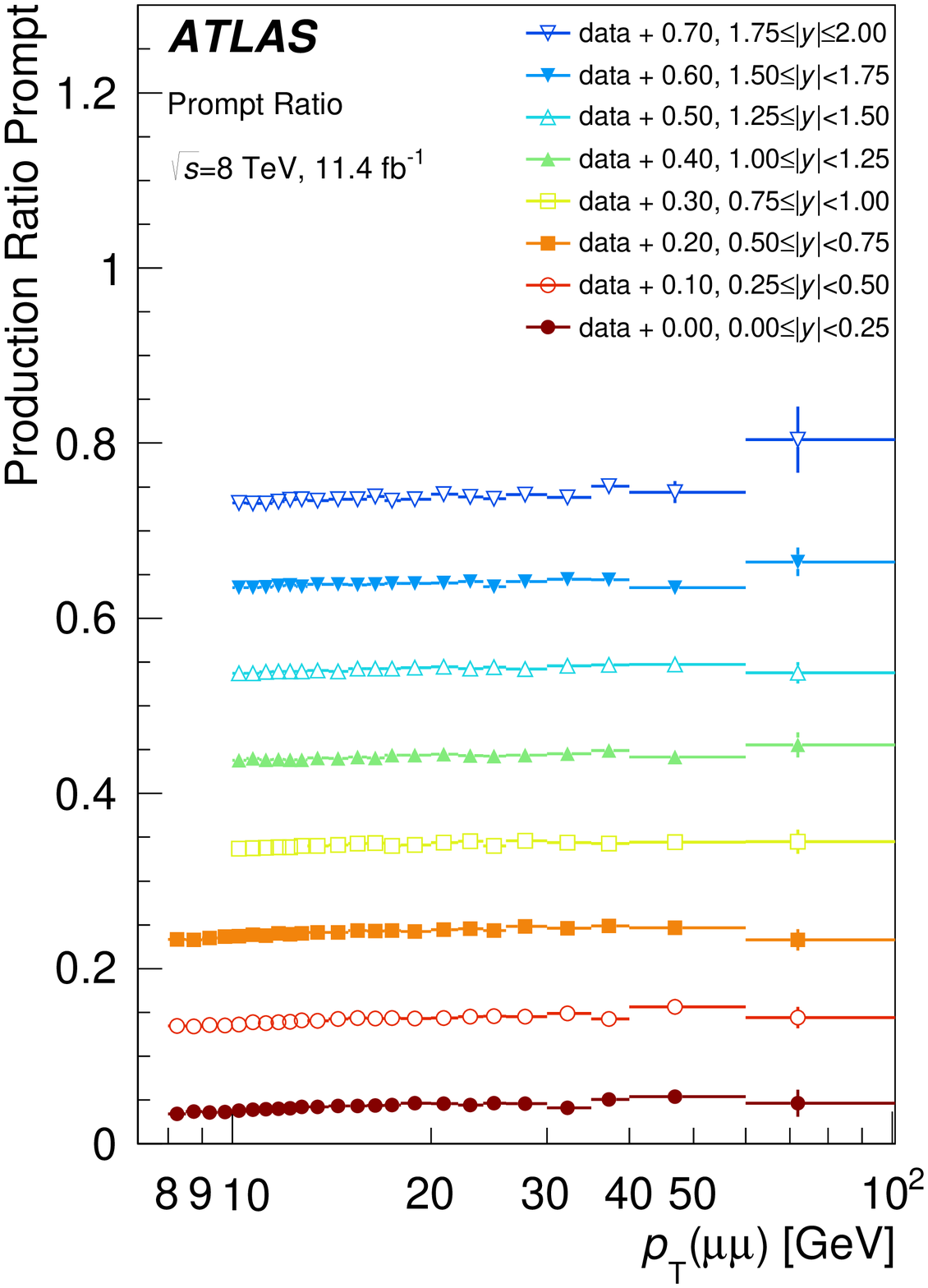}
    \includegraphics[width=0.44\textwidth]{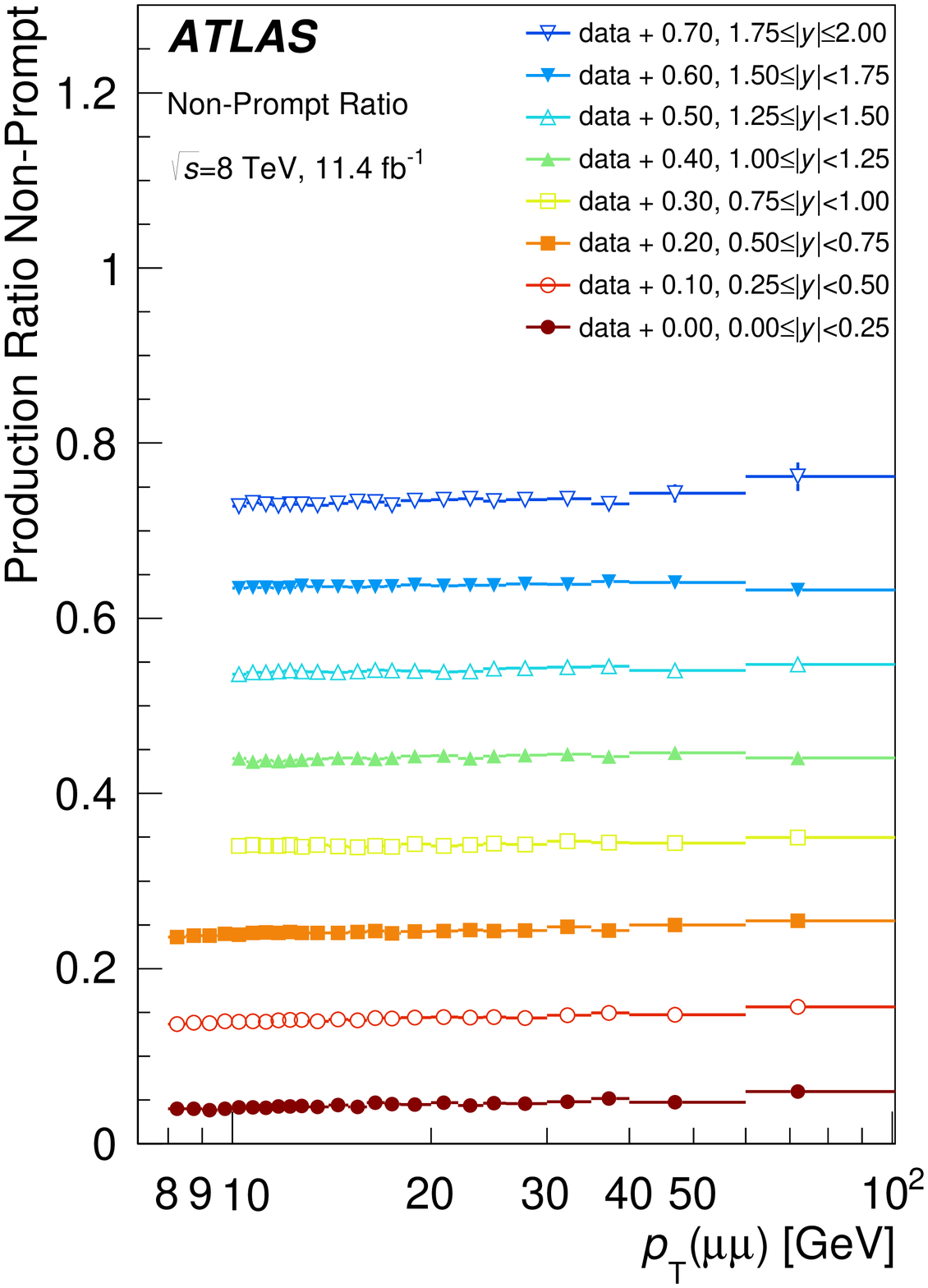}\hfil
    \caption{The ratio of \psiprime\ to \jpsi\ production times dimuon branching fraction for prompt (left) and non-prompt (right) processes
    as a function of $\pt(\mu\mu)$  for each of the slices of rapidity. 
    For each increasing rapidity slice, an additional factor of 0.1 is applied to the plotted points for visual clarity.
    The top (bottom) row shows the 7~\TeV\ (8~\TeV) results.
      The centre of each bin on the horizontal axis represents the mean of the weighted $\pt$ distribution. The
      horizontal error bars represent the range of $\pt$ for the bin, and the vertical error bar covers the statistical and systematic uncertainty.}
    \label{fig:res:PNP_Ratio}
  \end{center}
\end{figure} 

\clearpage

\item[Comparison with theory] \hfill

For prompt production, as shown in Figure~\ref{fig:xsecPtheoryRatio}, the ratio of the NLO NRQCD theory calculations~\cite{NRQCD1} 
to data, as a function of~$\pt$ and in slices of rapidity, is provided for \Jpsi\ and \psiprime\ at both the 7 and 8~\TeV\ centre-of-mass energies.
The theory predictions are 
based on the long-distance matrix elements (LDMEs) from Refs.~\cite{NRQCD1,Ma:2010vd}, with uncertainties originating from the
choice of scale, charm quark mass and LDMEs (see Refs.~\cite{NRQCD1,Ma:2010vd} for more details).
Figure~\ref{fig:xsecPtheoryRatio} shows fair agreement between the theoretical calculation and the data points for the whole $\pt$ range. 
The ratio of theory to data does not depend on rapidity.

\begin{figure} [!ht]
  \begin{center}
    \includegraphics[width=0.44\textwidth]{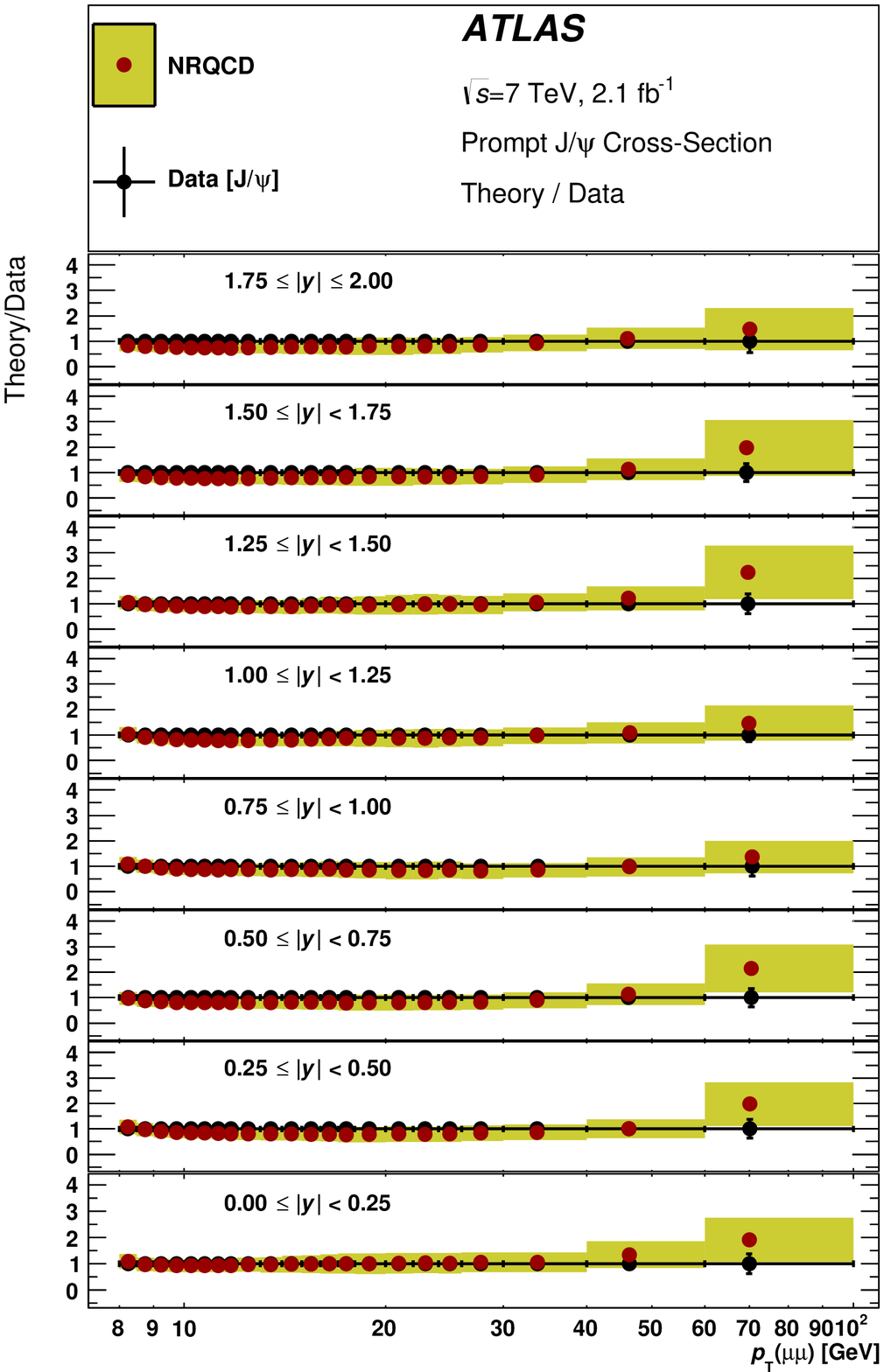} 
    \includegraphics[width=0.44\textwidth]{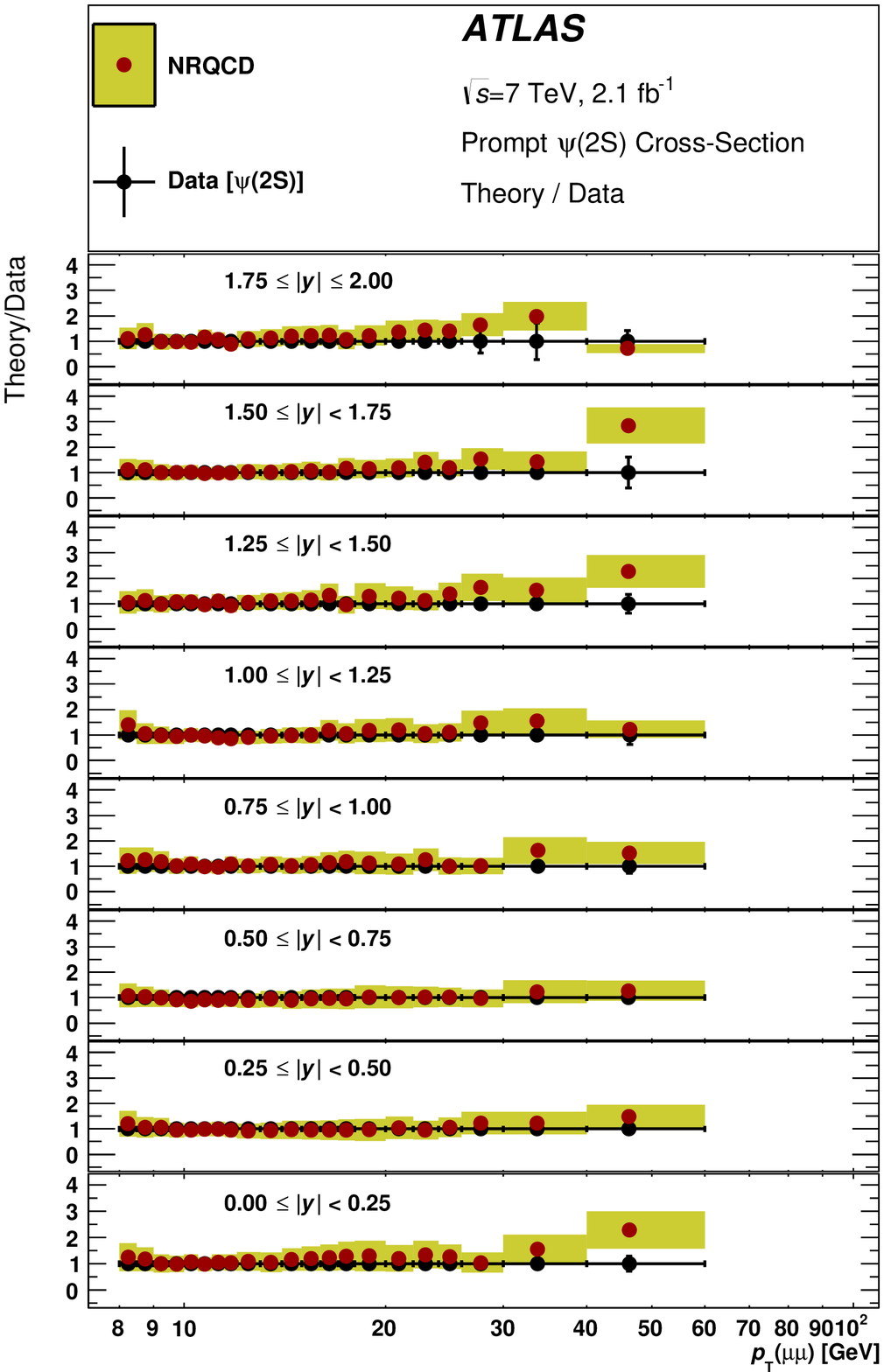}\hfil\\
    \includegraphics[width=0.44\textwidth]{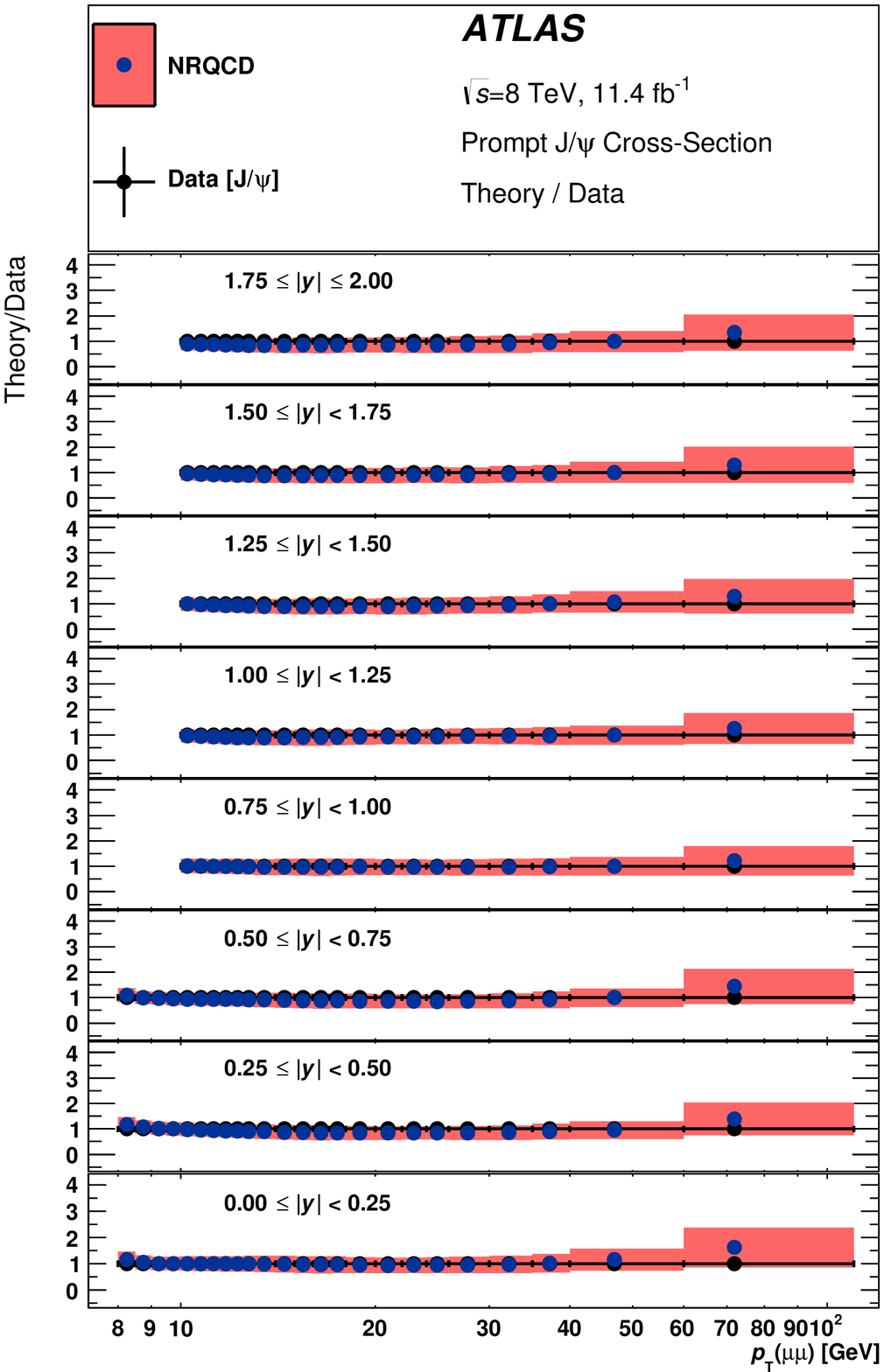}
    \includegraphics[width=0.44\textwidth]{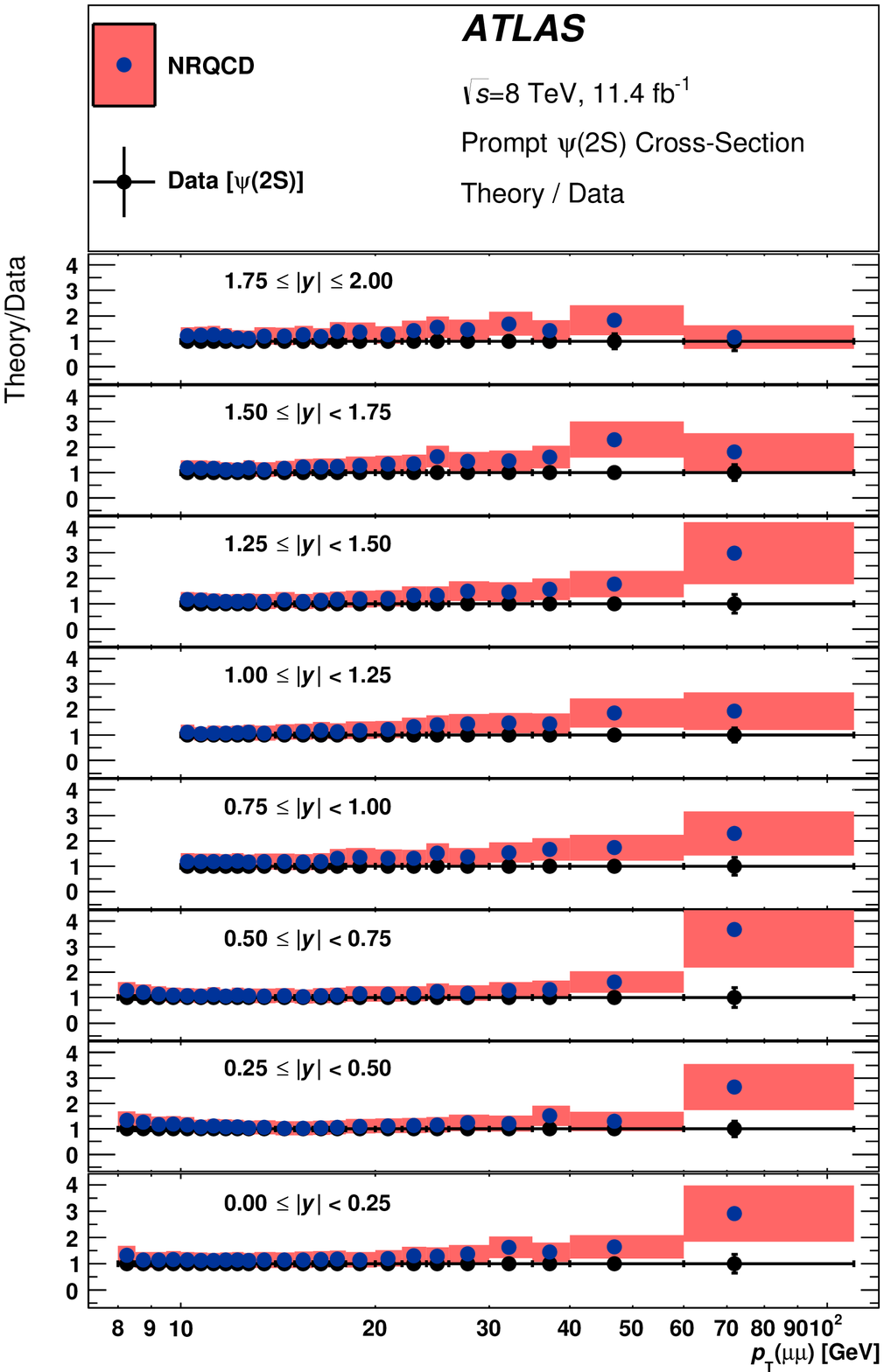}\hfil
    \caption{The ratios of the NRQCD theoretical predictions to data are presented for the differential prompt cross-section of \jpsi\ (left) and \psiprime\ (right) as a function 
    of $\pt(\mu\mu)$ for each rapidity slice. 
    The top (bottom) row shows the 7~\TeV\ (8~\TeV) results.
    The error on the data is the relative error of each data point, while the error bars on the theory prediction are the relative error of each theory point.}
    \label{fig:xsecPtheoryRatio}
  \end{center}
\end{figure} 

For non-prompt $\psi$ production, comparisons are made to FONLL theoretical predictions~\cite{FONLL_2001,Cacciari:2012ny}, which describe the production of 
$b$-hadrons followed by their decay into $\psi+X$.
Figure~\ref{fig:xsecNPtheoryRatio} shows the ratios of $\jpsi$ and $\psiprime$ FONLL predictions to data, as a function of $\pt$ and in slices of rapidity, 
for centre-of-mass energies of~7 and 8~\TeV.
For $\jpsi$, agreement is generally good, but the theory predicts slightly harder $\pt$ spectra than observed in the data.
For $\psiprime$, the shapes of data and theory appear to be in satisfactory agreement, but the theory predicts higher yields than in the data.
There is no observed dependence on rapidity in the comparisons between theory and data for non-prompt \jpsi\ and \psiprime\ production.

\begin{figure} [!ht]
  \begin{center}
    \includegraphics[width=0.44\textwidth]{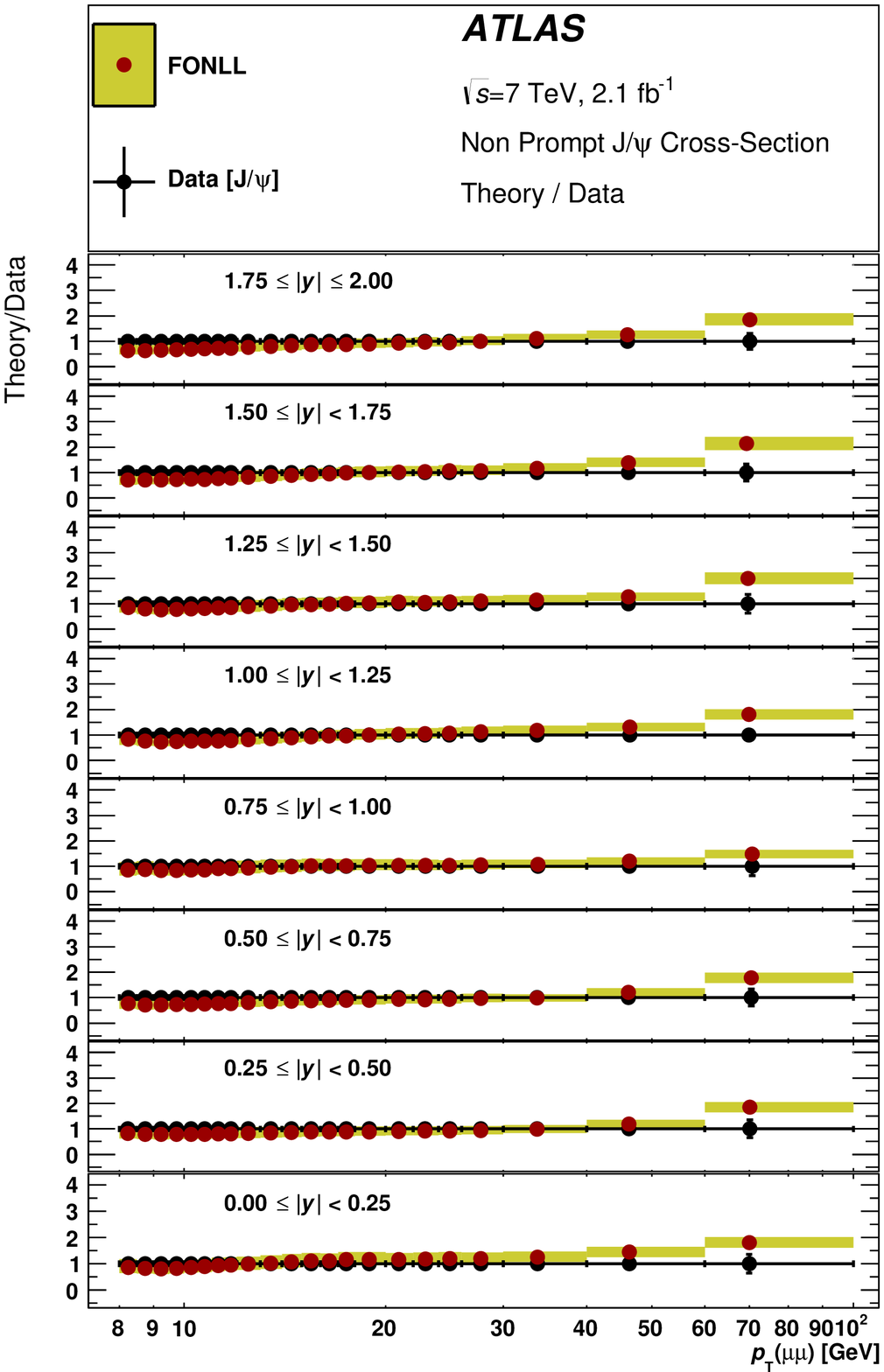} 
    \includegraphics[width=0.44\textwidth]{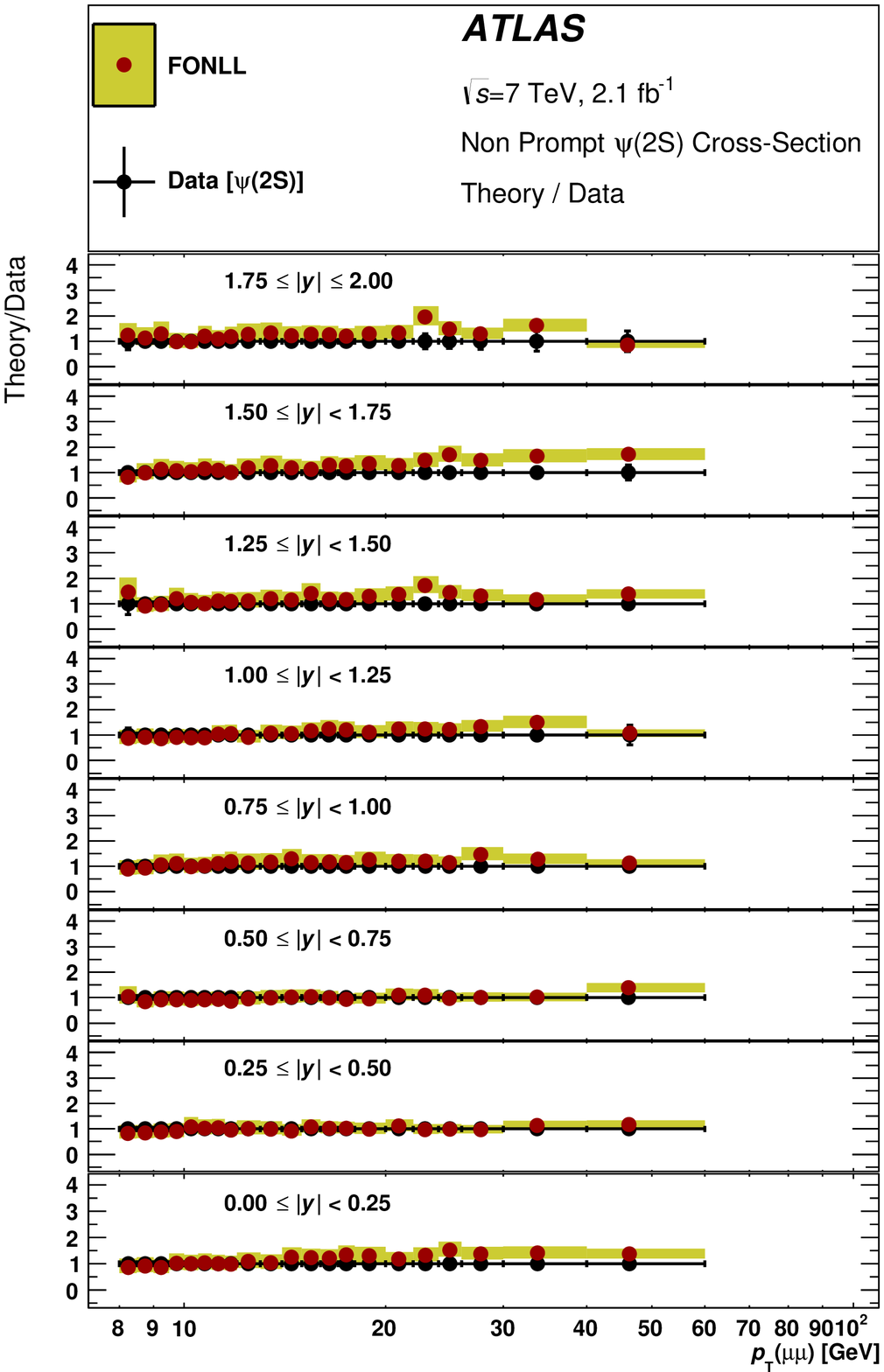}\hfil\\
    \includegraphics[width=0.44\textwidth]{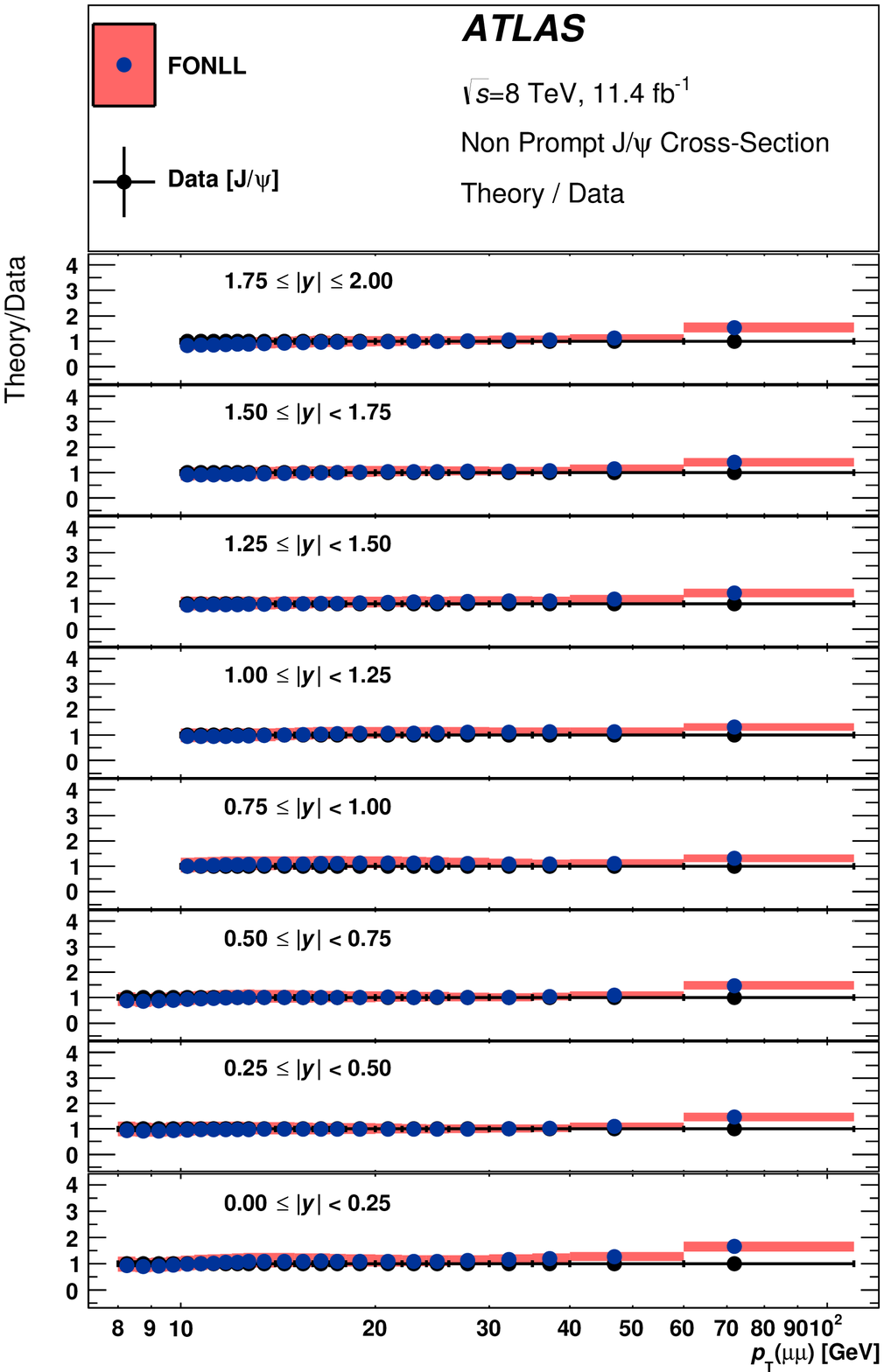}
    \includegraphics[width=0.44\textwidth]{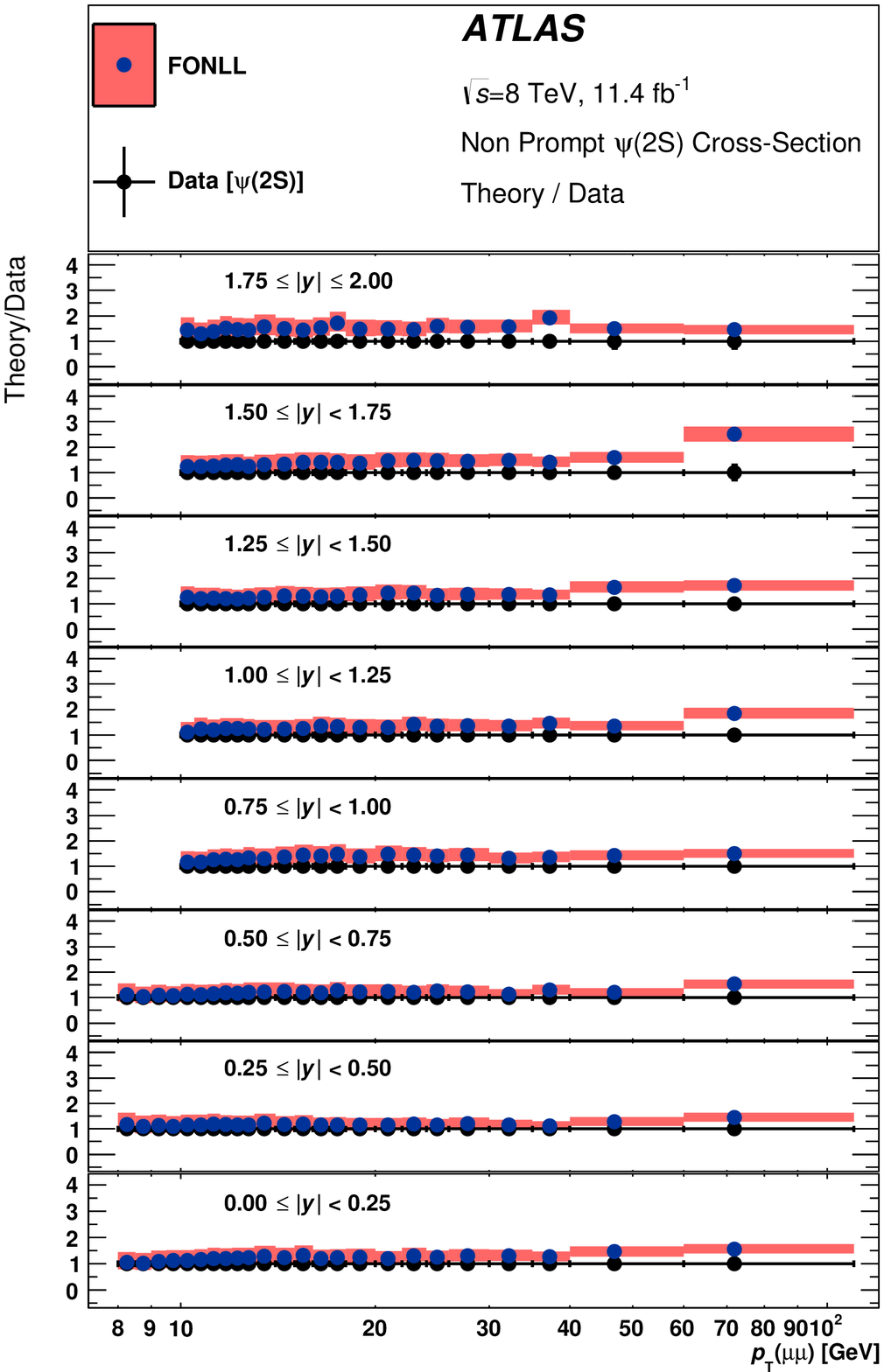}\hfil
    \caption{The ratio of the FONLL theoretical predictions to data are presented for the differential non-prompt 
    cross-section of \jpsi\ (left) and \psiprime\ (right) as a function 
    of $\pt(\mu\mu)$ for each rapidity slice. 
    The top (bottom) row shows the 7~\TeV\ (8~\TeV) results.
    The error on the data is the relative error of each data point, while the error bars on the theory prediction are the relative error of each theory point.}
    \label{fig:xsecNPtheoryRatio}
  \end{center}
\end{figure}

\item[Comparison of cross-sections 8~\TeV\  with 7~\TeV\ ] \hfill

It is interesting to compare the cross-section results between the two centre-of-mass energies, 
both for data and the theoretical predictions.

Figure~\ref{fig:xsecEnergyRatio} shows the 8~\TeV\  to 7~\TeV\  cross-section ratios of prompt and non-prompt $\jpsi$ and $\psiprime$ for both data sets.
For the theoretical ratios the uncertainties are neglected here, since the high correlation between them results in large cancellations.

Due to a finer granularity in $\pt$ for the 8~\TeV\  data, a weighted average of the 8~\TeV\ results is taken across equivalent intervals of the 7~\TeV\ data to enable direct comparisons.
Both data and theoretical predictions agree 
that the ratios become larger with increasing $\pt$,
however at the lower edge of the $\pT$ range the data tends to be slightly below theory.

\begin{figure} [!ht]
  \begin{center}
    \includegraphics[width=0.44\textwidth]{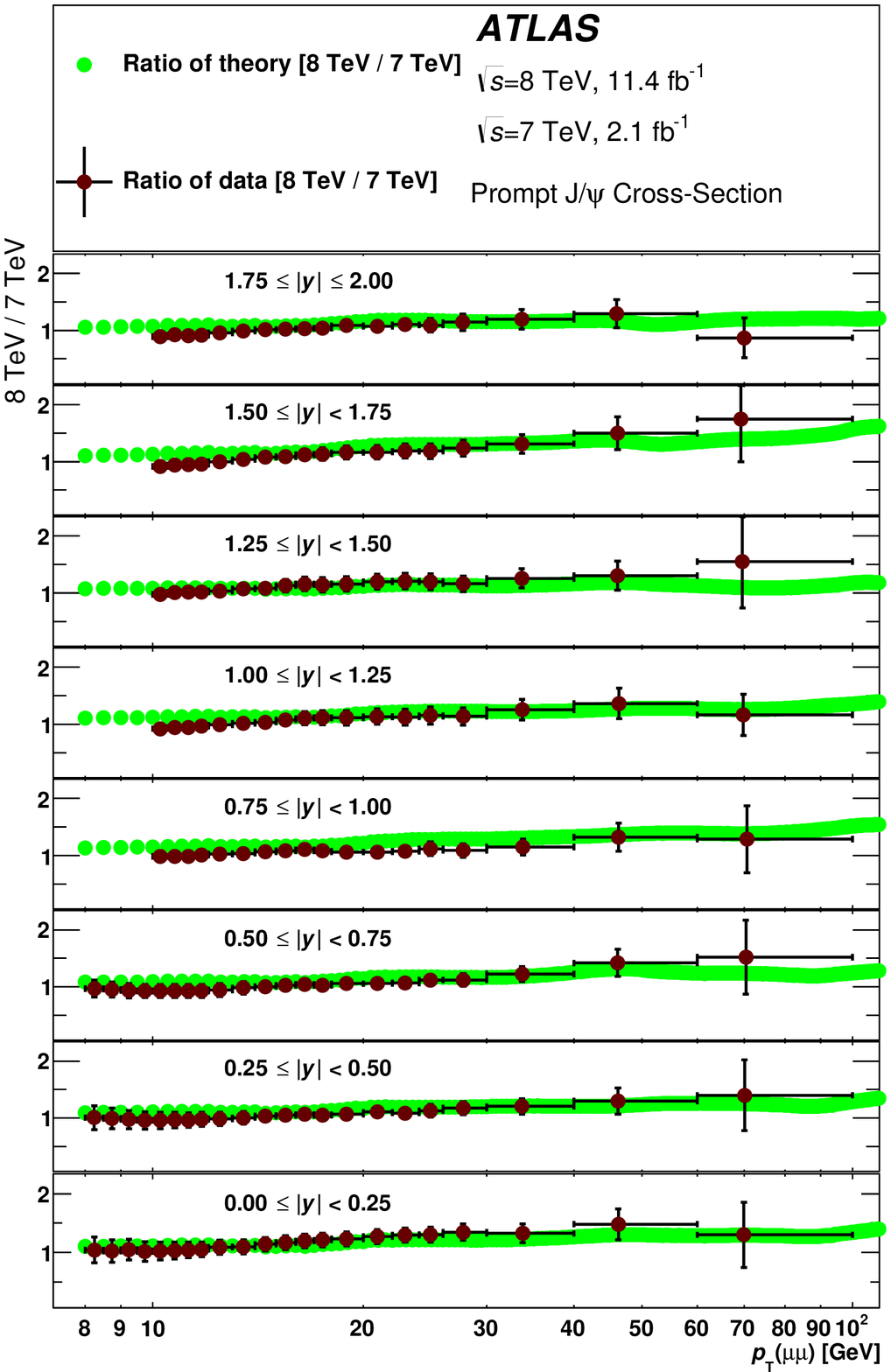} 
    \includegraphics[width=0.44\textwidth]{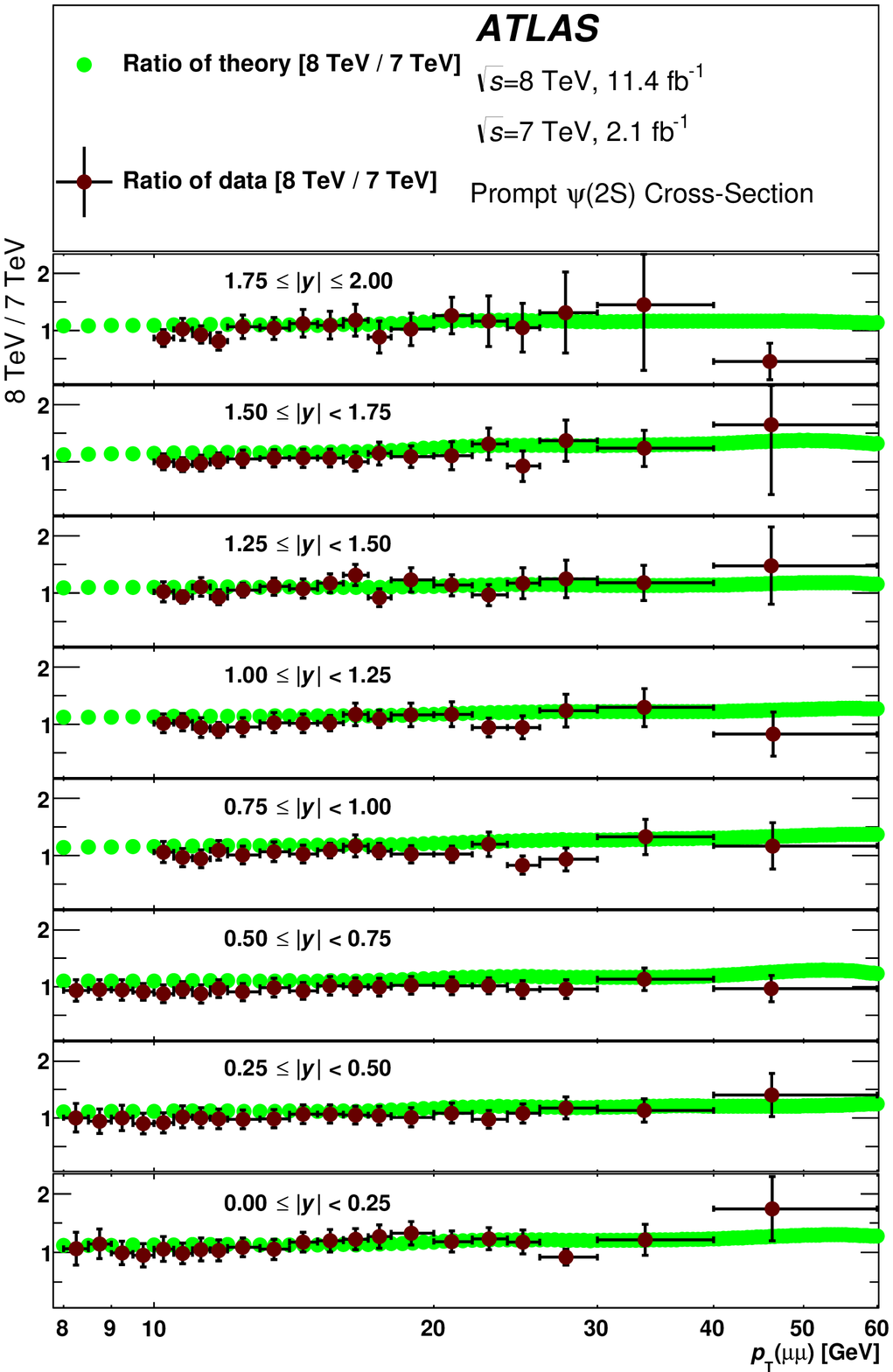}\hfil\\
    \includegraphics[width=0.44\textwidth]{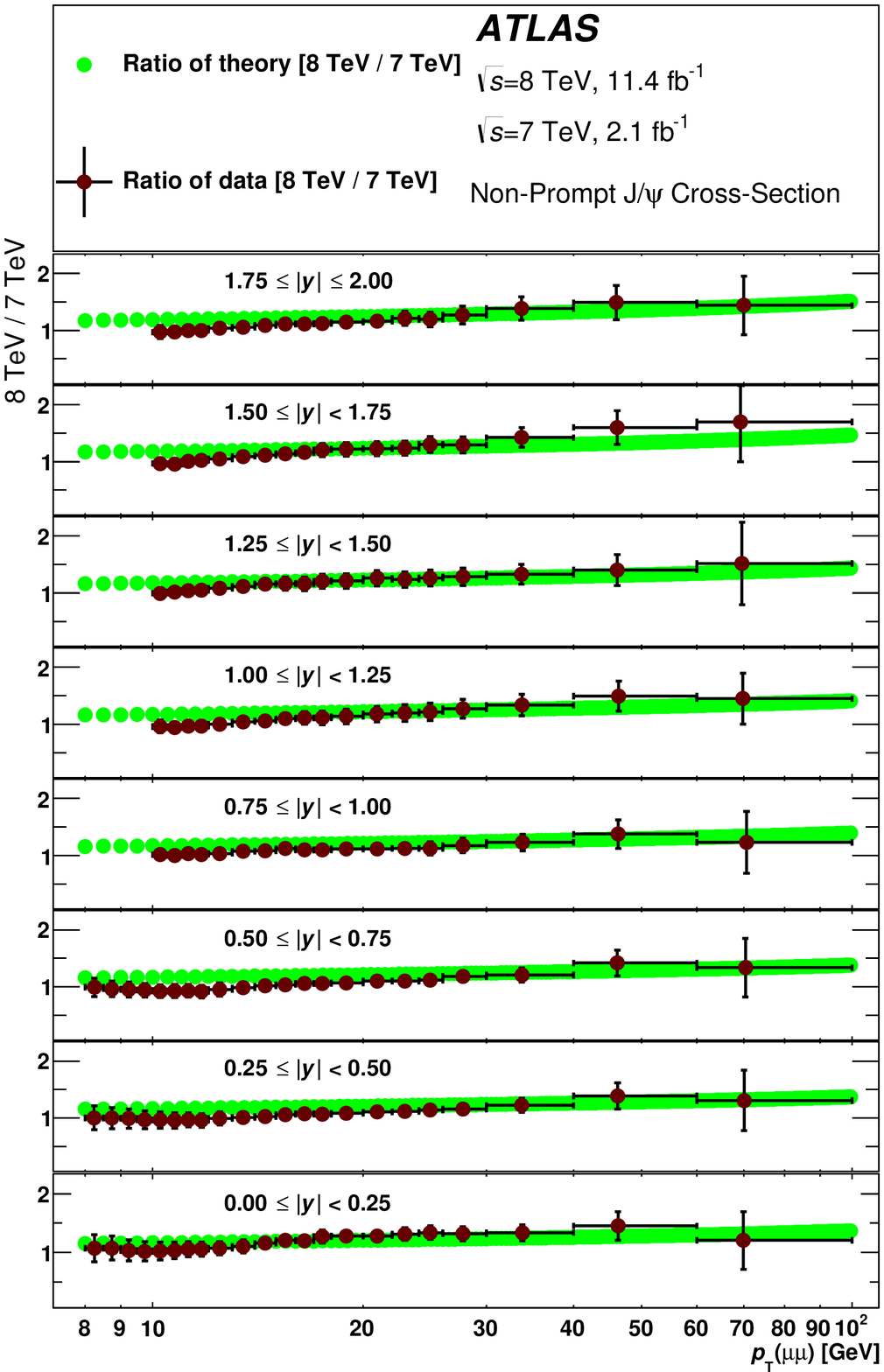}
    \includegraphics[width=0.44\textwidth]{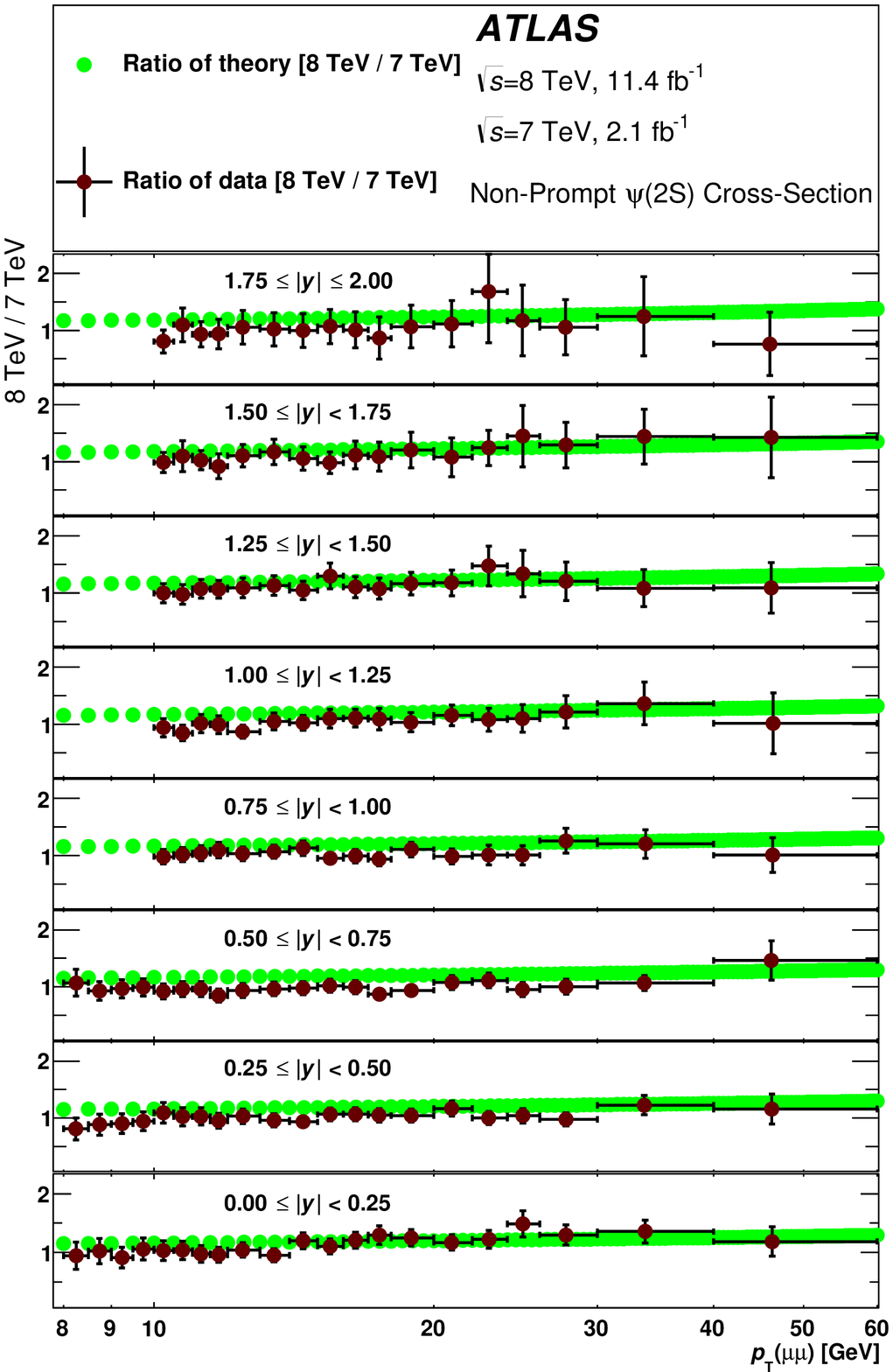}\hfil
    \caption{The ratio of the 8~\TeV\ and 7~\TeV\ differential cross-sections
	    are presented for prompt~(top) and non-prompt~(bottom) $\jpsi$~(left) and 
	    $\psiprime$~(right) for both data~(red points with error bars) and theoretical predictions~(green points).
	    The theoretical predictions used are NRQCD for prompt and FONLL for non-prompt production.
	    The uncertainty on the data ratio does not account for possible correlations between 7 and 8 \TeV\  data, and no uncertainty is shown 
	    for the ratio of theory predictions.
}
    \label{fig:xsecEnergyRatio}
  \end{center}
\end{figure}

\end{description}

%% file: Conclusions.tex
\section{Summary and conclusions}
\label{sec:conclusion}

The prompt and non-prompt production cross-sections, the non-prompt production fraction of the $\jpsi$ and $\psiprime$
decaying into two muons, the ratio of prompt $\psiprime$ to prompt $\jpsi$ production, and the ratio of non-prompt $\psiprime$ to non-prompt $\jpsi$ production
were measured in the rapidity range
$|y|<2.0$ for transverse momenta between $8$ and $110$~\GeV. This measurement was carried out using \lumiA (\lumiB) of $pp$
collision data at a centre-of-mass energy of $7$ \TeV\ ($8$ \TeV) recorded by the ATLAS experiment at the LHC. It is the latest in a series of 
related measurements of the production of charmonium states
made by ATLAS.
In line with previous measurements, the central values were obtained 
assuming isotropic $\psi \to \mu\mu$ decays.
Correction factors for these cross-sections, computed for a number of extreme spin-alignment scenarios, are between $-35\%$ and $+100\%$ at the lowest transverse momenta studied, and between $-14\%$ and $+9\%$ at the highest transverse momenta, depending on the specific scenario.

The ATLAS measurements presented here extend the range of existing measurements to higher transverse momenta, and to a higher collision energy of $\sqrt{s} = 8$ \TeV, and, in overlapping phase-space regions, are consistent with previous measurements made by ATLAS and other LHC experiments.
For the prompt production mechanism, the predictions from the NRQCD model, which includes colour-octet contributions with various matrix elements 
tuned to earlier collider data, are found to be in good agreement with the observed data points. 
For the non-prompt production, the fixed-order next-to-leading-logarithm calculations reproduce the data reasonably well, with a slight overestimation of the differential cross-sections at the highest 
transverse momenta reached in this analysis.

%% file: Acknowledgements.tex

We thank CERN for the very successful operation of the LHC, as well as the
support staff from our institutions without whom ATLAS could not be
operated efficiently.

We acknowledge the support of ANPCyT, Argentina; YerPhI, Armenia; ARC, Australia; BMWFW and FWF, Austria; ANAS, Azerbaijan; SSTC, Belarus; CNPq and FAPESP, Brazil; NSERC, NRC and CFI, Canada; CERN; CONICYT, Chile; CAS, MOST and NSFC, China; COLCIENCIAS, Colombia; MSMT CR, MPO CR and VSC CR, Czech Republic; DNRF and DNSRC, Denmark; IN2P3-CNRS, CEA-DSM/IRFU, France; GNSF, Georgia; BMBF, HGF, and MPG, Germany; GSRT, Greece; RGC, Hong Kong SAR, China; ISF, I-CORE and Benoziyo Center, Israel; INFN, Italy; MEXT and JSPS, Japan; CNRST, Morocco; FOM and NWO, Netherlands; RCN, Norway; MNiSW and NCN, Poland; FCT, Portugal; MNE/IFA, Romania; MES of Russia and NRC KI, Russian Federation; JINR; MESTD, Serbia; MSSR, Slovakia; ARRS and MIZ\v{S}, Slovenia; DST/NRF, South Africa; MINECO, Spain; SRC and Wallenberg Foundation, Sweden; SERI, SNSF and Cantons of Bern and Geneva, Switzerland; MOST, Taiwan; TAEK, Turkey; STFC, United Kingdom; DOE and NSF, United States of America. In addition, individual groups and members have received support from BCKDF, the Canada Council, CANARIE, CRC, Compute Canada, FQRNT, and the Ontario Innovation Trust, Canada; EPLANET, ERC, FP7, Horizon 2020 and Marie Sk{\l}odowska-Curie Actions, European Union; Investissements d'Avenir Labex and Idex, ANR, R{\'e}gion Auvergne and Fondation Partager le Savoir, France; DFG and AvH Foundation, Germany; Herakleitos, Thales and Aristeia programmes co-financed by EU-ESF and the Greek NSRF; BSF, GIF and Minerva, Israel; BRF, Norway; the Royal Society and Leverhulme Trust, United Kingdom.

The crucial computing support from all WLCG partners is acknowledged
gratefully, in particular from CERN and the ATLAS Tier-1 facilities at
TRIUMF (Canada), NDGF (Denmark, Norway, Sweden), CC-IN2P3 (France),
KIT/GridKA (Germany), INFN-CNAF (Italy), NL-T1 (Netherlands), PIC (Spain),
ASGC (Taiwan), RAL (UK) and BNL (USA) and in the Tier-2 facilities
worldwide.

%% file: SpinAlignmentFactors.tex
\FloatBarrier
\section{Spin-alignment correction factors}
\label{sec:spincorrection}

The measurement presented here assumes an unpolarized spin-alignment hypothesis for determining the correction factor. In principle, the polarization may be non-zero and may vary with $\pt$.
In order to correct these measurements when well-measured $\jpsi$ and $\psiprime$ polarizations are determined, a set of correction factors are provided 
in Tables~\ref{tab:sa_long_jpsi}--\ref{tab:sa_offN_psi2s} for the 7~\TeV\ data, and in the Tables~\ref{tab:sa_long_jpsi8}--\ref{tab:sa_offN_psi2s8} for the 8~\TeV\ data.
These tables are created by altering the spin-alignment hypothesis for either the $\jpsi$ or $\psiprime$ meson and then determining the ratio of the mean sum-of-weights of the new hypotheses to the original flat hypothesis.
The mean weight is calculated from all the events in each dimuon $\pt$ and rapidity analysis bin, selecting those dimuons within $\pm2\sigma$ of the $\apsi$ fitted mean mass position.
The choice of spin-alignment hypothesis for each $\apsi$ meson has negligible effect on the results of the other $\apsi$ meson, and therefore these possible permutations are not considered.
The definitions of each of the spin-alignment scenarios, which are given in the caption to the table, are defined in Table~\ref{tab:spin}.

\begin{table}[htp]
             \caption{Mean weight correction factor for $\jpsi$ under the ``longitudinal'' spin-alignment hypothesis for 7 \TeV.} 
             \begin{tiny} 
             \begin{center} 
\begin{tabular}{|l|c|c|c|c|c|c|c|c|} 
 \hline 
 & \multicolumn{8}{c |}{Absolute Rapidity Range} \\ \hline
$\pt$~[\GeV] & $0.00$--$0.25$ & $0.25$--$0.50$ & $0.50$--$0.75$ & $0.75$--$1.00$ & $1.00$--$1.25$ & $1.25$--$1.50$ & $1.50$--$1.75$ & $1.75$--$2.00$ \\ \hline
$8.00$--$8.50$ & 0.666 & 0.672 & 0.674 & 0.680 & 0.688 & 0.690 & 0.690 & 0.690  \\
$8.50$--$9.00$ & 0.670 & 0.674 & 0.678 & 0.685 & 0.689 & 0.694 & 0.694 & 0.698  \\
$9.00$--$9.50$ & 0.673 & 0.676 & 0.680 & 0.687 & 0.693 & 0.697 & 0.698 & 0.700  \\
$9.50$--$10.00$ & 0.675 & 0.678 & 0.683 & 0.689 & 0.694 & 0.697 & 0.701 & 0.703  \\
$10.00$--$10.50$ & 0.679 & 0.681 & 0.687 & 0.692 & 0.697 & 0.699 & 0.702 & 0.706  \\
$10.50$--$11.00$ & 0.682 & 0.686 & 0.691 & 0.696 & 0.700 & 0.702 & 0.704 & 0.708  \\
\hline
\end{tabular} 
             \end{center} 
             \end{tiny} 
             \label{tab:sa_long_jpsi} 
             \end{table}

\begin{table}[htp]
             \caption{Mean weight correction factor for $\jpsi$ under the ``transverse zero'' spin-alignment hypothesis for 7 \TeV.} 
             \begin{tiny} 
             \begin{center} 
             \begin{tabular}{|l|c|c|c|c|c|c|c|c|} 
 \hline 
 & \multicolumn{8}{c |}{Absolute Rapidity Range} \\ \hline
$\pt$~[\GeV] & $0.00$--$0.25$ & $0.25$--$0.50$ & $0.50$--$0.75$ & $0.75$--$1.00$ & $1.00$--$1.25$ & $1.25$--$1.50$ & $1.50$--$1.75$ & $1.75$--$2.00$ \\ \hline
$8.00$--$8.50$ & 1.336 & 1.324 & 1.315 & 1.309 & 1.299 & 1.297 & 1.296 & 1.298  \\
$8.50$--$9.00$ & 1.329 & 1.323 & 1.310 & 1.300 & 1.291 & 1.284 & 1.280 & 1.284  \\
$9.00$--$9.50$ & 1.326 & 1.315 & 1.303 & 1.295 & 1.289 & 1.281 & 1.279 & 1.276  \\
$9.50$--$10.00$ & 1.317 & 1.311 & 1.300 & 1.289 & 1.284 & 1.276 & 1.276 & 1.272  \\
$10.00$--$10.50$ & 1.310 & 1.304 & 1.297 & 1.290 & 1.280 & 1.276 & 1.273 & 1.269  \\
$10.50$--$11.00$ & 1.302 & 1.298 & 1.291 & 1.285 & 1.276 & 1.271 & 1.267 & 1.268  \\
\hline
\end{tabular} 
             \end{center} 
             \end{tiny} 
             \label{tab:sa_trp0_jpsi} 
             \end{table}

\begin{table}[htp]
             \caption{Mean weight correction factor for $\jpsi$ under the ``transverse positive'' spin-alignment transverse positive hypothesis for 7 \TeV.} 
             \begin{tiny} 
             \begin{center} 
             \begin{tabular}{|l|c|c|c|c|c|c|c|c|} 
 \hline 
 & \multicolumn{8}{c |}{Absolute Rapidity Range} \\ \hline
$\pt$~[\GeV] & $0.00$--$0.25$ & $0.25$--$0.50$ & $0.50$--$0.75$ & $0.75$--$1.00$ & $1.00$--$1.25$ & $1.25$--$1.50$ & $1.50$--$1.75$ & $1.75$--$2.00$ \\ \hline
$8.00$--$8.50$ & 1.693 & 1.694 & 1.700 & 1.711 & 1.727 & 1.720 & 1.720 & 1.747  \\
$8.50$--$9.00$ & 1.561 & 1.564 & 1.564 & 1.568 & 1.568 & 1.568 & 1.571 & 1.673  \\
$9.00$--$9.50$ & 1.468 & 1.468 & 1.465 & 1.466 & 1.470 & 1.466 & 1.471 & 1.519  \\
$9.50$--$10.00$ & 1.418 & 1.416 & 1.417 & 1.417 & 1.421 & 1.417 & 1.423 & 1.453  \\
$10.00$--$10.50$ & 1.383 & 1.383 & 1.387 & 1.389 & 1.390 & 1.389 & 1.391 & 1.406  \\
$10.50$--$11.00$ & 1.360 & 1.362 & 1.365 & 1.364 & 1.364 & 1.362 & 1.362 & 1.380  \\
\hline
\end{tabular} 
             \end{center} 
             \end{tiny} 
             \label{tab:sa_trpp_jpsi} 
             \end{table}

\begin{table}[htp]
             \caption{Mean weight correction factor for $\jpsi$ under the ``transverse negative'' spin-alignment hypothesis for 7 \TeV.} 
             \begin{tiny} 
             \begin{center} 
             \begin{tabular}{|l|c|c|c|c|c|c|c|c|} 
 \hline 
 & \multicolumn{8}{c |}{Absolute Rapidity Range} \\ \hline
$\pt$~[\GeV] & $0.00$--$0.25$ & $0.25$--$0.50$ & $0.50$--$0.75$ & $0.75$--$1.00$ & $1.00$--$1.25$ & $1.25$--$1.50$ & $1.50$--$1.75$ & $1.75$--$2.00$ \\ \hline
$8.00$--$8.50$ & 1.030 & 1.020 & 1.004 & 0.995 & 0.992 & 0.981 & 0.973 & 0.949  \\
$8.50$--$9.00$ & 1.157 & 1.148 & 1.134 & 1.113 & 1.101 & 1.092 & 1.079 & 1.055  \\
$9.00$--$9.50$ & 1.207 & 1.196 & 1.176 & 1.161 & 1.147 & 1.138 & 1.130 & 1.107  \\
$9.50$--$10.00$ & 1.231 & 1.219 & 1.202 & 1.186 & 1.174 & 1.162 & 1.158 & 1.138  \\
$10.00$--$10.50$ & 1.243 & 1.231 & 1.217 & 1.202 & 1.189 & 1.181 & 1.175 & 1.158  \\
$10.50$--$11.00$ & 1.246 & 1.239 & 1.228 & 1.213 & 1.200 & 1.191 & 1.186 & 1.174  \\
\hline
\end{tabular} 
             \end{center} 
             \end{tiny} 
             \label{tab:sa_trpm_jpsi} 
             \end{table}

\begin{table}[htp]
             \caption{Mean weight correction factor for $\jpsi$ under the ``off-($\lambda_{\theta}$--$\lambda_{\phi}$)-plane positive'' spin-alignment hypothesis for 7 \TeV.} 
             \begin{tiny} 
             \begin{center} 
             \begin{tabular}{|l|c|c|c|c|c|c|c|c|} 
 \hline 
 & \multicolumn{8}{c |}{Absolute Rapidity Range} \\ \hline
$\pt$~[\GeV] & $0.00$--$0.25$ & $0.25$--$0.50$ & $0.50$--$0.75$ & $0.75$--$1.00$ & $1.00$--$1.25$ & $1.25$--$1.50$ & $1.50$--$1.75$ & $1.75$--$2.00$ \\ \hline
$8.00$--$8.50$ & 1.015 & 1.047 & 1.073 & 1.094 & 1.113 & 1.120 & 1.124 & 1.122  \\
$8.50$--$9.00$ & 1.020 & 1.058 & 1.087 & 1.110 & 1.125 & 1.134 & 1.142 & 1.144  \\
$9.00$--$9.50$ & 1.019 & 1.056 & 1.084 & 1.107 & 1.127 & 1.138 & 1.144 & 1.145  \\
$9.50$--$10.00$ & 1.017 & 1.053 & 1.081 & 1.105 & 1.122 & 1.129 & 1.140 & 1.142  \\
$10.00$--$10.50$ & 1.017 & 1.049 & 1.077 & 1.100 & 1.115 & 1.125 & 1.132 & 1.136  \\
$10.50$--$11.00$ & 1.014 & 1.048 & 1.075 & 1.095 & 1.109 & 1.118 & 1.124 & 1.130  \\
\hline
\end{tabular} 
             \end{center} 
             \end{tiny} 
             \label{tab:sa_offP_jpsi} 
             \end{table}

\begin{table}[htp]
             \caption{Mean weight correction factor for $\jpsi$ under the ``off-($\lambda_{\theta}$--$\lambda_{\phi}$)-plane negative'' spin-alignment hypothesis for 7 \TeV.} 
             \begin{tiny} 
             \begin{center} 
             \begin{tabular}{|l|c|c|c|c|c|c|c|c|} 
 \hline 
 & \multicolumn{8}{c |}{Absolute Rapidity Range} \\ \hline
$\pt$~[\GeV] & $0.00$--$0.25$ & $0.25$--$0.50$ & $0.50$--$0.75$ & $0.75$--$1.00$ & $1.00$--$1.25$ & $1.25$--$1.50$ & $1.50$--$1.75$ & $1.75$--$2.00$ \\ \hline
$8.00$--$8.50$ & 0.984 & 0.956 & 0.932 & 0.920 & 0.912 & 0.904 & 0.898 & 0.905  \\
$8.50$--$9.00$ & 0.981 & 0.948 & 0.925 & 0.910 & 0.898 & 0.892 & 0.886 & 0.891  \\
$9.00$--$9.50$ & 0.983 & 0.950 & 0.925 & 0.910 & 0.901 & 0.893 & 0.889 & 0.888  \\
$9.50$--$10.00$ & 0.983 & 0.951 & 0.929 & 0.912 & 0.903 & 0.894 & 0.892 & 0.891  \\
$10.00$--$10.50$ & 0.985 & 0.953 & 0.932 & 0.918 & 0.907 & 0.900 & 0.896 & 0.894  \\
$10.50$--$11.00$ & 0.984 & 0.957 & 0.936 & 0.922 & 0.910 & 0.904 & 0.900 & 0.899  \\
\hline
\end{tabular} 
             \end{center} 
             \end{tiny} 
             \label{tab:sa_offN_jpsi} 
             \end{table}

\begin{table}[htp]
             \caption{Mean weight correction factor for $\psiprime$ under the ``longitudinal'' spin-alignment hypothesis for 7 \TeV.} 
             \begin{tiny} 
             \begin{center} 
             \begin{tabular}{|l|c|c|c|c|c|c|c|c|} 
 \hline 
 & \multicolumn{8}{c |}{Absolute Rapidity Range} \\ \hline
$\pt$~[\GeV] & $0.00$--$0.25$ & $0.25$--$0.50$ & $0.50$--$0.75$ & $0.75$--$1.00$ & $1.00$--$1.25$ & $1.25$--$1.50$ & $1.50$--$1.75$ & $1.75$--$2.00$ \\ \hline
$8.00$--$8.50$ & 0.670 & 0.678 & 0.685 & 0.692 & 0.701 & 0.707 & 0.713 & 0.709  \\
$8.50$--$9.00$ & 0.676 & 0.681 & 0.688 & 0.698 & 0.703 & 0.709 & 0.712 & 0.713  \\
$9.00$--$9.50$ & 0.678 & 0.683 & 0.691 & 0.700 & 0.708 & 0.713 & 0.717 & 0.718  \\
$9.50$--$10.00$ & 0.680 & 0.684 & 0.693 & 0.699 & 0.708 & 0.710 & 0.720 & 0.722  \\
$10.00$--$10.50$ & 0.684 & 0.687 & 0.695 & 0.704 & 0.707 & 0.713 & 0.720 & 0.725  \\
$10.50$--$11.00$ & 0.687 & 0.691 & 0.698 & 0.705 & 0.712 & 0.714 & 0.719 & 0.728  \\
\hline
\end{tabular} 
             \end{center} 
             \end{tiny} 
             \label{tab:sa_long_psi2s} 
             \end{table}

\begin{table}[htp]
             \caption{Mean weight correction factor for $\psiprime$ under the ``transverse zero'' spin-alignment hypothesis for 7 \TeV.} 
             \begin{tiny} 
             \begin{center} 
             \begin{tabular}{|l|c|c|c|c|c|c|c|c|} 
 \hline 
 & \multicolumn{8}{c |}{Absolute Rapidity Range} \\ \hline
$\pt$~[\GeV] & $0.00$--$0.25$ & $0.25$--$0.50$ & $0.50$--$0.75$ & $0.75$--$1.00$ & $1.00$--$1.25$ & $1.25$--$1.50$ & $1.50$--$1.75$ & $1.75$--$2.00$ \\ \hline
$8.00$--$8.50$ & 1.328 & 1.311 & 1.300 & 1.284 & 1.274 & 1.267 & 1.261 & 1.265  \\
$8.50$--$9.00$ & 1.318 & 1.309 & 1.293 & 1.279 & 1.268 & 1.263 & 1.252 & 1.259  \\
$9.00$--$9.50$ & 1.317 & 1.303 & 1.287 & 1.273 & 1.267 & 1.256 & 1.249 & 1.250  \\
$9.50$--$10.00$ & 1.310 & 1.301 & 1.286 & 1.275 & 1.262 & 1.255 & 1.248 & 1.247  \\
$10.00$--$10.50$ & 1.303 & 1.294 & 1.283 & 1.271 & 1.265 & 1.257 & 1.248 & 1.243  \\
$10.50$--$11.00$ & 1.295 & 1.289 & 1.279 & 1.271 & 1.259 & 1.254 & 1.246 & 1.240  \\
\hline
\end{tabular} 
             \end{center} 
             \end{tiny} 
             \label{tab:sa_trp0_psi2s} 
             \end{table}

\begin{table}[htp]
             \caption{Mean weight correction factor for $\psiprime$ under the ``transverse positive'' spin-alignment hypothesis for 7 \TeV.} 
             \begin{tiny} 
             \begin{center} 
             \begin{tabular}{|l|c|c|c|c|c|c|c|c|} 
 \hline 
 & \multicolumn{8}{c |}{Absolute Rapidity Range} \\ \hline
$\pt$~[\GeV] & $0.00$--$0.25$ & $0.25$--$0.50$ & $0.50$--$0.75$ & $0.75$--$1.00$ & $1.00$--$1.25$ & $1.25$--$1.50$ & $1.50$--$1.75$ & $1.75$--$2.00$ \\ \hline
$8.00$--$8.50$ & 2.009 & 2.007 & 1.986 & 1.994 & 1.964 & 1.936 & 1.949 & 1.967  \\
$8.50$--$9.00$ & 1.614 & 1.617 & 1.617 & 1.613 & 1.618 & 1.624 & 1.606 & 1.872  \\
$9.00$--$9.50$ & 1.504 & 1.502 & 1.496 & 1.493 & 1.500 & 1.499 & 1.494 & 1.741  \\
$9.50$--$10.00$ & 1.445 & 1.443 & 1.440 & 1.445 & 1.440 & 1.441 & 1.436 & 1.621  \\
$10.00$--$10.50$ & 1.404 & 1.401 & 1.403 & 1.401 & 1.412 & 1.406 & 1.400 & 1.507  \\
$10.50$--$11.00$ & 1.374 & 1.377 & 1.378 & 1.378 & 1.377 & 1.378 & 1.372 & 1.447  \\
\hline
\end{tabular} 
             \end{center} 
             \end{tiny} 
             \label{tab:sa_trpp_psi2s} 
             \end{table}

\begin{table}[htp]
             \caption{Mean weight correction factor for $\psiprime$ under the ``transverse negative'' spin-alignment hypothesis for 7 \TeV.} 
             \begin{tiny} 
             \begin{center} 
             \begin{tabular}{|l|c|c|c|c|c|c|c|c|} 
 \hline 
 & \multicolumn{8}{c |}{Absolute Rapidity Range} \\ \hline
$\pt$~[\GeV] & $0.00$--$0.25$ & $0.25$--$0.50$ & $0.50$--$0.75$ & $0.75$--$1.00$ & $1.00$--$1.25$ & $1.25$--$1.50$ & $1.50$--$1.75$ & $1.75$--$2.00$ \\ \hline
$8.00$--$8.50$ & 0.998 & 0.986 & 0.970 & 0.957 & 0.949 & 0.941 & 0.935 & 0.883  \\
$8.50$--$9.00$ & 1.115 & 1.102 & 1.084 & 1.062 & 1.047 & 1.039 & 1.025 & 0.959  \\
$9.00$--$9.50$ & 1.169 & 1.154 & 1.131 & 1.110 & 1.096 & 1.084 & 1.075 & 1.007  \\
$9.50$--$10.00$ & 1.200 & 1.185 & 1.163 & 1.144 & 1.126 & 1.114 & 1.105 & 1.047  \\
$10.00$--$10.50$ & 1.216 & 1.200 & 1.181 & 1.161 & 1.148 & 1.137 & 1.127 & 1.075  \\
$10.50$--$11.00$ & 1.222 & 1.212 & 1.196 & 1.178 & 1.161 & 1.152 & 1.143 & 1.097  \\
\hline
\end{tabular} 
             \end{center} 
             \end{tiny} 
             \label{tab:sa_trpm_psi2s} 
             \end{table}

\begin{table}[htp]
             \caption{Mean weight correction factor for $\psiprime$ under the ``off-($\lambda_{\theta}$--$\lambda_{\phi}$)-plane positive'' spin-alignment hypothesis for 7 \TeV.} 
             \begin{tiny} 
             \begin{center} 
             \begin{tabular}{|l|c|c|c|c|c|c|c|c|} 
 \hline 
 & \multicolumn{8}{c |}{Absolute Rapidity Range} \\ \hline
$\pt$~[\GeV] & $0.00$--$0.25$ & $0.25$--$0.50$ & $0.50$--$0.75$ & $0.75$--$1.00$ & $1.00$--$1.25$ & $1.25$--$1.50$ & $1.50$--$1.75$ & $1.75$--$2.00$ \\ \hline
$8.00$--$8.50$ & 1.017 & 1.052 & 1.081 & 1.100 & 1.118 & 1.123 & 1.129 & 1.106  \\
$8.50$--$9.00$ & 1.023 & 1.064 & 1.094 & 1.118 & 1.136 & 1.146 & 1.151 & 1.132  \\
$9.00$--$9.50$ & 1.021 & 1.062 & 1.093 & 1.119 & 1.140 & 1.150 & 1.153 & 1.139  \\
$9.50$--$10.00$ & 1.019 & 1.060 & 1.092 & 1.119 & 1.135 & 1.144 & 1.152 & 1.146  \\
$10.00$--$10.50$ & 1.020 & 1.057 & 1.088 & 1.112 & 1.132 & 1.140 & 1.146 & 1.145  \\
$10.50$--$11.00$ & 1.017 & 1.055 & 1.085 & 1.108 & 1.124 & 1.134 & 1.139 & 1.141  \\
\hline
\end{tabular} 
             \end{center} 
             \end{tiny} 
             \label{tab:sa_offP_psi2s} 
             \end{table}

\begin{table}[htp]
             \caption{Mean weight correction factor for $\psiprime$ under the ``off-($\lambda_{\theta}$--$\lambda_{\phi}$)-plane negative'' spin-alignment hypothesis for 7 \TeV.} 
             \begin{tiny} 
             \begin{center} 
             \begin{tabular}{|l|c|c|c|c|c|c|c|c|} 
 \hline 
 & \multicolumn{8}{c |}{Absolute Rapidity Range} \\ \hline
$\pt$~[\GeV] & $0.00$--$0.25$ & $0.25$--$0.50$ & $0.50$--$0.75$ & $0.75$--$1.00$ & $1.00$--$1.25$ & $1.25$--$1.50$ & $1.50$--$1.75$ & $1.75$--$2.00$ \\ \hline
$8.00$--$8.50$ & 0.983 & 0.950 & 0.931 & 0.916 & 0.908 & 0.902 & 0.902 & 0.911  \\
$8.50$--$9.00$ & 0.979 & 0.944 & 0.919 & 0.904 & 0.892 & 0.887 & 0.882 & 0.898  \\
$9.00$--$9.50$ & 0.981 & 0.943 & 0.919 & 0.901 & 0.894 & 0.886 & 0.883 & 0.891  \\
$9.50$--$10.00$ & 0.981 & 0.945 & 0.922 & 0.903 & 0.894 & 0.886 & 0.885 & 0.891  \\
$10.00$--$10.50$ & 0.982 & 0.948 & 0.925 & 0.910 & 0.897 & 0.891 & 0.888 & 0.890  \\
$10.50$--$11.00$ & 0.982 & 0.951 & 0.929 & 0.913 & 0.901 & 0.895 & 0.891 & 0.892  \\
\hline
\end{tabular} 
             \end{center} 
             \end{tiny} 
             \label{tab:sa_offN_psi2s} 
             \end{table}

\clearpage

\begin{table}[htp]
             \caption{Mean weight correction factor for $\jpsi$ under the ``longitudinal'' spin-alignment hypothesis for 8 \TeV.
             Those intervals not measured in the analysis at low $\pt$, high rapidity are also excluded here. } 
             \begin{tiny} 
             \begin{center} 
             \begin{tabular}{|l|c|c|c|c|c|c|c|c|} 
 \hline 
 & \multicolumn{8}{c |}{Absolute Rapidity Range} \\ \hline
$\pt$~[\GeV] & $0.00$--$0.25$ & $0.25$--$0.50$ & $0.50$--$0.75$ & $0.75$--$1.00$ & $1.00$--$1.25$ & $1.25$--$1.50$ & $1.50$--$1.75$ & $1.75$--$2.00$ \\ \hline
       8.00 --  8.50 &  0.672 & 0.674 & 0.678  & -- & -- & -- & -- & --  \\ 
       8.50 --  9.00 &  0.670 & 0.673 & 0.678  & -- & -- & -- & -- & --  \\ 
       9.00 --  9.50 &  0.671 & 0.674 & 0.679  & -- & -- & -- & -- & --  \\ 
       9.50 -- 10.00 &  0.674 & 0.676 & 0.681  & -- & -- & -- & -- & --  \\ 
      10.00 -- 10.50 &  0.676 & 0.678 & 0.683 & 0.686 & 0.691 & 0.694 & 0.695 & 0.696 \\ 
      10.50 -- 11.00 &  0.680 & 0.681 & 0.686 & 0.689 & 0.693 & 0.696 & 0.697 & 0.698 \\
\hline
\end{tabular} 
             \end{center} 
             \end{tiny} 
             \label{tab:sa_long_jpsi8} 
             \end{table}

\begin{table}[htp]
             \caption{Mean weight correction factor for $\jpsi$ under the ``transverse zero'' spin-alignment hypothesis for 8 \TeV.
             Those intervals not measured in the analysis at low $\pt$, high rapidity are also excluded here.} 
             \begin{tiny} 
             \begin{center} 
             \begin{tabular}{|l|c|c|c|c|c|c|c|c|} 
 \hline 
 & \multicolumn{8}{c |}{Absolute Rapidity Range} \\ \hline
$\pt$~[\GeV] & $0.00$--$0.25$ & $0.25$--$0.50$ & $0.50$--$0.75$ & $0.75$--$1.00$ & $1.00$--$1.25$ & $1.25$--$1.50$ & $1.50$--$1.75$ & $1.75$--$2.00$ \\ \hline
       8.00 --  8.50 &  1.326 & 1.321 & 1.311  & -- & -- & -- & -- & --  \\ 
       8.50 --  9.00 &  1.326 & 1.320 & 1.309  & -- & -- & -- & -- & --  \\ 
       9.00 --  9.50 &  1.322 & 1.316 & 1.306  & -- & -- & -- & -- & --  \\ 
       9.50 -- 10.00 &  1.317 & 1.312 & 1.302  & -- & -- & -- & -- & --  \\ 
      10.00 -- 10.50 &  1.311 & 1.306 & 1.297 & 1.291 & 1.283 & 1.278 & 1.275 & 1.273 \\ 
      10.50 -- 11.00 &  1.304 & 1.300 & 1.292 & 1.286 & 1.279 & 1.274 & 1.272 & 1.269 \\
\hline
\end{tabular} 
             \end{center} 
             \end{tiny} 
             \label{tab:sa_trp0_jpsi8} 
             \end{table}

\begin{table}[htp]
             \caption{Mean weight correction factor for $\jpsi$ under the ``transverse positive'' spin-alignment hypothesis for 8 \TeV.
             Those intervals not measured in the analysis at low $\pt$, high rapidity are also excluded here.} 
             \begin{tiny} 
             \begin{center} 
             \begin{tabular}{|l|c|c|c|c|c|c|c|c|} 
 \hline 
 & \multicolumn{8}{c |}{Absolute Rapidity Range} \\ \hline
$\pt$~[\GeV] & $0.00$--$0.25$ & $0.25$--$0.50$ & $0.50$--$0.75$ & $0.75$--$1.00$ & $1.00$--$1.25$ & $1.25$--$1.50$ & $1.50$--$1.75$ & $1.75$--$2.00$ \\ \hline
       8.00 --  8.50 &  1.926 & 1.933 & 1.930   & -- & -- & -- & -- & --  \\ 
       8.50 --  9.00 &  1.555 & 1.558 & 1.559   & -- & -- & -- & -- & --  \\ 
       9.00 --  9.50 &  1.463 & 1.464 & 1.465   & -- & -- & -- & -- & --  \\ 
       9.50 -- 10.00 &  1.416 & 1.418 & 1.418   & -- & -- & -- & -- & --  \\ 
      10.00 -- 10.50 &  1.386 & 1.388 & 1.387 & 1.390 & 1.390 & 1.390 & 1.391 & 1.411 \\ 
      10.50 -- 11.00 &  1.363 & 1.365 & 1.365 & 1.367 & 1.367 & 1.366 & 1.368 & 1.382 \\
\hline
\end{tabular} 
             \end{center} 
             \end{tiny} 
             \label{tab:sa_trpp_jpsi8} 
             \end{table}

\begin{table}[htp]
             \caption{Mean weight correction factor for $\jpsi$ under the ``transverse negative'' spin-alignment hypothesis for 8 \TeV.
             Those intervals not measured in the analysis at low $\pt$, high rapidity are also excluded here.} 
             \begin{tiny} 
             \begin{center} 
             \begin{tabular}{|l|c|c|c|c|c|c|c|c|} 
 \hline 
 & \multicolumn{8}{c |}{Absolute Rapidity Range} \\ \hline
$\pt$~[\GeV] & $0.00$--$0.25$ & $0.25$--$0.50$ & $0.50$--$0.75$ & $0.75$--$1.00$ & $1.00$--$1.25$ & $1.25$--$1.50$ & $1.50$--$1.75$ & $1.75$--$2.00$ \\ \hline
       8.00 --  8.50 &  1.026 & 1.017 & 1.005  & -- & -- & -- & -- & --  \\ 
       8.50 --  9.00 &  1.157 & 1.145 & 1.129  & -- & -- & -- & -- & --  \\ 
       9.00 --  9.50 &  1.207 & 1.196 & 1.178  & -- & -- & -- & -- & --  \\ 
       9.50 -- 10.00 &  1.231 & 1.220 & 1.203  & -- & -- & -- & -- & --  \\ 
      10.00 -- 10.50 &  1.244 & 1.234 & 1.218 & 1.204 & 1.192 & 1.182 & 1.177 & 1.161 \\ 
      10.50 -- 11.00 &  1.250 & 1.241 & 1.227 & 1.214 & 1.202 & 1.193 & 1.188 & 1.175 \\
\hline
\end{tabular} 
             \end{center} 
             \end{tiny} 
             \label{tab:sa_trpm_jpsi8} 
             \end{table}

\begin{table}[htp]
             \caption{Mean weight correction factor for $\jpsi$ under the ``off-($\lambda_{\theta}$--$\lambda_{\phi}$)-plane positive'' spin-alignment hypothesis for 8 \TeV.
             Those intervals not measured in the analysis at low $\pt$, high rapidity are also excluded here.} 
             \begin{tiny} 
             \begin{center} 
             \begin{tabular}{|l|c|c|c|c|c|c|c|c|} 
 \hline 
 & \multicolumn{8}{c |}{Absolute Rapidity Range} \\ \hline
$\pt$~[\GeV] & $0.00$--$0.25$ & $0.25$--$0.50$ & $0.50$--$0.75$ & $0.75$--$1.00$ & $1.00$--$1.25$ & $1.25$--$1.50$ & $1.50$--$1.75$ & $1.75$--$2.00$ \\ \hline
       8.00 --  8.50 &  1.016 & 1.048 & 1.074  & -- & -- & -- & -- & --  \\ 
       8.50 --  9.00 &  1.019 & 1.056 & 1.087  & -- & -- & -- & -- & --  \\ 
       9.00 --  9.50 &  1.019 & 1.055 & 1.086  & -- & -- & -- & -- & --  \\ 
       9.50 -- 10.00 &  1.018 & 1.053 & 1.083  & -- & -- & -- & -- & --  \\ 
      10.00 -- 10.50 &  1.017 & 1.051 & 1.079 & 1.101 & 1.117 & 1.127 & 1.134 & 1.138 \\ 
      10.50 -- 11.00 &  1.016 & 1.048 & 1.075 & 1.096 & 1.110 & 1.120 & 1.126 & 1.131 \\
\hline
\end{tabular} 
             \end{center} 
             \end{tiny} 
             \label{tab:sa_offP_jpsi8} 
             \end{table}

\begin{table}[htp]
             \caption{Mean weight correction factor for $\jpsi$ under the ``off-($\lambda_{\theta}$--$\lambda_{\phi}$)-plane negative'' spin-alignment hypothesis for 8 \TeV.
             Those intervals not measured in the analysis at low $\pt$, high rapidity are also excluded here.} 
             \begin{tiny} 
             \begin{center} 
             \begin{tabular}{|l|c|c|c|c|c|c|c|c|} 
 \hline 
 & \multicolumn{8}{c |}{Absolute Rapidity Range} \\ \hline
$\pt$~[\GeV] & $0.00$--$0.25$ & $0.25$--$0.50$ & $0.50$--$0.75$ & $0.75$--$1.00$ & $1.00$--$1.25$ & $1.25$--$1.50$ & $1.50$--$1.75$ & $1.75$--$2.00$ \\ \hline
       8.00 --  8.50 &  0.985 & 0.957 & 0.936  & -- & -- & -- & -- & --  \\ 
       8.50 --  9.00 &  0.982 & 0.950 & 0.926  & -- & -- & -- & -- & --  \\ 
       9.00 --  9.50 &  0.982 & 0.950 & 0.926  & -- & -- & -- & -- & --  \\ 
       9.50 -- 10.00 &  0.983 & 0.952 & 0.929  & -- & -- & -- & -- & --  \\ 
      10.00 -- 10.50 &  0.984 & 0.954 & 0.932 & 0.916 & 0.905 & 0.898 & 0.894 & 0.891 \\ 
      10.50 -- 11.00 &  0.985 & 0.956 & 0.935 & 0.919 & 0.909 & 0.903 & 0.899 & 0.895 \\
\hline
\end{tabular} 
             \end{center} 
             \end{tiny} 
             \label{tab:sa_offN_jpsi8} 
             \end{table}

\begin{table}[htp]
             \caption{Mean weight correction factor for $\psiprime$ under the ``longitudinal'' spin-alignment hypothesis for 8 \TeV.
             Those intervals not measured in the analysis at low $\pt$, high rapidity are also excluded here.} 
             \begin{tiny} 
             \begin{center} 
             \begin{tabular}{|l|c|c|c|c|c|c|c|c|} 
 \hline 
 & \multicolumn{8}{c |}{Absolute Rapidity Range} \\ \hline
$\pt$~[\GeV] & $0.00$--$0.25$ & $0.25$--$0.50$ & $0.50$--$0.75$ & $0.75$--$1.00$ & $1.00$--$1.25$ & $1.25$--$1.50$ & $1.50$--$1.75$ & $1.75$--$2.00$ \\ \hline
       8.00 --  8.50 &  0.672 & 0.677 & 0.686  & -- & -- & -- & -- & --  \\ 
       8.50 --  9.00 &  0.674 & 0.680 & 0.689  & -- & -- & -- & -- & --  \\ 
       9.00 --  9.50 &  0.677 & 0.682 & 0.691  & -- & -- & -- & -- & --  \\ 
       9.50 -- 10.00 &  0.680 & 0.684 & 0.692  & -- & -- & -- & -- & --  \\ 
      10.00 -- 10.50 &  0.683 & 0.688 & 0.695 & 0.702 & 0.709 & 0.713 & 0.717 & 0.721 \\ 
      10.50 -- 11.00 &  0.687 & 0.692 & 0.698 & 0.705 & 0.710 & 0.715 & 0.718 & 0.722 \\
\hline
\end{tabular} 
             \end{center} 
             \end{tiny} 
             \label{tab:sa_long_psi2s8} 
             \end{table}

\begin{table}[htp]
             \caption{Mean weight correction factor for $\psiprime$ under the ``transverse zero'' spin-alignment hypothesis for 8 \TeV.
             Those intervals not measured in the analysis at low $\pt$, high rapidity are also excluded here.} 
             \begin{tiny} 
             \begin{center} 
             \begin{tabular}{|l|c|c|c|c|c|c|c|c|} 
 \hline 
 & \multicolumn{8}{c |}{Absolute Rapidity Range} \\ \hline
$\pt$~[\GeV] & $0.00$--$0.25$ & $0.25$--$0.50$ & $0.50$--$0.75$ & $0.75$--$1.00$ & $1.00$--$1.25$ & $1.25$--$1.50$ & $1.50$--$1.75$ & $1.75$--$2.00$ \\ \hline
       8.00 --  8.50 &  1.325 & 1.316 & 1.301  & -- & -- & -- & -- & --  \\ 
       8.50 --  9.00 &  1.321 & 1.311 & 1.295  & -- & -- & -- & -- & --  \\ 
       9.00 --  9.50 &  1.316 & 1.307 & 1.291  & -- & -- & -- & -- & --  \\ 
       9.50 -- 10.00 &  1.310 & 1.301 & 1.288  & -- & -- & -- & -- & --  \\ 
      10.00 -- 10.50 &  1.303 & 1.295 & 1.283 & 1.272 & 1.261 & 1.254 & 1.249 & 1.244 \\ 
      10.50 -- 11.00 &  1.296 & 1.289 & 1.278 & 1.267 & 1.259 & 1.252 & 1.247 & 1.241 \\
\hline
\end{tabular} 
             \end{center} 
             \end{tiny} 
             \label{tab:sa_trp0_psi2s8} 
             \end{table}

\begin{table}[htp]
             \caption{Mean weight correction factor for $\psiprime$ under the ``transverse positive'' spin-alignment hypothesis for 8 \TeV.
             Those intervals not measured in the analysis at low $\pt$, high rapidity are also excluded here.} 
             \begin{tiny} 
             \begin{center} 
             \begin{tabular}{|l|c|c|c|c|c|c|c|c|} 
 \hline 
 & \multicolumn{8}{c |}{Absolute Rapidity Range} \\ \hline
$\pt$~[\GeV] & $0.00$--$0.25$ & $0.25$--$0.50$ & $0.50$--$0.75$ & $0.75$--$1.00$ & $1.00$--$1.25$ & $1.25$--$1.50$ & $1.50$--$1.75$ & $1.75$--$2.00$ \\ \hline
       8.00 --  8.50 &  2.029 & 2.023 & 2.022  & -- & -- & -- & -- & --  \\ 
       8.50 --  9.00 &  1.620 & 1.620 & 1.618  & -- & -- & -- & -- & --  \\ 
       9.00 --  9.50 &  1.504 & 1.504 & 1.502  & -- & -- & -- & -- & --  \\ 
       9.50 -- 10.00 &  1.444 & 1.444 & 1.443  & -- & -- & -- & -- & --  \\ 
      10.00 -- 10.50 &  1.405 & 1.405 & 1.404 & 1.404 & 1.402 & 1.401 & 1.400 & 1.500 \\ 
      10.50 -- 11.00 &  1.377 & 1.377 & 1.376 & 1.375 & 1.375 & 1.373 & 1.373 & 1.443 \\
\hline
\end{tabular} 
             \end{center} 
             \end{tiny} 
             \label{tab:sa_trpp_psi2s8} 
             \end{table}

\begin{table}[htp]
             \caption{Mean weight correction factor for $\psiprime$ under the ``transverse negative'' spin-alignment hypothesis for 8 \TeV.
             Those intervals not measured in the analysis at low $\pt$, high rapidity are also excluded here.} 
             \begin{tiny} 
             \begin{center} 
             \begin{tabular}{|l|c|c|c|c|c|c|c|c|} 
 \hline 
 & \multicolumn{8}{c |}{Absolute Rapidity Range} \\ \hline
$\pt$~[\GeV] & $0.00$--$0.25$ & $0.25$--$0.50$ & $0.50$--$0.75$ & $0.75$--$1.00$ & $1.00$--$1.25$ & $1.25$--$1.50$ & $1.50$--$1.75$ & $1.75$--$2.00$ \\ \hline
       8.00 --  8.50 &  0.995 & 0.986 & 0.970   & -- & -- & -- & -- & --  \\ 
       8.50 --  9.00 &  1.116 & 1.102 & 1.081   & -- & -- & -- & -- & --  \\ 
       9.00 --  9.50 &  1.170 & 1.156 & 1.133   & -- & -- & -- & -- & --  \\ 
       9.50 -- 10.00 &  1.199 & 1.185 & 1.163   & -- & -- & -- & -- & --  \\ 
      10.00 -- 10.50 &  1.215 & 1.202 & 1.182 & 1.163 & 1.146 & 1.135 & 1.127 & 1.075 \\ 
      10.50 -- 11.00 &  1.225 & 1.212 & 1.194 & 1.175 & 1.161 & 1.150 & 1.142 & 1.098 \\
\hline
\end{tabular} 
             \end{center} 
             \end{tiny} 
             \label{tab:sa_trpm_psi2s8} 
             \end{table}

\begin{table}[htp]
             \caption{Mean weight correction factor for $\psiprime$ under the ``off-($\lambda_{\theta}$--$\lambda_{\phi}$)-plane positive'' spin-alignment hypothesis for 8 \TeV.
             Those intervals not measured in the analysis at low $\pt$, high rapidity are also excluded here.} 
             \begin{tiny} 
             \begin{center} 
             \begin{tabular}{|l|c|c|c|c|c|c|c|c|} 
 \hline 
 & \multicolumn{8}{c |}{Absolute Rapidity Range} \\ \hline
$\pt$~[\GeV] & $0.00$--$0.25$ & $0.25$--$0.50$ & $0.50$--$0.75$ & $0.75$--$1.00$ & $1.00$--$1.25$ & $1.25$--$1.50$ & $1.50$--$1.75$ & $1.75$--$2.00$ \\ \hline
       8.00 --  8.50 &  1.018 & 1.053 & 1.081  & -- & -- & -- & -- & --  \\ 
       8.50 --  9.00 &  1.021 & 1.062 & 1.095  & -- & -- & -- & -- & --  \\ 
       9.00 --  9.50 &  1.021 & 1.062 & 1.096  & -- & -- & -- & -- & --  \\ 
       9.50 -- 10.00 &  1.020 & 1.060 & 1.094  & -- & -- & -- & -- & --  \\ 
      10.00 -- 10.50 &  1.020 & 1.057 & 1.089 & 1.114 & 1.130 & 1.140 & 1.146 & 1.145 \\ 
      10.50 -- 11.00 &  1.018 & 1.055 & 1.085 & 1.108 & 1.124 & 1.133 & 1.139 & 1.142 \\
\hline
\end{tabular} 
             \end{center} 
             \end{tiny} 
             \label{tab:sa_offP_psi2s8} 
             \end{table}

\begin{table}[htp]
             \caption{Mean weight correction factor for $\psiprime$ under the ``off-($\lambda_{\theta}$--$\lambda_{\phi}$)-plane negative'' spin-alignment hypothesis for 8 \TeV.
             Those intervals not measured in the analysis at low $\pt$, high rapidity are also excluded here.} 
             \begin{tiny} 
             \begin{center} 
             \begin{tabular}{|l|c|c|c|c|c|c|c|c|} 
 \hline 
 & \multicolumn{8}{c |}{Absolute Rapidity Range} \\ \hline
$\pt$~[\GeV] & $0.00$--$0.25$ & $0.25$--$0.50$ & $0.50$--$0.75$ & $0.75$--$1.00$ & $1.00$--$1.25$ & $1.25$--$1.50$ & $1.50$--$1.75$ & $1.75$--$2.00$ \\ \hline
       8.00 --  8.50 &  0.983 & 0.952 & 0.931  & -- & -- & -- & -- & --  \\ 
       8.50 --  9.00 &  0.980 & 0.945 & 0.920  & -- & -- & -- & -- & --  \\ 
       9.00 --  9.50 &  0.980 & 0.945 & 0.919  & -- & -- & -- & -- & --  \\ 
       9.50 -- 10.00 &  0.981 & 0.946 & 0.921  & -- & -- & -- & -- & --  \\ 
      10.00 -- 10.50 &  0.981 & 0.949 & 0.924 & 0.908 & 0.897 & 0.891 & 0.887 & 0.888 \\ 
      10.50 -- 11.00 &  0.982 & 0.951 & 0.928 & 0.912 & 0.901 & 0.895 & 0.891 & 0.890 \\
\hline
\end{tabular} 
             \end{center} 
             \end{tiny} 
             \label{tab:sa_offN_psi2s8} 
             \end{table}

\clearpage

%% file: atlas_authlist.tex
\begin{flushleft}
{\Large The ATLAS Collaboration}

\bigskip

G.~Aad$^\textrm{\scriptsize 87}$,
B.~Abbott$^\textrm{\scriptsize 115}$,
J.~Abdallah$^\textrm{\scriptsize 153}$,
O.~Abdinov$^\textrm{\scriptsize 11}$,
R.~Aben$^\textrm{\scriptsize 109}$,
M.~Abolins$^\textrm{\scriptsize 92}$,
O.S.~AbouZeid$^\textrm{\scriptsize 160}$,
H.~Abramowicz$^\textrm{\scriptsize 155}$,
H.~Abreu$^\textrm{\scriptsize 154}$,
R.~Abreu$^\textrm{\scriptsize 118}$
% ...
\bigskip
\\
% ...
$^{11}$ Institute of Physics, Azerbaijan Academy of Sciences, Baku, Azerbaijan\\
% ...
$^{87}$ CPPM, Aix-Marseille Universit{\'e} and CNRS/IN2P3, Marseille, France\\
% ...
$^{92}$ Department of Physics and Astronomy, Michigan State University, East Lansing MI, United States of America\\
% ...
$^{109}$ Nikhef National Institute for Subatomic Physics and University of Amsterdam, Amsterdam, Netherlands\\
% ...
$^{115}$ Homer L. Dodge Department of Physics and Astronomy, University of Oklahoma, Norman OK, United States of America\\
% ...
$^{118}$ Center for High Energy Physics, University of Oregon, Eugene OR, United States of America\\
% ...
$^{153}$ Institute of Physics, Academia Sinica, Taipei, Taiwan\\
$^{154}$ Department of Physics, Technion: Israel Institute of Technology, Haifa, Israel\\
$^{155}$ Raymond and Beverly Sackler School of Physics and Astronomy, Tel Aviv University, Tel Aviv, Israel\\
% ...
$^{160}$ Department of Physics, University of Toronto, Toronto ON, Canada\\
% ...
\end{flushleft}
